\pdfoutput=1
\documentclass{mynature}

\usepackage{chapterbib}
\usepackage{amsmath}
\usepackage{amssymb}

\usepackage{graphics} 
\usepackage{graphicx} 
\usepackage{times} 
\usepackage{mathptmx}
\usepackage{typearea}

\typearea{14}

\onecolumn

\usepackage{setspace}

\def\del#1{{}}

\renewcommand\refname{References}
\renewcommand{\thefootnote}{\fnsymbol{footnote}}

\SetSymbolFont{symbols}{bold}{OMS}{cmsy}{b}{n}
\DeclareSymbolFont{bmisymbols}{OML}{cmm}{b}{it}
\DeclareMathSymbol{\balpha}{0}{bmisymbols}{"0B}
\DeclareMathSymbol{\bbeta}{0}{bmisymbols}{"0C}
\DeclareMathSymbol{\bgamma}{0}{bmisymbols}{"0D}
\DeclareMathSymbol{\bdelta}{0}{bmisymbols}{"0E}
\DeclareMathSymbol{\bepsilon}{0}{bmisymbols}{"0F}
\DeclareMathSymbol{\bzeta}{0}{bmisymbols}{"10}
\DeclareMathSymbol{\boldeta}{0}{bmisymbols}{"11}
\DeclareMathSymbol{\btheta}{0}{bmisymbols}{"12}
\DeclareMathSymbol{\biota}{0}{bmisymbols}{"13}
\DeclareMathSymbol{\bkappa}{0}{bmisymbols}{"14}
\DeclareMathSymbol{\blambda}{0}{bmisymbols}{"15}
\DeclareMathSymbol{\bmu}{0}{bmisymbols}{"16}
\DeclareMathSymbol{\bnu}{0}{bmisymbols}{"17}
\DeclareMathSymbol{\bxi}{0}{bmisymbols}{"18}
\DeclareMathSymbol{\bpi}{0}{bmisymbols}{"19}
\DeclareMathSymbol{\brho}{0}{bmisymbols}{"1A}
\DeclareMathSymbol{\bsigma}{0}{bmisymbols}{"1B}
\DeclareMathSymbol{\btau}{0}{bmisymbols}{"1C}
\DeclareMathSymbol{\bupsilon}{0}{bmisymbols}{"1D}
\DeclareMathSymbol{\bphi}{0}{bmisymbols}{"1E}
\DeclareMathSymbol{\bchi}{0}{bmisymbols}{"1F}
\DeclareMathSymbol{\bpsi}{0}{bmisymbols}{"20}
\DeclareMathSymbol{\bomega}{0}{bmisymbols}{"21}
\DeclareMathSymbol{\bvarepsilon}{0}{bmisymbols}{"22}
\DeclareMathSymbol{\bvartheta}{0}{bmisymbols}{"23}
\DeclareMathSymbol{\bvarpi}{0}{bmisymbols}{"24}
\DeclareMathSymbol{\bvarrho}{0}{bmisymbols}{"25}
\DeclareMathSymbol{\bvarsigma}{0}{bmisymbols}{"26}
\DeclareMathSymbol{\bvarphi}{0}{bmisymbols}{"27}

\newcommand{\mathbfit}[1]{\textbf{\textit{#1}}}

\newcommand{\rmn}{\mathrm}
\newcommand{\dd}{\mathrm{d}}

\newcommand{\bl}{\begin{large}}
\newcommand{\el}{\end{large}}

\newcommand{\M}{{\mathcal M}}

\newcommand{\e}{\mathrm{e}}

\newcommand{\eps}{\varepsilon}
\newcommand{\vecbf}{\mathbfit}
\newcommand{\vel}{\upsilon}

\newcommand {\apgt} {\ {\raise-.5ex\hbox{$\buildrel>\over\sim$}}\ }
\newcommand {\aplt} {\ {\raise-.5ex\hbox{$\buildrel<\over\sim$}}\ } 

\usepackage{sectsty}

\begin{document}


\bibliographystyle{naturemag}

\onecolumn
\typearea{12}

\section*{\LARGE  Detecting the orientation of magnetic fields 
 in galaxy clusters}


\baselineskip20pt

\noindent{\sffamily\large  
Christoph~Pfrommer$^1$, %
L.~Jonathan~Dursi$^{2,1}$ \\

\noindent%
{\normalsize\it%
$^{1}${Canadian Institute for Theoretical Astrophysics, University of Toronto, Toronto, Ontario, M5S~3H8, Canada}\\
$^{2}${SciNet Consortium, University of Toronto, Toronto, Ontario, M5T~1W5, Canada}\\
}
}

\baselineskip26pt 
\setlength{\parskip}{12pt}
\setlength{\parindent}{22pt}%

\begin{large} {\bf Clusters of galaxies, filled with hot magnetized
    plasma, are the largest bound objects in existence and an
    important touchstone in understanding the formation of structures
    in our Universe. In such clusters, thermal conduction follows
    field lines, so magnetic fields strongly shape the cluster's
    thermal history; that some have not since cooled and collapsed is
    a mystery. In a seemingly unrelated puzzle, recent observations of
    Virgo cluster spiral galaxies imply ridges of strong, coherent
    magnetic fields offset from their centre. Here we demonstrate,
    using three-dimensional magnetohydrodynamical simulations, that
    such ridges are easily explained by galaxies sweeping up field
    lines as they orbit inside the cluster. This magnetic drape is
    then lit up with cosmic rays from the galaxies' stars, generating
    coherent polarized emission at the galaxies' leading edges. This
    immediately presents a technique for probing local orientations
    and characteristic length scales of cluster magnetic fields. The
    first application of this technique, mapping the field of the
    Virgo cluster, gives a startling result: outside a central region,
    the magnetic field is preferentially oriented radially as
    predicted by the magnetothermal instability. Our results strongly
    suggest a mechanism for maintaining some clusters in a
    `non-cooling-core' state. }
\end{large}
\newpage

\begin{figure}
\begin{center}
\includegraphics[width=2in]{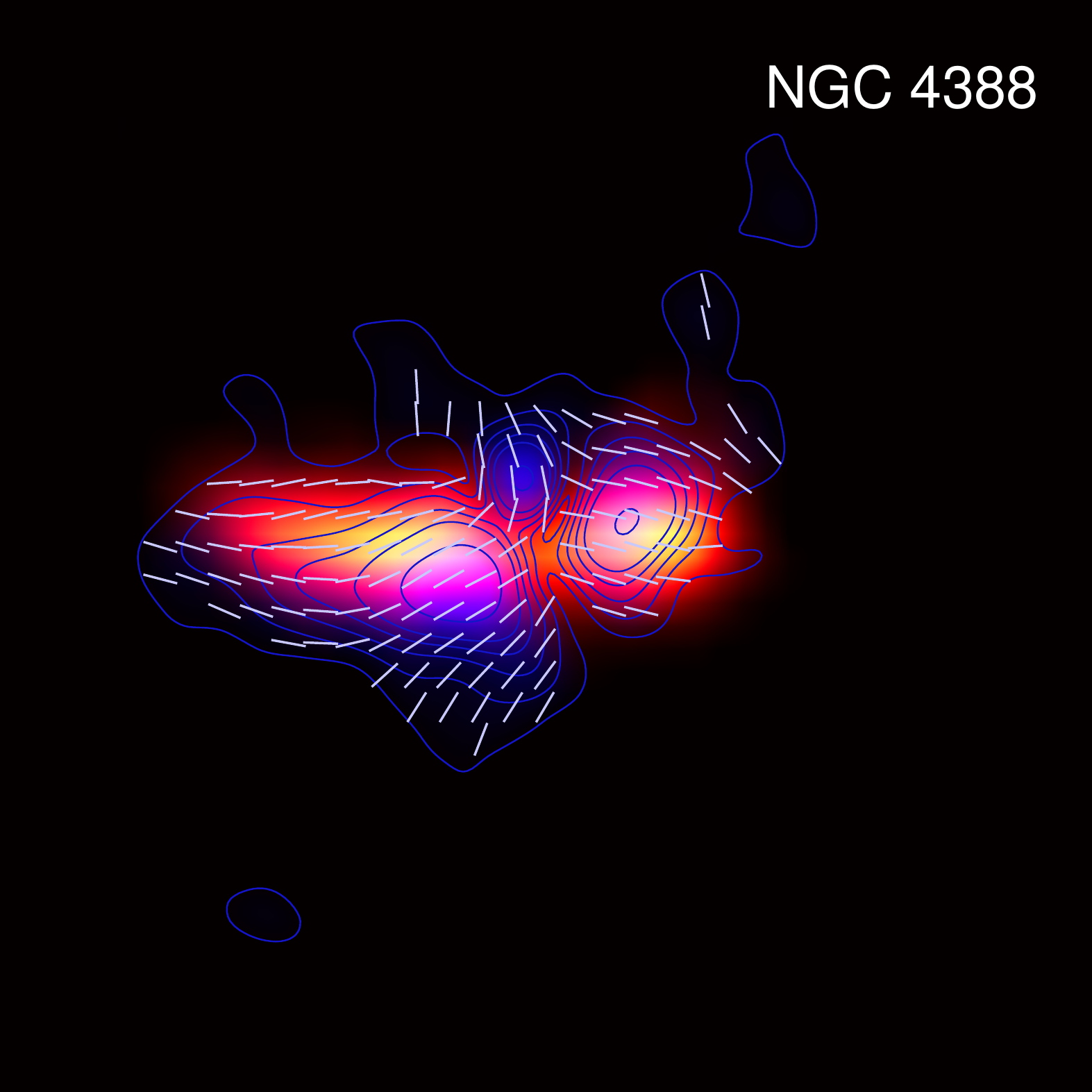}
\includegraphics[width=2in]{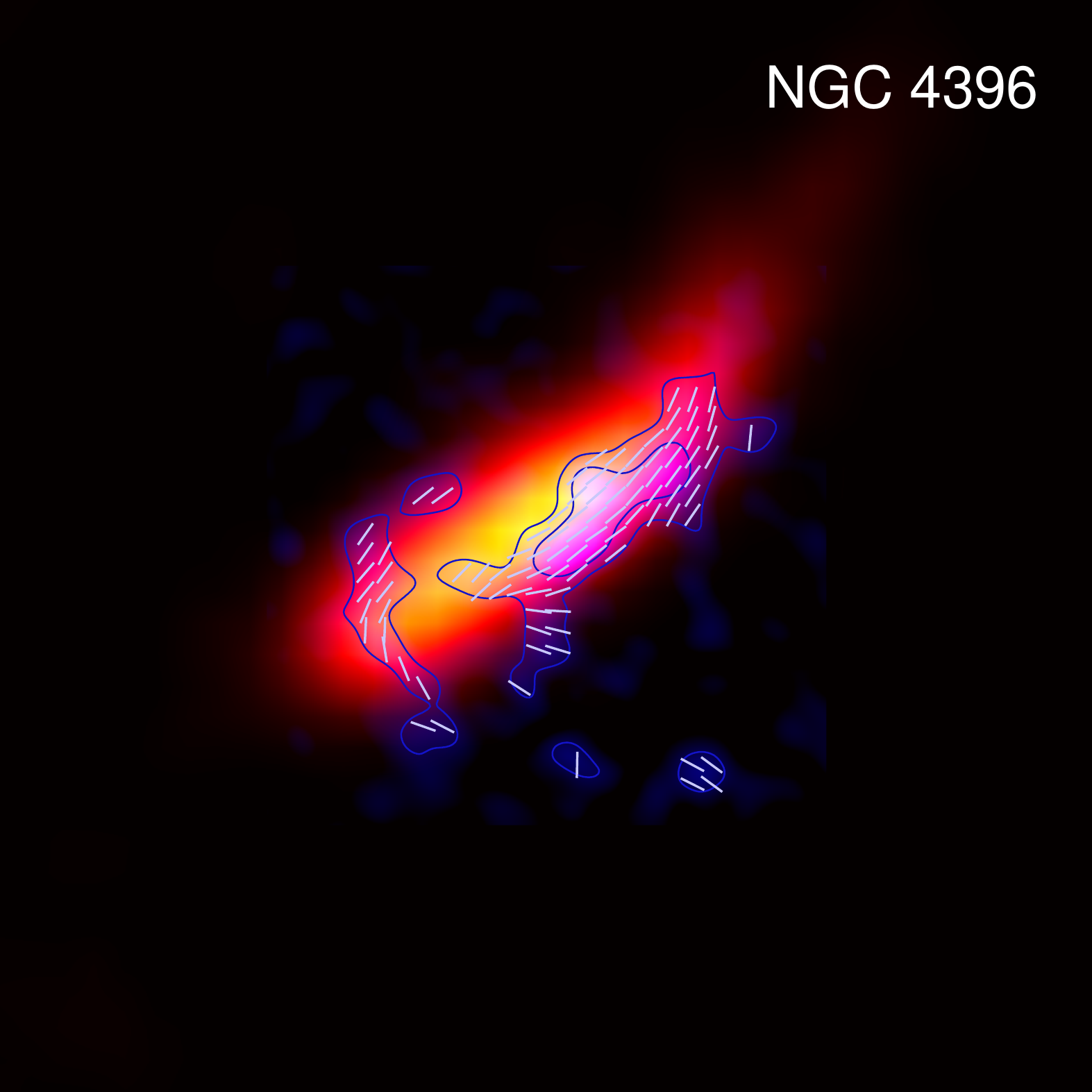}
\includegraphics[width=2in]{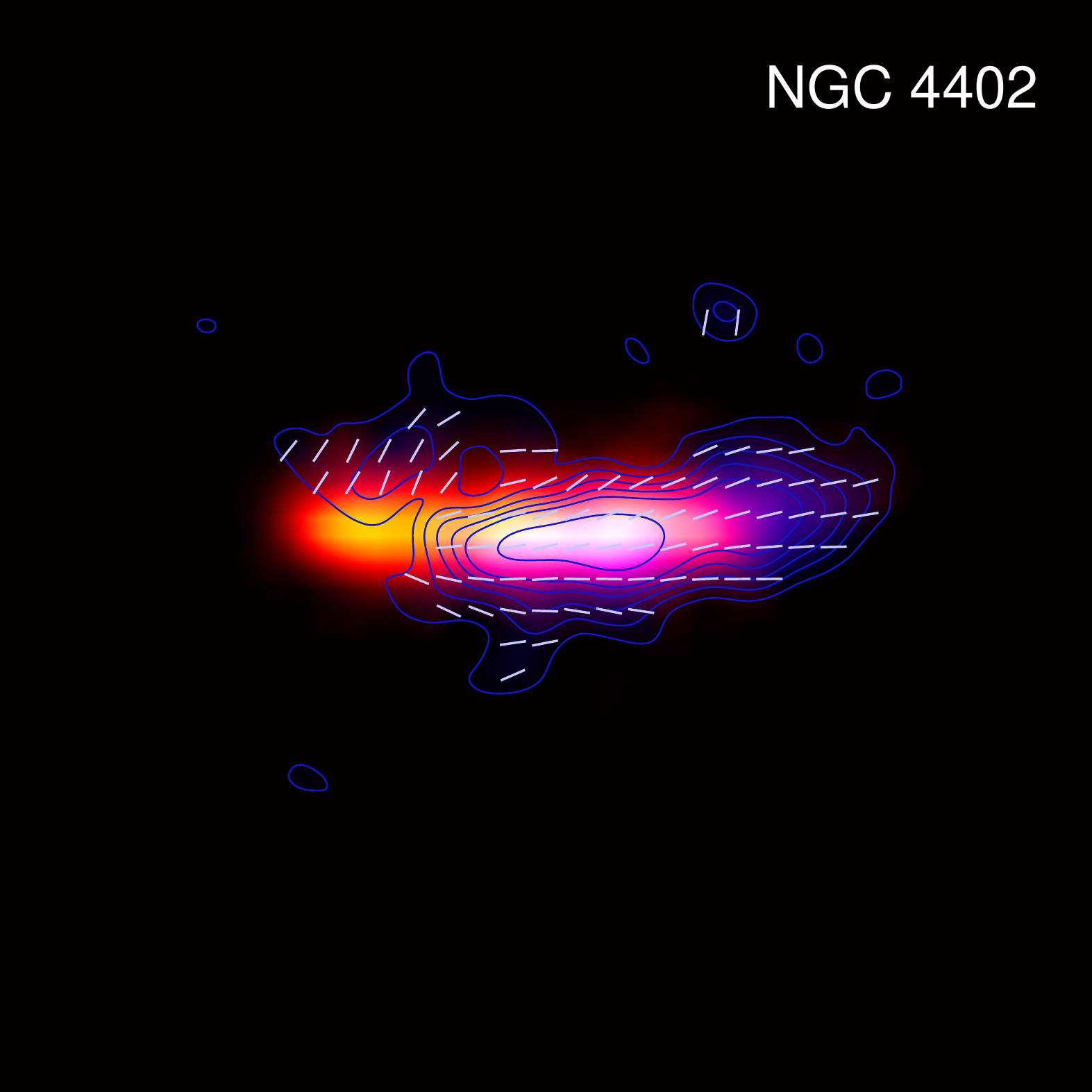}\\[.25em]
\includegraphics[width=2in]{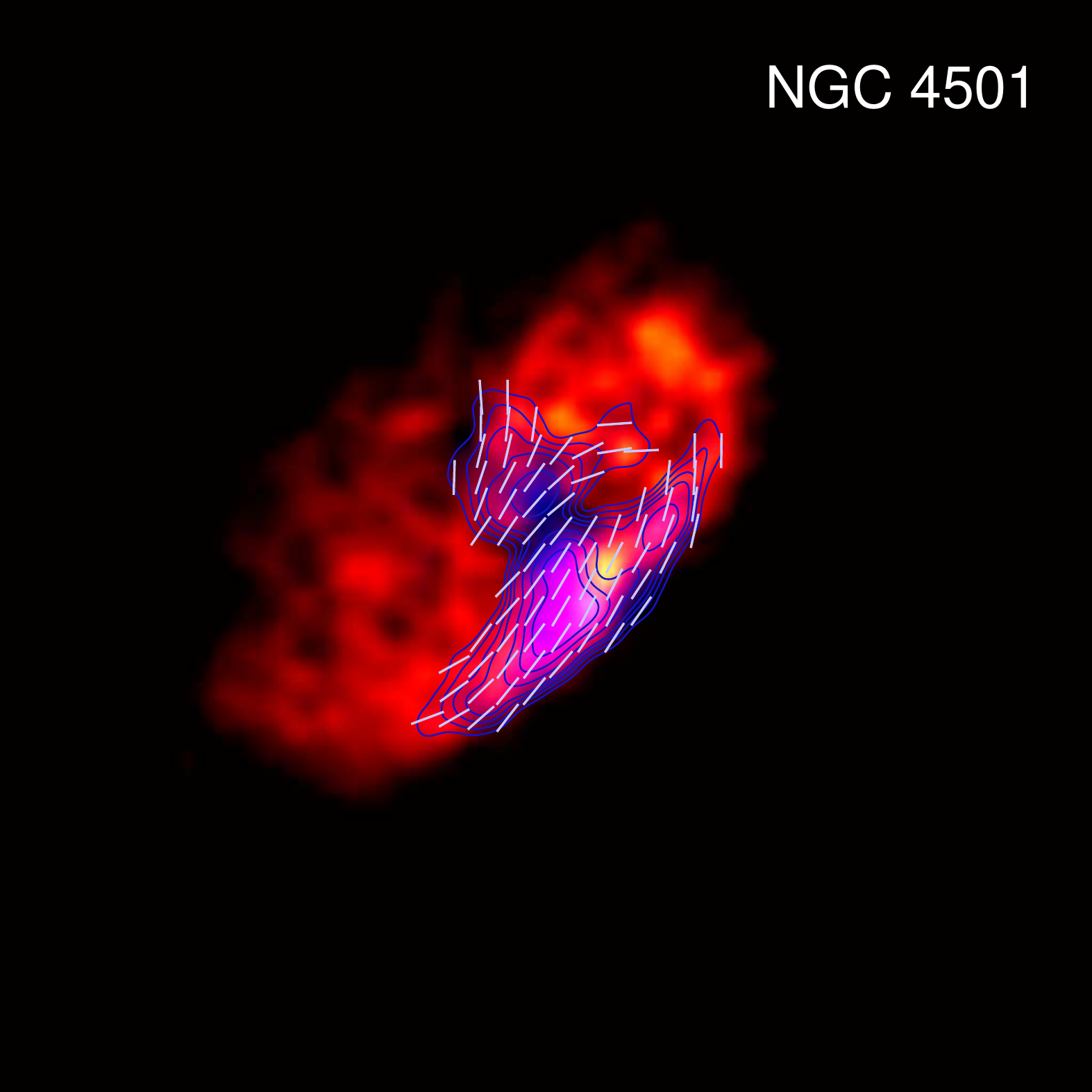}
\includegraphics[width=2in]{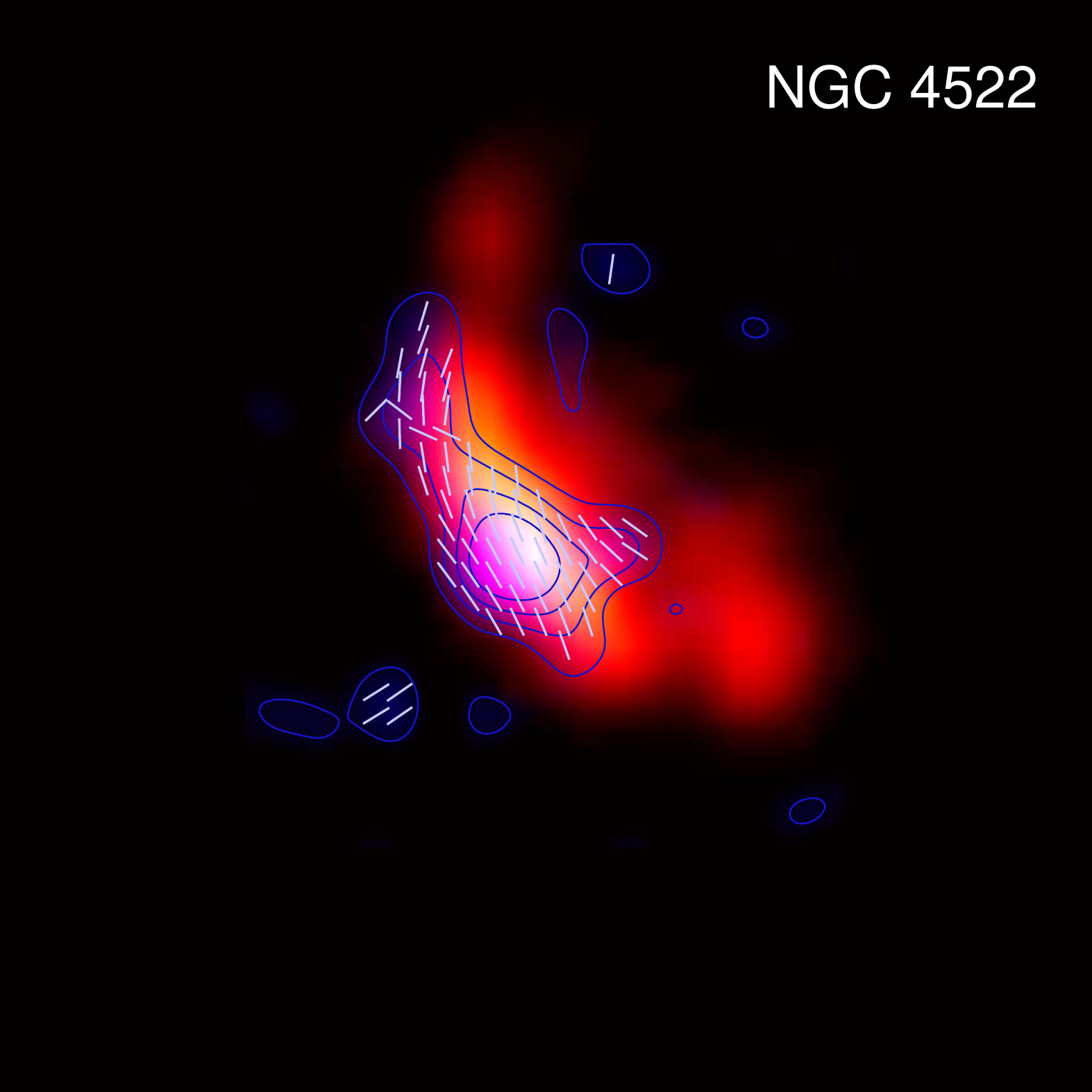}
\includegraphics[width=2in]{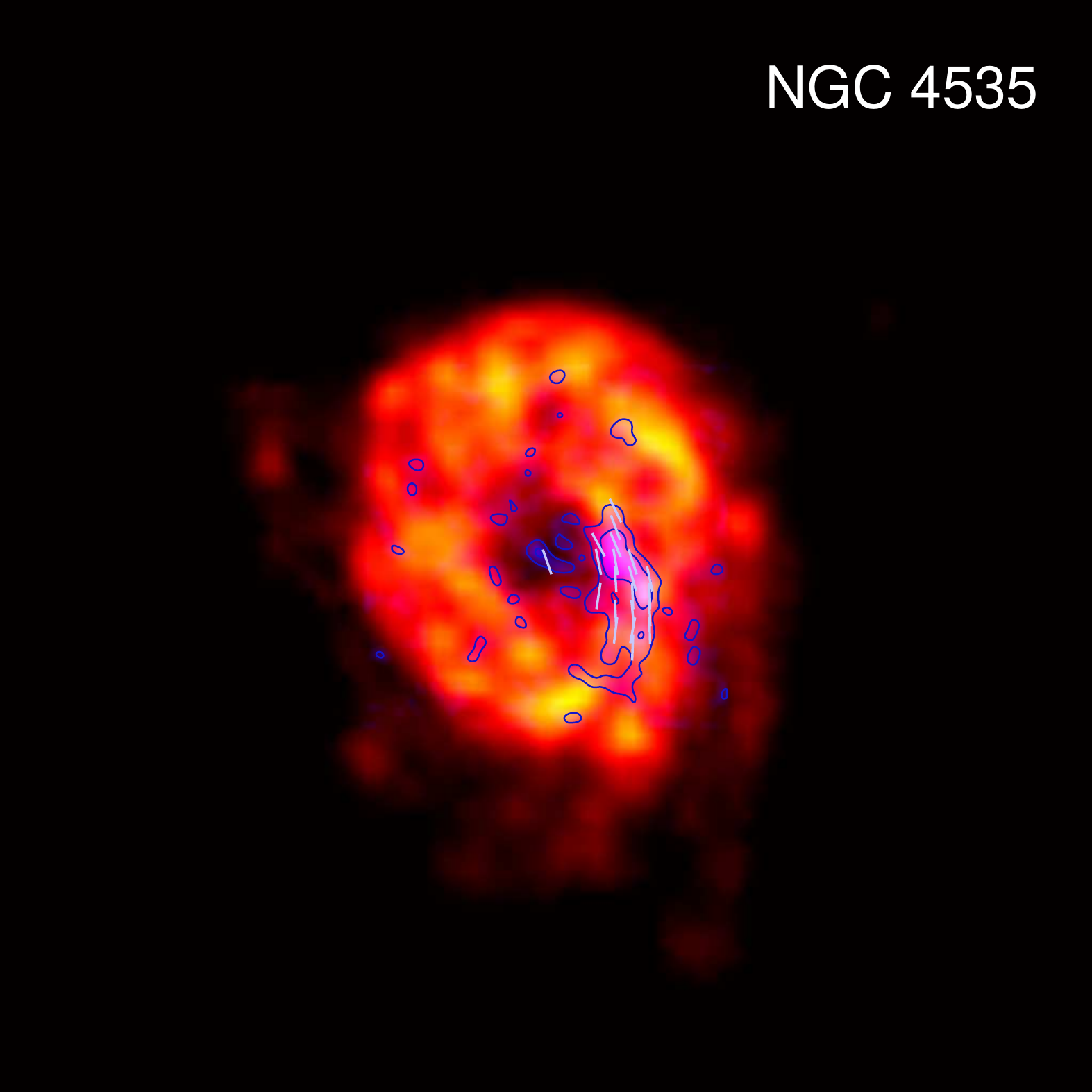}\\[.25em]
\includegraphics[width=2in]{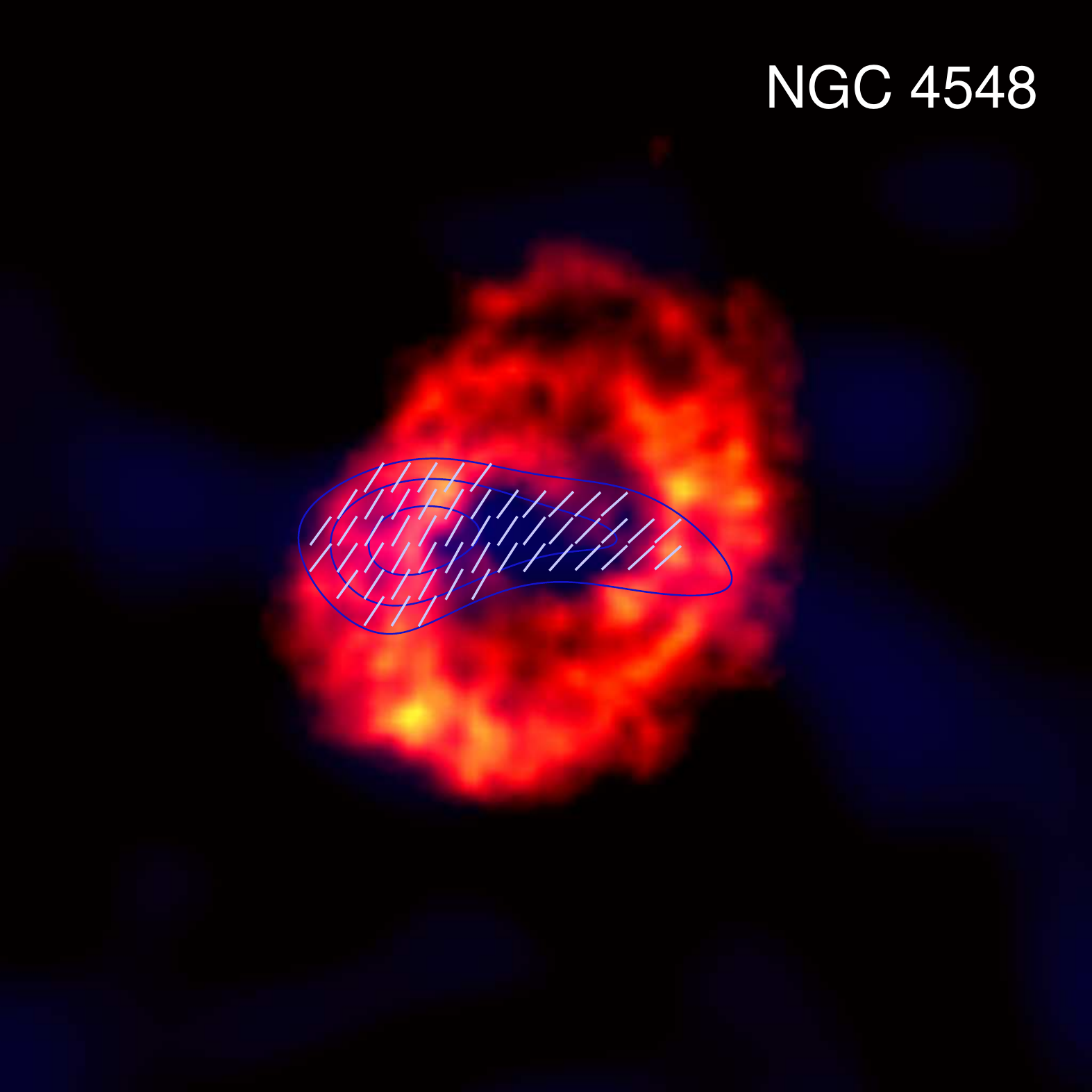}
\includegraphics[width=2in]{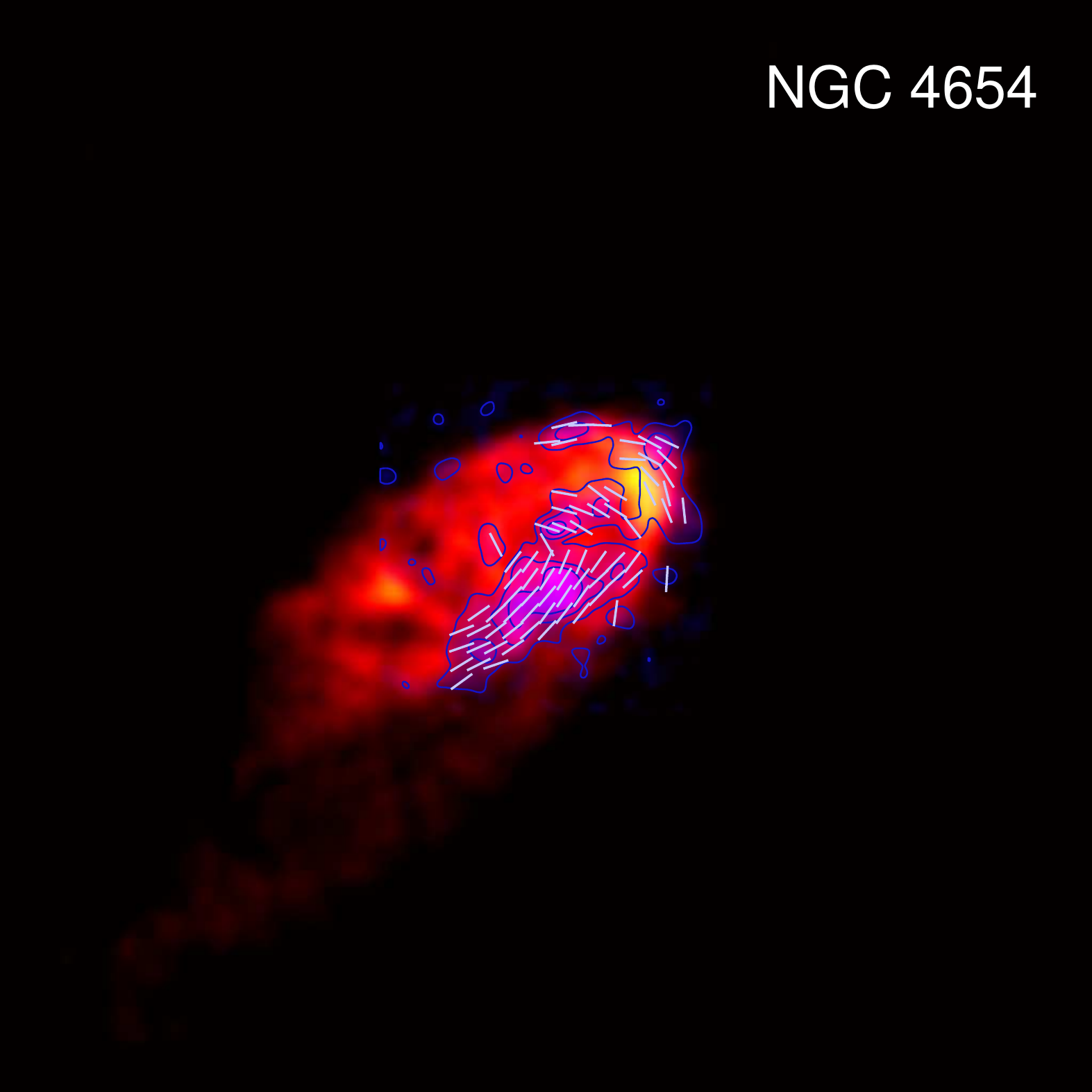}
\hspace{2in}
\caption{ \baselineskip22pt {\bf Polarized radio ridges in comparison with the H
    {\sc i} emission.}  Six-centimetre polarized intensity contours and blue
  colour\cite{2004AJ....127.3375V, 2007A&A...464L..37V, 2007A&A...471...93W,
    2010A&A...512A..36V} overlaid on H {\sc i} intensity
  data\cite{2009AJ....138.1741C} (red-to-yellow). The short lines represent the
  polarization vectors rotated by 90 degrees, hence delineating the orientation
  of the magnetic field in the draping layer uncorrected for Faraday rotation
  (hereafter referred to as `B-vectors'). The contour levels are
  $(4,8,12,16,20,30,40,50)\times\xi \mu$Jy ($\xi=5$ for NGC~4522; $\xi=8$ for
  NGC~4388, NGC~4396, NGC~4402, NGC~4654; $\xi=10$ for NGC~4535; $\xi=11$ for
  NGC~4501) and $(4,5,6)\times 90 \mu$Jy for NGC~4548.  We employed an additive
  colour scheme: if the polarized intensity were perfectly correlated with the H
  {\sc i}, the two colours should add up to `white' at the region of maximum
  emission. The fact that they do not in most cases is one reason for an
  extragalactic origin of the radio emission (see text for more detail).}
\label{Fig1}
\end{center}
\end{figure}

\bl Recent high-resolution radio continuum observations of cluster spirals in
Virgo show strongly asymmetric distributions of polarized intensity with
elongated ridges located in the outer galactic disk\cite{2004AJ....127.3375V,
  2007A&A...464L..37V, 2007A&A...471...93W, 2010A&A...512A..36V} as shown in
Fig.~\ref{Fig1}. The polarization angle is observed to be coherent across the
entire galaxy.  The origin and persistence of these polarization ridges poses a
puzzle as these unusual features are not found in field spiral galaxies where
the polarization is generally relatively symmetric and strongest in the
inter-arm regions\cite{2001SSRv...99..243B}.\el

\begin{figure}
\begin{center}
\includegraphics[width=6in]{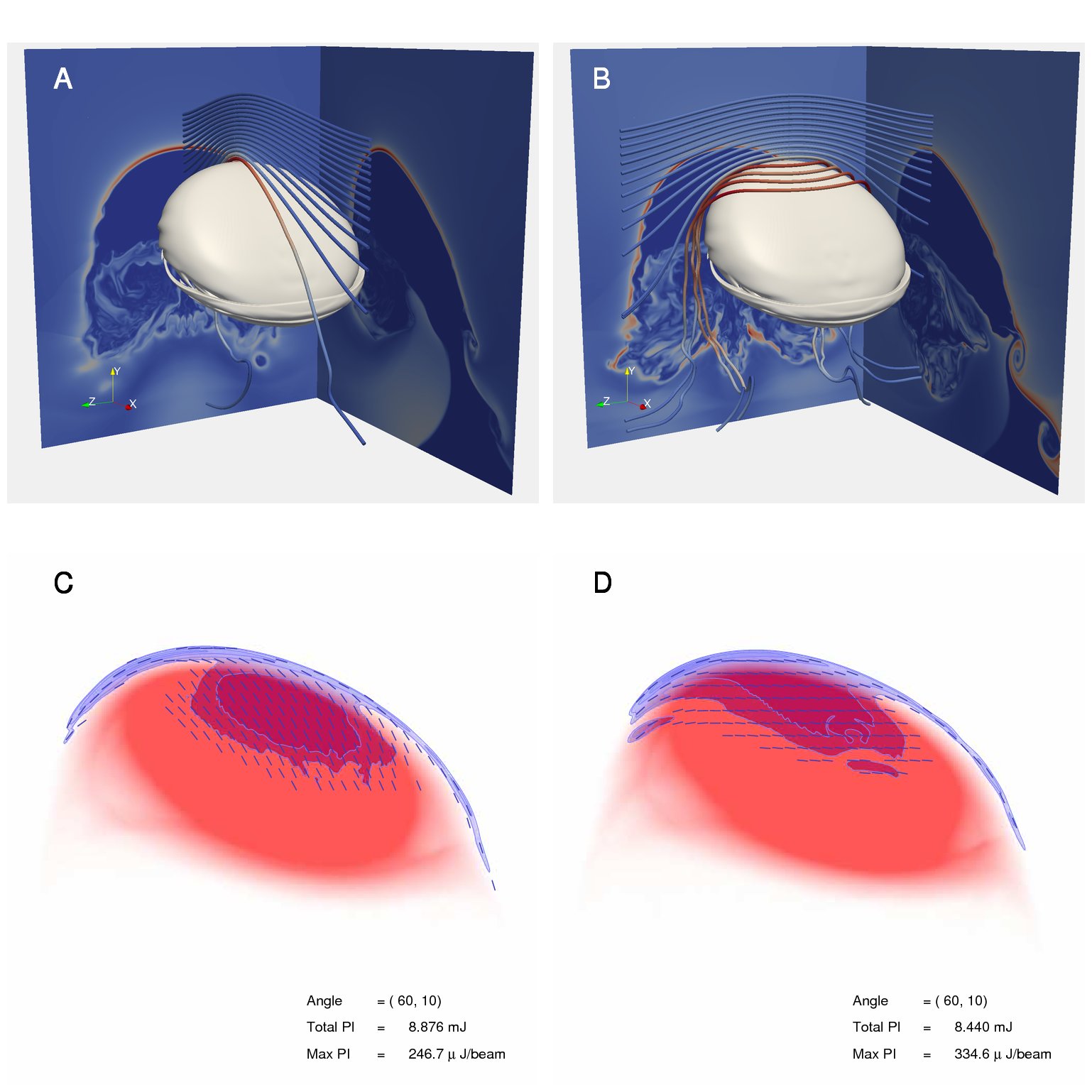}
\caption{ \baselineskip22pt {\bf Simulations of spiral galaxies interacting with
    a uniform cluster magnetic field.}  {\em (Top)} The grey isosurface
  represents an upwards moving galaxy in an Athena
  simulation\cite{StoneEtAl2008,2005JCoPh.205..509G,2008JCoPh.227.4123G}.
  Representative magnetic field lines and cut planes through the stagnation line
  of the flow are coloured by the magnitude of the local magnetic field
  strength.  {\em (Bottom)} Projected density (red) and a simulated
  $6\,\rmn{cm}$ synchrotron emission map, with polarized intensity shown in blue
  contours (using the same contour levels and beam sizes of $18''\times 18''$ as
  recent observational work\cite{2008A&A...483...89V}; for details, see
  Supplementary Information).  Lines indicate local polarization direction
  rotated by ninety degrees (`B-vectors'), to indicate the local magnetic field
  orientation.  The left and right panels differ only in the initial magnetic
  field orientation; this difference clearly shows up in the simulated radio
  observations.}
\label{Fig2}
\end{center}
\end{figure}

\bl We propose a new model that explains this riddle self-consistently, and has
significant consequences for the understanding of galaxy
clusters\cite{2005RMP...77...207V}; the model is illustrated in Fig.~\ref{Fig2}.
A spiral galaxy orbiting through the very weakly magnetized intra-cluster plasma
necessarily sweeps up enough magnetic field around its dense interstellar medium
(ISM) to build up a dynamically important sheath.  This `magnetic draping'
effect is well known and understood in space science.  It has been observed
extensively around Mars\cite{2004mmis.book.....W},
comets\cite{2004inco.book.....B}, Venus\cite{2005JGRA..11001209B},
Earth\cite{2005AnGeo..23..885C}, a moon of Saturn\cite{2006JGRA..11110220N}, and
even around the Sun's coronal mass ejections\cite{2006JGRA..11109108L}.  The
magnetic field amplification comes solely from redirecting fluid motions and
from the slowing of flow in the boundary layer; it is not from compression of
fluid, and indeed happens even in incompressible flows\cite{lyutikovdraping}.
The layer's strength is set by a competition between `ploughing up' and slipping
around of field lines, yielding a magnetic energy density that is comparable to
the ram pressure seen by the moving galaxy\cite{lyutikovdraping,
  2008ApJ...677..993D}, and is stable against the shear that creates
it\cite{2007ApJ...670..221D}.  The magnetic energy density in the so-called
draping layer can be amplified by a factor of 100 or even more compared to the
value of the ambient field in the cluster.  For typical conditions in the
intra-cluster medium (ICM) of $n_\rmn{icm}\simeq 10^{-4}\,\mbox{cm}^{-3}$ and
galaxy velocities $\vel_\rmn{gal}\simeq1000\,\rmn{km~s}^{-1}$, this leads to a
maximum field strength in the draping layer of
$B\simeq\sqrt{16\pi\,\rho_\rmn{icm}^{}
  \vel_\rmn{gal}^2}\simeq7\,\mu\rmn{G}=7\times10^{-10}\,\rmn{T}$.\el

\bl The ram pressure felt by the galaxy as it moves through the ICM displaces
and strips some of the outermost layers of ISM gas in the galaxy; but the stars,
being small and massive, are largely unaffected.  Thus the stars lead the
galactic gas at the leading edge of the galaxy, crossing the boundary between
ISM and ICM, as is seen in observations\cite{2004AJ....127.3375V,
  2007A&A...464L..37V, 2007A&A...471...93W}, and so overlap with the magnetic
drape.  As in the bulk of the galaxy, and in our own, these stars produce
energetic particles; once these stars end their life in a supernova they drive
shock waves into the ambient medium that accelerates electrons to relativistic
energies\cite{1999ApJ...525..357S, 2006ApJ...648L..33V}.  These so-called cosmic
ray electrons are then constrained to gyrate around the field lines of the
magnetic drape, which results in radio synchrotron emission in the draped
region, tracing out the field lines there. \el

\bl The size and shape of this synchrotron-illuminated region is determined by
the transport of the cosmic rays.  The cosmic rays diffuse along field lines,
smoothing out emission; but they are largely constrained to stay on any given
line, and thus are advected by the lines as they are dragged over the galaxy by
the ambient flow.  In the draping boundary layer, because of the magnetic back
reaction, the flow speed is much smaller than the velocity of the galaxy.  These
cosmic ray electrons emit synchrotron radiation until they have lost enough
energy to no longer be visible.  In a magnetic field of $B=7\,\mu\rmn{G}$,
cosmic ray electrons with an energy of $E=5$~GeV or equivalently a Lorentz
factor of $\gamma=10^4$ radiate synchrotron emission at a frequency of
$\nu=5$~GHz; where the polarized radio ridges are observed.  The synchrotron
cooling timescale $\tau_\rmn{syn}\simeq5\times 10^7\,\rmn{yr}$ of these
electrons yields a finite width of the polarization ridge, $L_\rmn{max}\simeq
\eta\vel_\rmn{drape}\tau_\rmn{syn}\simeq10\,\rmn{kpc}$, that is set by the
advection velocity in the drape, $\vel_\rmn{drape}\simeq100\,\rmn{km~s}^{-1}$
and a geometric factor $\eta\simeq2$ that accounts for an extended cosmic ray
electron injection region into the drape (consistent with NGC~4501).  For a
conservative supernova rate of one per century, we show that the different
supernova remnants easily overlap within a synchrotron cooling timescale. This
implies a smooth distribution of cosmic ray electrons that follows that of the
star light, which is also consistent with the synchrotron emission in our
Galaxy\cite{2009ApJS..180..265G}; for details, see Supplementary Information.
\el

\bl Figure~\ref{Fig2} shows this process of draping magnetic field lines at a
galaxy in our simulations with a homogeneous field of two different initial
field orientations. During the draping process, the intra-cluster magnetic field
is dynamically projected onto the contact surface between the galaxy's ISM and
the intra-cluster plasma that is advected around the galaxy.  Outside the
draping sheath in the upper hemisphere along the direction of motion, the smooth
flow pattern resembles that of an almost perfect potential flow
solution\cite{2008ApJ...677..993D}. This great degree of regularity of the
magnetic field in the upstream of the galaxy is then reflected in the resulting
projection of magnetic field in the draping layer. In particular, it varies
significantly and fairly straightforwardly with different ICM field orientations
with respect to the directions of motion. The regularity of the draped field
implies then a coherent synchrotron polarization pattern across the entire
galaxy (bottom panels in Fig.~\ref{Fig2}).  Thus in the case of known proper
motion of the galaxy and with the aid of three-dimensional (3D)
magneto-hydrodynamical simulations to correctly model the geometry, it is
possible to unambiguously infer the orientation of the 3D ICM magnetic field
that the galaxy is moving through (our new method that uses only observables
will be demonstrated in the following). We note that this method provides
information complementary to the Faraday rotation measure, which gives the
integral of the field component along the line-of-sight.  The main complication
is that the synchrotron emission maps out the magnetic field component only in
the plane of the sky, leading to a geometric bias; however, this would be a
serious problem only if it led to significant ambiguities---if different field
configurations and motions led to similar morphologies of polarized synchrotron
emission.  We demonstrate with representative figures in the Supplementary
Information that this seems not to be the case by running a grid of simulations
covering a wide parameter space with differing galactic inclinations, magnetic
tilts, and viewing angles. \el

\begin{figure}
\begin{center}
\includegraphics[width=4in]{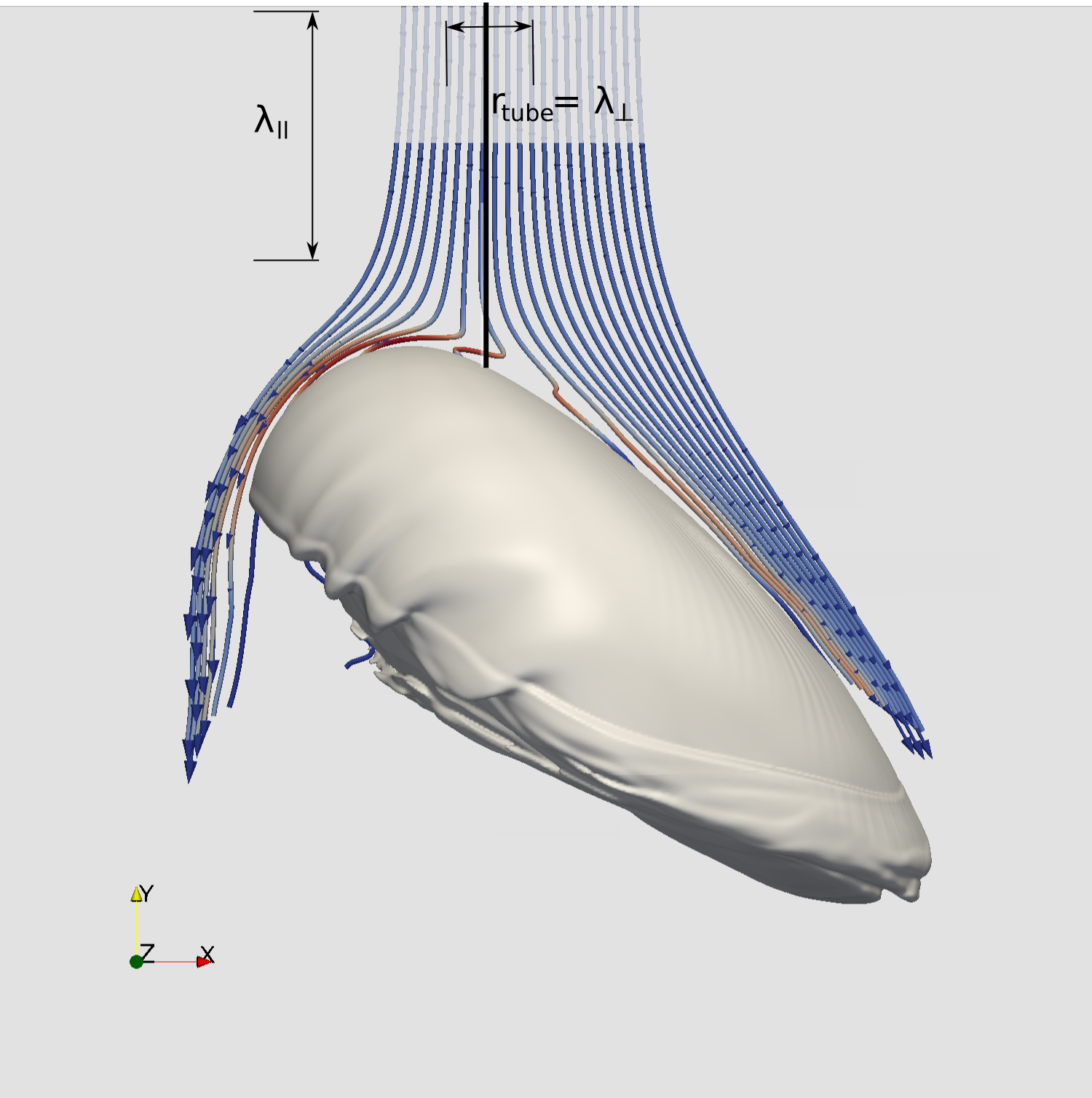}
\caption{ \baselineskip22pt {\bf Streamlines in the rest frame of the galaxy.}
  The arrows indicate the magnitude of the flow and the colours the magnetic
  energy density that they encounter. As the uniform flow approaches the galaxy
  it decelerates and gets deflected.  Only those streamlines initially in a
  narrow tube of radius $\lambda_\perp$ from the stagnation line become part of
  the draping layer.  The streamlines that do not intersect the tube get
  deflected away from the galaxy, become never part of the drape and eventually
  get accelerated as they have less volume to transport the plasma past the
  galaxy (known as the Bernoulli effect).  As we will see in Fig.~4, magnetic
  field must be coherent along the direction of motion on a scale greater than
  $2\,\lambda_\parallel$ for a significant draping signal to be seen.  We note
  that the streamlines are almost identical for our simulations with different
  realizations of the magnetic field.  Note a visible kink feature in some
  draping-layer stream lines as the solution changes from the hydrodynamic
  potential flow solution to the draped layer, where the back reaction plays a
  significant role.  }
\label{Fig3}
\end{center}
\end{figure}

\bl In Fig.~\ref{Fig2}, we showed draping of uniform fields---fields with an
infinite correlation length.  In a turbulent fluid like the ICM, such regularity
of fields is not expected.  To study the effects of a varying field, we first
look at the physics of the draping process by considering the streamlines around
the galaxy as shown in Fig.~\ref{Fig3}.  The boundary layer between the galaxy
and the ICM consists of fluid following streamlines very near the stagnation
line with an impact parameter that is smaller than a critical value of
$p_\rmn{cr}=R/\sqrt{3\beta\M^2}\simeq R/15\simeq1.3$~kpc. Here $R\simeq20$~kpc
is an effective curvature radius over the solid angle of this `tube' of
streamlines which we assume to be equal to the radius of the galaxy, and we
adopted our simulation values of $\beta=100$ for the ratio of
thermal-to-magnetic energy density in the ICM, and
$\M=\vel_\rmn{gal}/c_\rmn{icm}\simeq1$ for the sonic Mach number of the galaxy
that is defined as the galaxy's velocity $\vel_\rmn{gal}$ in units of the ICM
sound speed $c_\rmn{icm}$. Fluid further away from the stagnation line than this
critical impact parameter is deflected away from the galaxy and never becomes
part of the draping boundary layer.  Thus, only fields with correlation lengths
$\lambda_B \apgt 2\,p_\rmn{cr}$ transverse to the direction of motion could
participate in the draping process; for details, see Supplementary
Information.\el

\bl In Fig.~\ref{Fig4}, we see a loss of magnetic draping synchrotron signal
long before this for fields on even larger scales.  Since variations of the
magnetic field in the direction of motion matter most for the synchrotron
signal, we consider the simplest case---a uniform field with an orientation
that rotates as one moves `upwards' and thereby forms a helical structure. By
varying the wavelength of this helix, we see that the magnetic coherence length
needs to be at least of order the galaxy's size for polarized emission to be
significant. Otherwise, the rapid change in field orientation leads to
depolarization of the emission although there is no strong evidence of numerical
reconnection.  Thus, the fact that polarized emission is seen in the drape
suggests that field coherence lengths are at least galaxy-sized.  Note that if
the magnetic field coherence length is comparable to the galaxy scale, then the
change of orientation of field vectors imprints as a change of the polarization
vectors along the vertical direction of the ridge showing a
`polarization-twist'. This is demonstrated in Fig.~\ref{Fig4}.  The pile-up of
field lines in the drape and the reduced speed of the boundary flow means that a
length scale across the draping layer $L_\rmn{drape}$ corresponds to a larger
length scale of the unperturbed magnetic field ahead of the galaxy
$L_\rmn{coh}\simeq\eta L_\rmn{drape}\vel_\rmn{gal}/\vel_\rmn{drape}=
\eta\tau_\rmn{syn}\vel_\rmn{gal}$.  The finite lifetime of cosmic ray electrons
and the non-observation of a polarization-twist in the data limits the coherence
length to be $>100$~kpc (for NGC~4501). Radio observations at lower frequencies
will enable us to study even larger length scales as the lifetime of lower
energy electrons, which emit at these lower radio frequencies, is longer.\el

\bl Figure~\ref{Fig5} compares two observations of these polarization ridges to
two mock observations of our simulations that are also shown with 3D volume
renderings. We simulated our galaxy that encountered a homogeneous field with
varying inclinations, and changed the magnetic tilt with respect to the plane of
symmetry as well as the viewing angle to obtain the best match with the
observations. The impressive concordance of the overall magnitude as well as the
morphology of the polarized intensities and B-vectors in these cases is a strong
argument in favour of our model.  Our model naturally predicts coherence of the
polarization orientation across the entire galaxy as well as
sometimes---dependent on the viewing angle of the galaxy---a coherent
polarization pattern at the galaxy's side with B-vectors in the direction of
motion (see NGC 4654). Additionally for inclined spirals, our model predicts the
polarized synchrotron intensity to lead the column density of the gas of neutral
hydrogen atoms (H {\sc i}) as well as slightly trail the optical and far
infra-red (FIR) emission of the stars, both of which are observed in the
data\cite{2008A&A...483...89V, 2007A&A...464L..37V}. The stars have a
characteristic displacement from the gas distribution depending on the strength
of the ram pressure whereas the thickness of the draping layer is set by the
curvature radius of the gas at the stagnation point and the alfv\'enic Mach
number.  As the FIR and radio emission in galaxies are tightly coupled by the
nearly universal FIR-radio correlation of normal
spirals\cite{1985A&A...147L...6D, 1985ApJ...298L...7H}, our model predicts a
radio deficit relative to the infra-red emission just upstream the polarization
ridge---in agreement with recent findings\cite{2009ApJ...694.1435M}. \el

\bl We see, then, that magnetic draping---an effect well understood and
frequently observed in a solar system context---can easily reproduce the
observed polarization ridges seen in Virgo galaxies.  Draping is of course not
the only way to generate significant regions of polarized synchrotron radiation;
but for any other effect to dominate the emission in the ridge, it would have to
represent coherent action on galactic scales (to match the observed coherence of
polarization vectors over the entire galaxy), and not be limited to the disk of
the galaxy (as some ridges are observed to be significantly extra-planar, and
others to significantly lead the H {\sc i} or even H$\alpha$ emission from the disk.)
We also note that, apart from the problem of extraplanar emission, ram pressure
compressing the galaxy's ISM would, by energy conservation at the stagnation
line, imply compression to a number density of only $n_\rmn{ism}\simeq
n_\rmn{icm}\,\vel_\rmn{gal}^2/c_\rmn{ism}^2\simeq 1\,\rmn{particle~cm}^{-3}$,
where we adopted typical conditions in the intra-cluster medium and a sound
speed in the ISM of $c_\rmn{ism}\simeq 10\,\rmn{km~s}^{-1}$. Since this number
density is at the interstellar mean, we do not expect large compression effects
in the ISM (consistent with the observed H {\sc i} distribution) and hence only very
moderate amplifications of the interstellar magnetic field.  Thus the observed
properties of the ridges are impossible to explain using purely galactic
magnetic field, although it has been attempted\cite{2008A&A...483...89V}; see
the Supplementary Information for more detail.  \el

\bl We now explain the method of how we can infer the orientation of cluster
magnetic fields by using the observation of polarized radio ridges.  We use the
morphology of the H {\sc i} and (if available) the total synchrotron emission to obtain
an estimate of the galaxy's velocity component on the sky.  If the galaxy is
inclined with the plane of the sky, we determine the projected stagnation point
by localizing the `H {\sc i} hot spot' and drop a perpendicular to the edge of the
galaxy which then points in the opposite direction from the ram-pressure stripped
tail (see Fig.~\ref{Fig5}). If the galaxy is edge-on, we additionally use the
location and morphology of the polarized radio emission as an independent
estimate while keeping in mind the potential biases that are associated with it.
The galaxy's redshift gives an indication about the velocity component along the
line-of-sight. For well resolved galaxies, we then compare the data to our mock
observations where we iteratively varied galactic inclination, magnetic tilt,
and viewing angle so that they matched the H {\sc i} morphology and polarized
intensity.  Preserving the field line mapping from our simulated polarized
intensity map to the upstream orientation of the field in our simulation enables
us to infer the an approximate 3D orientation of the upstream magnetic field. In
the case of edge-on galaxies (or lower quality data) numerical resolution
considerations limited us from running the appropriate galaxy models. Instead,
we determine the orientation of the B-vectors in a region around this stagnation
point and identify this with the (projected) orientation of the ambient field
before it got swept up.  \el

\bl With this new tool at hand, we are now able to measure the geometry of the
magnetic field in the Virgo galaxy cluster and find it to be preferentially
radially aligned outside a central region (see Fig.~\ref{Fig6})---in stark
disagreement with the usual expectation of turbulent fields.  The alignment of
the field in the plane of the sky is significantly more radial than expected
from random chance.  Considering the sum of deviations from radial alignment
gives a chance coincidence of less than 1.7\%. (We point out a major caveat as
the statistical analysis presented here does not include systematic
uncertainties. In particular, line-of-sight effects could introduce a larger
systematic scatter, which is however impossible to address which such a small
observational sample at hand.) In addition, the three galaxy pairs that are
close in the sky show a significant alignment of the magnetic field.  The
isotropic distribution with respect to the centre (M87) is difficult to explain
with the past activity of the active galactic nucleus in M87 and the spherical
geometry argues against primordial fields. In contrast, this finding is very
suggestive that the magneto-thermal instability is operating; at these distances
outside the cluster centre it encounters a decreasing temperature profile which
is the necessary condition for it to operate\cite{2000ApJ...534..420B}. In the
low-collisionality plasma of a galaxy cluster, the heat flux is forced to follow
field lines as the collisional mean free path is much larger than the electron
Larmor radius\cite{1965RvPP....1..205B}. On displacing a fluid element on a
horizontal field line upwards in the cluster potential, it is conductively
heated from the hotter part below, gains energy, and continues to
rise---displacing it downwards causes it to be conductively cooled from the
cooler part on top and it continues to sink deeper in the gravitational
field. As a result, the magnetic field will reorder itself until it is
preferentially radial\cite{2007ApJ...664..135P, 2008ApJ...688..905P} if the
temperature gradient as the source of free energy is maintained by constant
heating through AGN feedback or shocks driven by gravitational infall. Numerical
cosmological simulations suggest that the latter is expected to maintain the
temperature gradient as it preferentially heats the central parts of a
cluster\cite{2006MNRAS.367..113P}. Even cosmological cluster simulations that
employ isotropic conduction at $1/3$ of the classical Spitzer value are not able
to establish an isothermal profile\cite{2004MNRAS.351..423J,
  2004ApJ...606L..97D}.  Our result of the global, predominantly radial field
orientation in Virgo strongly suggests that gravitational heating seems to
stabilize the temperature gradient in galaxy clusters even in the presence of
the magneto-thermal instability, hence confirming a prediction of the
hierarchical structure formation scenario.  These theoretical considerations
would imply efficient thermal conduction throughout the entire galaxy cluster
except for the very central regions of so-called cooling core clusters which
show an inverted temperature profile\cite{2001MNRAS.328L..37A}. Under these
conditions, a different instability, the so-called heat-flux-driven buoyancy
instability is expected to operate\cite{2008ApJ...673..758Q} which saturates by
re-arranging the magnetic fields to become approximately perpendicular to the
temperature gradient\cite{2008ApJ...677L...9P}. In principle our new method
would also be able to demonstrate its existence if the galaxies that are
close-by in projection to the cluster centre can be proven to be within the
cooling core region. In fact, NGC~4402 and NGC~4388 are observed in the vicinity
of M86\cite{2008ApJ...688..208R} and the inferred magnetic field orientations at
their position would be consistent with a saturated toroidal field from the
heat-flux-driven buoyancy instability. \el

\bl Our finding of the radial orientation of the cluster magnetic field at
intermediate radii would make it possible for conduction to stabilize cooling,
if the radial orientation continued into the cluster core. This could explain
the thermodynamic stability of these non-cool core clusters (NCCs); some of
which show no signs of merger events.  In fact, half of the entire population of
galaxy clusters in the {\em Chandra} archival sample show cooling times that are
longer than 2~Gyr and have high-entropy cores\cite{2009ApJS..182...12C}. This
cool-core/NCC bimodality is indeed real and not due to archival bias as a
complementary approach shows with a statistical
sample\cite{2009MNRAS.395..764S}.  The centres of these galaxy clusters show no
signs of cooling such as H$\alpha$ emission and an absence of blue light that
traces regions of newly born stars suggesting that conduction might be an
attractive solution to the problem of keeping them in a hot
state\cite{2008ApJ...681L...5V, 2009MNRAS.395..764S}. A global Lagrangian
stability analysis shows that there are stable solutions of NCCs that are
stabilized primarily by conduction\cite{2008ApJ...688..859G}. We emphasize that
the other half of the galaxy cluster population, in which the cores are cooler
and denser, also manage to avoid the cool core catastrophe. The absence of
catastrophic cooling in any of these clusters is also a major puzzle, but one
that cannot be solved by conduction alone.  We show in the Supplementary
Information that the Virgo cluster seems to be on its transition to a cool core
at the centre but still shows all signs of a NCC on large scales: if placed at a
redshift $z>0.13$ it would be indistinguishable from other NCCs in the
sample.\el

\bl The findings of a preferentially radially oriented field in the Virgo
cluster suggests an evolutionary sequence of galaxy clusters. After a merging
event, the injected turbulence decays on an eddy turnover time
$\tau_\rmn{eddy}\simeq L_\rmn{eddy}/\vel_\rmn{turb}\sim300\,\rmn{kpc}/
(300\,\rmn{km~s}^{-1})\sim1$~Gyr whereas the magneto-thermal instability grows on a
similar timescale of less than 1~Gyr and the magnetic field becomes radially
oriented\cite{2008ApJ...688..905P}. The accompanying efficient thermal
conduction stabilizes this cluster until a cooling instability in the centre
causes the cluster to enter a cooling core state---similar to Virgo now---and
possibly requires feedback by an active galactic nuclei to be
stabilized\cite{2001ApJ...554..261C, 2008ApJ...688..859G}. We note that this
work has severe implications for the next generation of cosmological cluster
simulations that need to include magnetic fields with anisotropic conduction to
realistically model the evolution and global stability of NCC systems. To date,
these simulated systems show little agreement with realistic
NCCs\cite{2007MNRAS.378..385P, 2007ApJ...668....1N}.  Interesting questions
arise from this work such as which are the specific processes that set the
central entropy excess: are these mergers or do we need some epoch of
pre-heating before the cluster formed? It will be exciting to see the presented
tool applied in other nearby clusters to scrutinize this picture. \el

\begin{large}
\bibliography{bibtex/paper}
\end{large}

\paragraphfont{\large}

\begin{large}
\vspace*{-0.75cm}\paragraph*{Correspondence} and requests for materials should
be addressed to CP~(email: pfrommer@cita.utoronto.ca).
\end{large}

\bl \vspace*{-0.75cm}\paragraph*{Acknowledgements} The authors wish to thank
A.~Chung, B.~Vollmer, and M.~We{\.z}gowiec for providing observational data and
acknowledge C.~Thompson, Y.~Lithwick, J.~Sievers, and M. Ruszkowski for
discussions during the preparation of this manuscript. We also wish to thank our
referees for insightful comments. CP gratefully acknowledges the financial
support of the National Science and Engineering Research Council of Canada.
Computations were performed on the GPC supercomputer at the SciNet HPC
Consortium. SciNet is funded by: the Canada Foundation for Innovation under the
auspices of Compute Canada; the Government of Ontario; Ontario Research Fund -
Research Excellence; and the University of Toronto.  3D renderings were
performed with Paraview.  \el

\bl \vspace*{-0.75cm}\paragraph*{Author Contributions} CP initiated this
project; performed the analytic estimates; developed an approach for comparing
to observations; presented and analyzed the observational data; measured
magnetic field angles and discussed their uncertainties; explored consequences
for cluster physics.  LJD carried out magneto-hydrodynamical simulations with
Athena, explored the parameter space, and visualized the simulations.  Both
authors contributed to the exploratory simulations with Flash, the development
of the synchrotron polarization model and post-processing, the statistical
analysis, the interpretation of the results, and writing of the paper.  \el

\begin{large}
\vspace*{-0.75cm}\paragraph*{Competing interests} The authors declare that they have no
competing financial interests.
\end{large}

\newpage

\begin{figure}
\begin{center}
\includegraphics[width=6in]{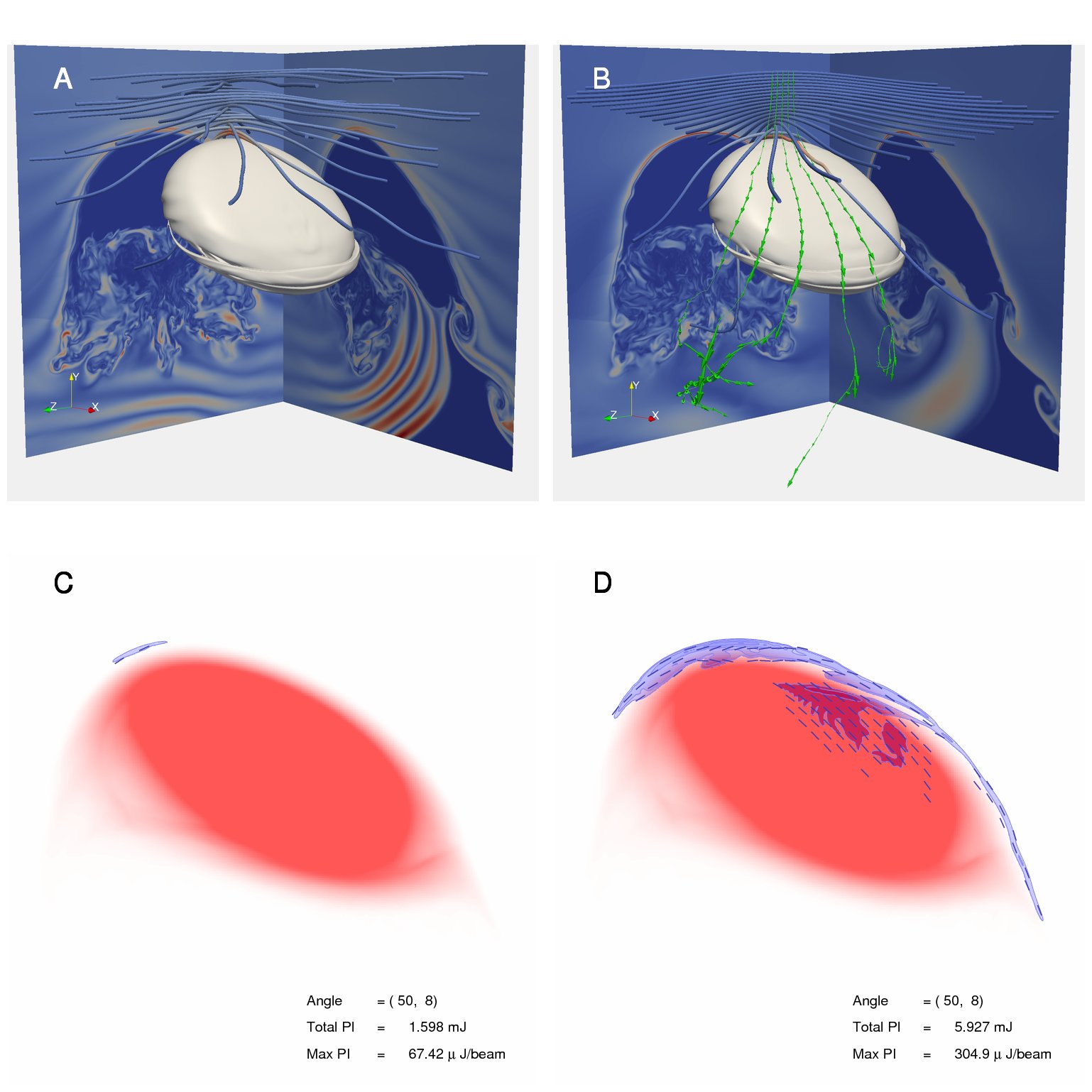}
\caption{ \baselineskip22pt {\bf Simulations of spiral galaxies interacting with a
  non-uniform cluster magnetic field.}  As in Fig.~\ref{Fig2}, but with cluster magnetic
  fields in a force-free helical configuration with a given coherence length in
  the direction of motion of the galaxy.  {\em (Left)} Wavelength of the helix is
  $14\,\mathrm{kpc}$, less than the radius of the galaxy.  {\em (Right)}
  Wavelength is $57\,\mathrm{kpc}$, larger than the diameter of the galaxy.  In
  {\bf (B)}, we also show representative streamlines in green, with arrow size
  indicating speed.  The flow speed is greatly reduced in the draping boundary
  layer. The radio emission is greatly reduced for {\bf (C)} because of
  depolarization by the superposition of different field orientations in the
  draping layer. In {\bf (D)}, note the increased complexity of the B-vectors
  compared to Fig.~\ref{Fig2}D which is due to the helical structure of
  the ambient field.}
  \label{Fig4}
\end{center}
\end{figure}

\begin{figure}
\begin{center}
\includegraphics[width=6in]{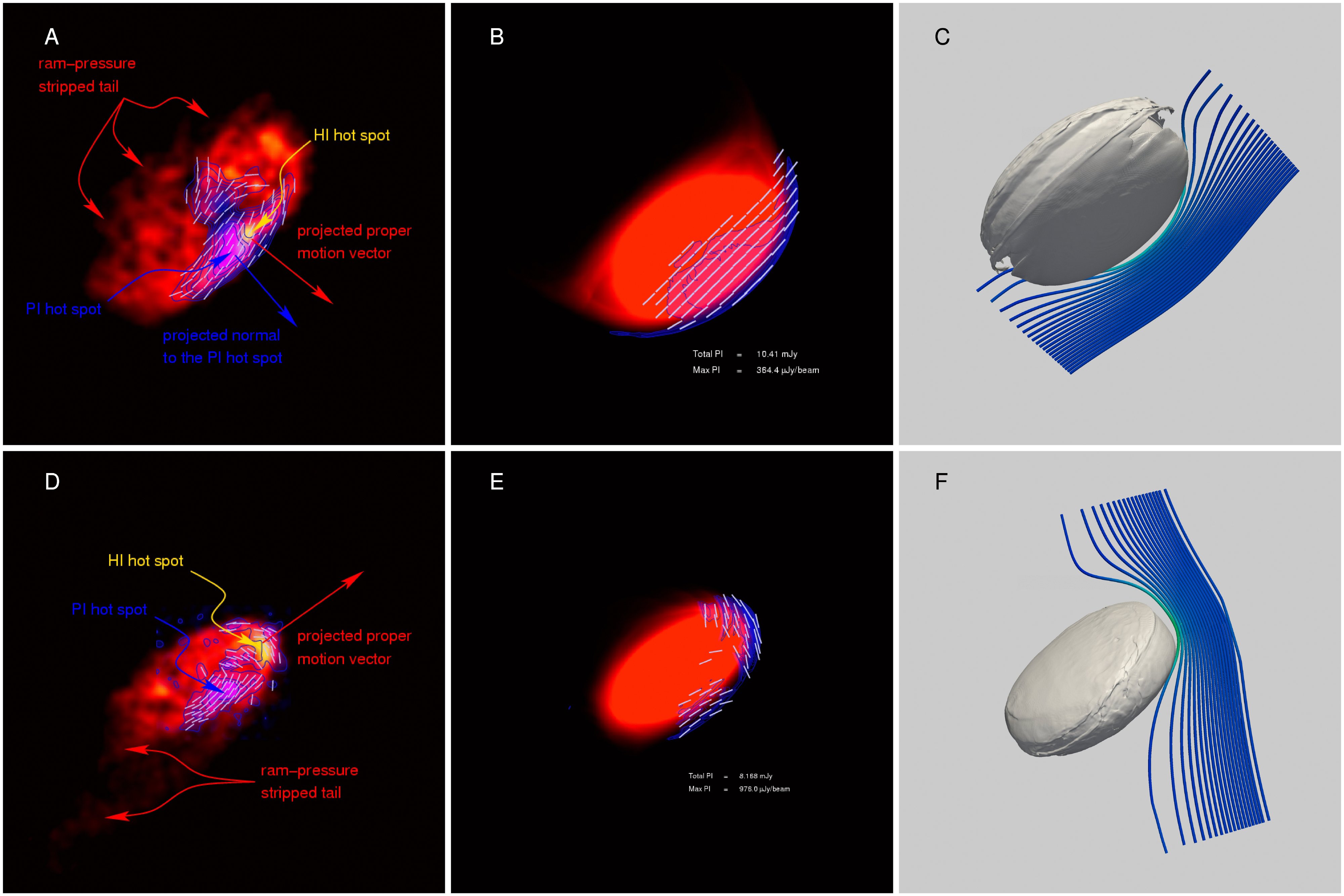}
\caption{ \baselineskip22pt {\bf Polarized radio emission: observations versus
    simulation.} {\em (Left)} H {\sc i} emission\cite{2009AJ....138.1741C} of two
  spirals ({\bf A:} NGC~4501, and {\bf D:} NGC~4654) tracing the neutral
  hydrogen distribution that is severely affected by the ram pressure resulting
  from the galaxies' motion in the intra-cluster plasma
  (red-to-yellow). Over-plotted are the polarized intensity (PI) of the radio
  synchrotron ridges\cite{2007A&A...464L..37V} at $6\,\rmn{cm}$ (blue and
  contours) with the B-vectors indicated in white.  {\em (Middle)} Simulated
  synchrotron maps that are selected from the uniform field models. {\em
    (Right)} Shown is a 3D volume rendering of our best matching galaxy model
  (shown with grey isosurfaces) and representative magnetic field lines (see
  Supplement for details). The misalignment of the H {\sc i} and PI hot spots for
  NGC~4654 might point to a geometric bias (upstream magnetic field has a
  non-negligible line-of-sight component) as our numerical model suggests; hence
  we omitted plotting the projected normal to the PI hot spot for this galaxy.}
  \label{Fig5}
\end{center}
\end{figure}

\begin{figure}
\begin{center}
\resizebox{16.0cm}{!}{\includegraphics{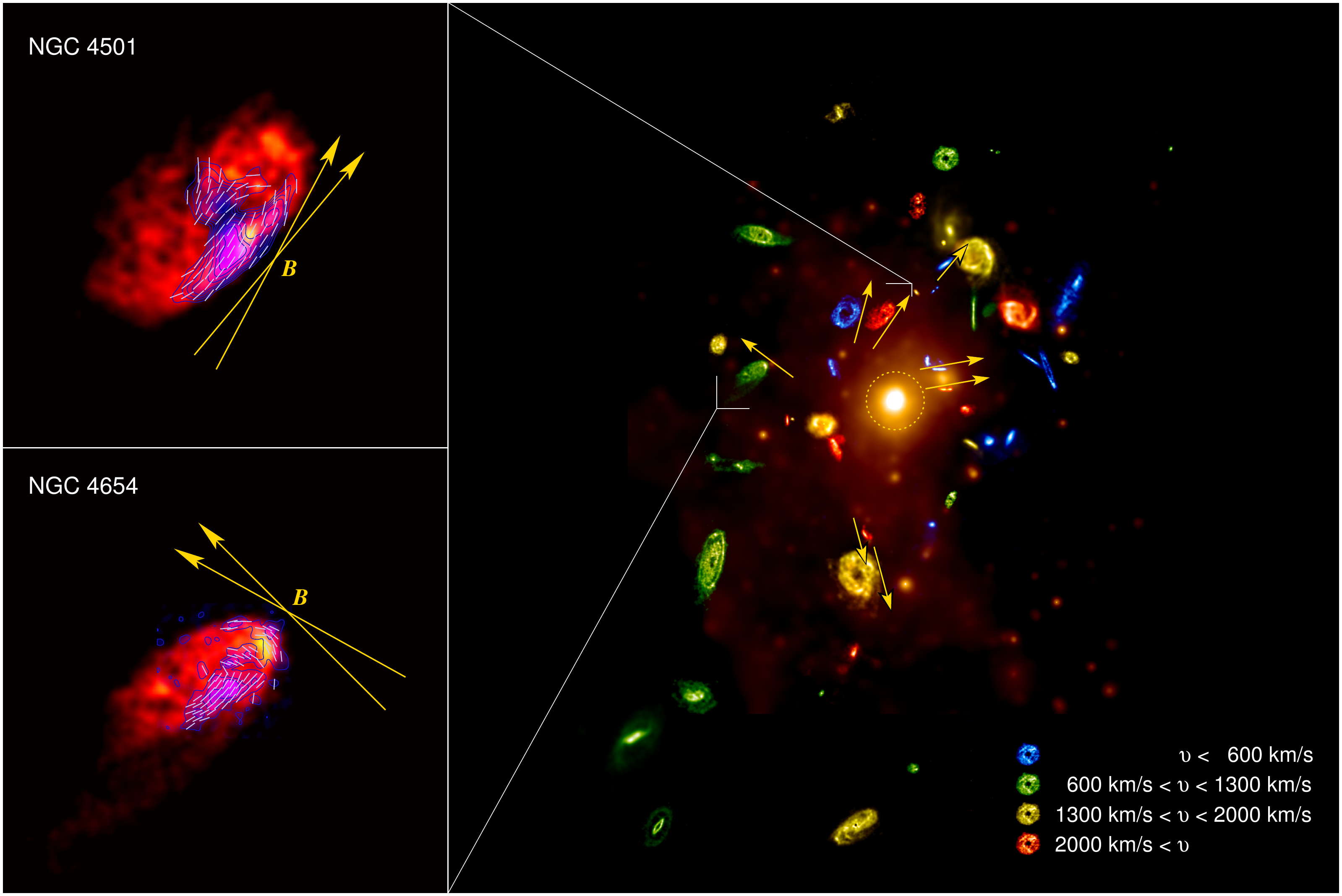}}%
\caption{ \baselineskip22pt {\bf Mapping out the magnetic field orientations in
    the Virgo galaxy cluster as inferred from the polarized synchrotron emission
    ridges of cluster spirals.} {\em (Left)} H {\sc i} intensity
  data\cite{2009AJ....138.1741C} (red-to-yellow) overplotted with $6\,\rmn{cm}$
  polarized intensity contours and B-vectors\cite{2007A&A...464L..37V} (blue).
  The direction of the B-vectors directly measures the projected local
  orientation of the cluster magnetic fields $\vecbf{B}$ (yellow) with the range
  of plausible field orientations indicated by the double-arrow. {\em (Right)}
  ROSAT X-ray image of the Virgo cluster\cite{1994Natur.368..828B} (orange
  colour scale). Over-plotted is the H {\sc i} emission of Virgo spirals in the
  VIVA survey\cite{2009AJ....138.1741C} that is colour coded according to their
  velocities, and the green colour scale coincides with the cluster's mean
  heliocentric velocity of 1100 km sec$^{-1}$ (after image by Chung {\em et
    al.}\cite{2009AJ....138.1741C}). Galaxies are magnified by a factor of 10.
  The yellow arrows indicate the projected orientation of the magnetic field in
  Virgo at the position of the galaxies that have high-resolution polarized
  radio data. Note that the arrowheads could also point in the opposite
  direction because of the $n\pi$-ambiguity of inferring magnetic orientations
  from synchrotron polarization. We find that the magnetic field is
  preferentially oriented radially which is very suggestive that the
  magneto-thermal instability is operating\cite{2000ApJ...534..420B,
    2008ApJ...688..905P} as it encounters the necessary conditions outside a
  spherical region with radius 200 kpc\cite{1994Natur.368..828B} (indicated with
  the yellow dotted circle).  }
\label{Fig6}
\end{center}
\end{figure}

\newpage


\paragraphfont{\small}
\sectionfont{\large}
\subsectionfont{\normalsize}

\def\captionfont{\small}

\def\@cite#1#2{[$^{\mbox{\scriptsize #1\if@tempswa , #2\fi}}$]}

\typearea{14}

\newcommand{\bra}{\langle}
\newcommand{\ket}{\rangle}

\renewcommand*{\figurename}{Supplementary Figure}
\renewcommand*{\refname}{Supplementary Information References}
\renewcommand{\thefootnote}{\fnsymbol{footnote}}

\renewcommand{\textfraction}{0}
\renewcommand{\floatpagefraction}{1.0}
\renewcommand{\topfraction}{1.0}
\renewcommand{\bottomfraction}{1.0}

\title{\vspace*{0.3cm} \ \\ {\Large \em Supplementary Information for the Nature Physics Article\\
    ``Detecting the orientation of magnetic fields in galaxy clusters''}
  \vspace*{0.3cm}}

\author{\centering{
{\small\sffamily%
Christoph~Pfrommer$^{1}$, %
L.~Jonathan~Dursi$^{2,1}$}}}

\date{}

\bibliographystyle{naturemag}

\baselineskip14pt 
\setlength{\parskip}{4pt}
\setlength{\parindent}{18pt}%

\twocolumn

\setlength{\footskip}{25pt}
\setlength{\textheight}{640pt}
\setcounter{figure}{0}

\maketitle

\renewcommand{\thefootnote}{\arabic{footnote}}
\footnotetext[1]{Canadian Institute for Theoretical Astrophysics, University of Toronto, Toronto, Ontario, M5S~3H8, Canada}
\footnotetext[2]{SciNet Consortium for High Performance Computing, University of Toronto, Toronto, Ontario, M5T~1W5, Canada}

\tableofcontents

\renewcommand{\thefootnote}{\fnsymbol{footnote}}

\small 

\clearpage

\section{Simulations}

\subsection{Modelling the galaxy}

In our model, we deliberately choose to neglect interstellar magnetic fields,
the galaxies' rotation, winds and outflows, or even the multiphase structure of
the galaxies for the following reasons: we wanted to study a clean controlled
experiment without complications such as numerical reconnection and the
mentioned properties do not have any immediate influence on the physics of
draping. Moreover, since there is no evidence in the data of these galaxies for
any outflows, we are save to neglect them.  All edge-on galaxies in our sample
show {\em extraplanar polarised emission} which is impossible to reconcile in a
model where the polarised emission is due to interstellar magnetic fields or
intracluster magnetic fields that interact with the neutral component of the
interstellar medium.  Similarly, the {\em coherence of the observed polarised
  emission across entire galaxies} requires a coherent process that works on
galactic scales: the strong turbulent magnetic field on small scales found in
the star forming regions is incapable of explaining this fact. Shearing motions
seeded by galactic rotation do not come into questions as they fail to explain
the extraplanar polarised emission as well as the fact that the polarised
synchrotron emission leads the HI distribution in some cases.  Hence we are
forced to consider coherence of the intracluster magnetic field on scales at
least as large as these galaxies under consideration.  We will discuss this in
more detail in Sect.~\ref{sec:obs}.

The magnetohydrodynamical (MHD) simulations described in this work
were performed using the Athena code\cite{GardinerStone2005,GardinerStone2008,StoneEtAl2008}, a
freely-available uniform-grid dimensionally unsplit MHD solver for
compressible astrophysical flows.  Our earlier
work\cite{2008ApJ...677..993D} used the {\sc Flash}
code\cite{flashcode,flashvalidation} which had the great advantage
of having an adaptive mesh; however for these simulations with
roughly sonic flow, dimensionally split solvers are susceptible
to the well-known stationary shock instability and thus {\sc Flash}
was unsuitable.   

Our simulations were of an ideal, perfectly conducting, adiabatic,
$\gamma=5/3$ fluid.  The simulations were in the frame of the galaxy,
with an incoming wind with velocity
$1000\,\mathrm{km}\,\mathrm{s}^{-1}$ of ambient fluid with $P = 1.67
\times 10^{-12}\,\mathrm{dyne}\,\mathrm{cm}^{-2}$ and $\rho =
m_\rmn{p}\, 10^{-4}\,\mathrm{cm}^{-3}$ which results in a number
density $n =\rho/(\mu\,m_\rmn{p})=1.7\times 10^{-4}\,\mathrm{cm}^{-3}$
assuming primordial element composition with a mean molecular weight
of $\mu=0.6$.  Because the magnetic draping is solely an effect
occurring at the wind/galaxy contact discontinuity, and does not
depend on the internal structure of the galaxy at all, the spiral
galaxy was simply modelled as a cold dense oblate ellipsoid of gas
with an axis ratio $q = 0.1$ initially centred at $(0,0,0)$ with no
self-gravity.  It was in pressure equilibrium with the ambient cluster
medium , had a gently-peaked density profile which smoothly matches on
to the ambient medium, $\rho(r_e) = \rho_0 ( \cos(\pi r_e / r_{e,0}) +
1 ) / 2 $ where $r_e$ is the elliptical radius, $r_{e,0}$ is the
elliptical radius of the galaxy, and $\rho_0$ is the central density
of the `galaxy'. 

\begin{figure*}
\begin{center}
\includegraphics[width=5in]{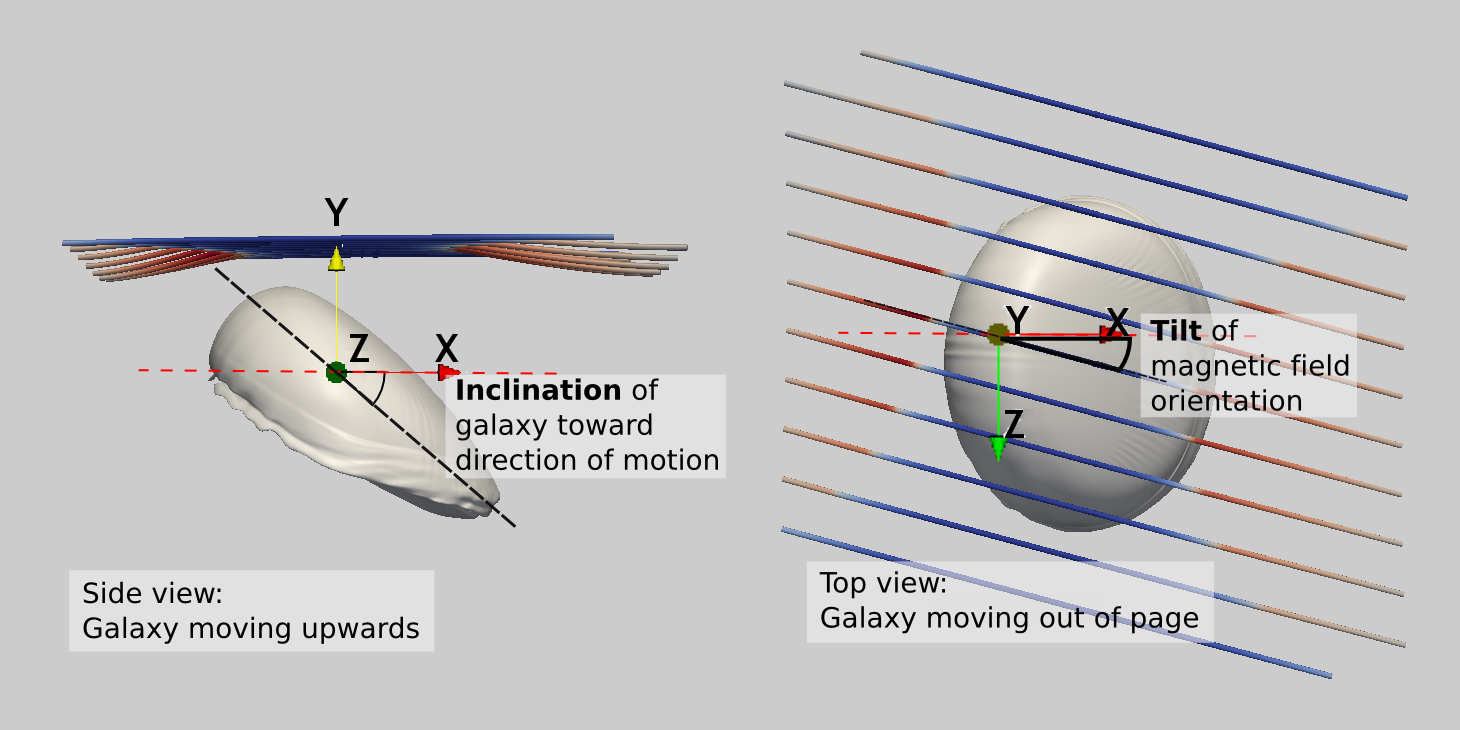} 
\end{center}
\caption{  
\label{fig:geometry-schematic}
{\bf Schematic of simulation geometry.} Shown is the orientation of the 
model galaxy (grey isosurface) and the magnetic fields, with respect to 
the simulation coordinate system.   The galaxy moves in the $y$ direction,
with an inclination along the $z$-axis, and the field is generally
taken to be uniform with some `tilt' angle along the $y$-axis from the $x$-axis.
} \end{figure*}

A schematic of the geometry modelled is shown in Figure~\ref{fig:geometry-schematic}.
The direction of motion for the galaxy was taken to be along positive $y$
direction, so that the wind seen by the galaxy moves in the negative $y$
direction.  For the simulations considered here, the normal of the galaxy's disk
was inclined at a $45^o$ angle to the direction of motion.  The disk of the
galaxy was taken to have a radius of $20\,\mathrm{kpc}$.  The simulation domain
was generally taken to be $[-28.3\,\mathrm{kpc},28.3\,\mathrm{kpc}]$$\times$
$[-39.6\,\mathrm{kpc},28.3,\mathrm{kpc}]$$\times$
$[-28.3\,\mathrm{kpc},28.3\,\mathrm{kpc}]$.  Because the draping occurs
super-alfv\'enincally at the contact discontinuity, varying domain sizes or
boundary conditions in early simulations were not seen to affect the structure
of the boundary layer at all.  The turbulent wake, very interesting in its own
right, did show some modest reaction to domain size and boundary conditions, but
is not the focus of this work.  The boundary conditions used here were fixed
inflow at velocity $\vel_w$ at the top ($+y$) boundary, and zero-gradient `outflow'
at the bottom ($-y$) and horizontal ($\pm x,z$) boundaries.

In initial experiments with the top boundary, rather than start the inflow
at near-sonic we allowed the flow to `ramp up' over some characteristic
time.   This had the important effect of reducing large transients in
small initial high Mach number simulations, but had no effect for these
runs, and was not used here.   The boundary condition is also responsible
for the magnetisation of the ambient medium; at time zero, there is no
magnetic field anywhere in the domain; the magnetic field is advected
in with the inflow from the $+y$ boundary.

\subsection{Magnetic field strength and resolution study}
\label{sec:sim-B}

The simulations here are scale-free and can be described solely in terms of
dimensionless quantities.  For modelling galaxies orbiting within a galaxy
cluster, our length scale (the size of a typical spiral galaxy) is more or less
a given.  The velocity scale is set to be roughly sonic (we choose here $\vel
\approx 1000\,\mathrm{km}\,\mathrm{s}^{-1}$), as the same gravitational
potential sets both orbit speeds and the gas thermodynamic profiles.  Our choice
implies somewhat subsonic motions, ${\cal{M}} = \vel_\rmn{gal}/c_\rmn{s} =
\vel_\rmn{gal}/ \sqrt{\gamma P/\rho} = \sqrt{3/5} \approx 0.77$. 

With length and velocity scales given, the timescale is then set, as well.
Remaining is only the magnetic field parameter describing the ambient magnetic
field strength in the cluster.  This can be given by the `plasma beta',
the ratio of ambient gas pressure to magnitude of the ambient magnetic field,
$\beta = P / \sqrt{B^2/(8 \pi)}$, or alternatively the Alfv\'enic Mach number,
${\cal{M}}_A = \vel_\rmn{gal}/\vel_A = \M\,\sqrt{\beta\gamma/2}$.

The plasma beta in the outskirts of a galaxy cluster such as Virgo is uncertain,
but is expected to be in the range $300\aplt \beta \aplt 10^5$. This can be seen
by taking a central magnetic field $B_0\simeq 8\,\mu$G of similar strength as it
is inferred in other clusters\cite{2002ARA&A..40..319C, 2002RvMP...74..775W,
  2004IJMPD..13.1549G, 2005A&A...434...67V, 2009arXiv0912.3930K} and using a
density of $n\simeq 10^{-4}\,\rmn{cm}^{-3}$ which is approximately a factor of
$\delta=n_0/n\simeq10^3$ lower compared to the central value
$n_0$\cite{2002A&A...386...77M}. We estimate its value at a cluster radius of
$\sim 600$~kpc (where we observe these galaxies) to $B\simeq B_0\,
\delta^{-\alpha_B} \simeq(0.02,\ldots,0.25)\,\mu$G, with $\alpha_B =
(0.5,\ldots,0.9)$ (see Fig.~\ref{fig:profiles}). The choice $\alpha_B=0.7$
corresponds to the flux freezing condition. We employ a plausible range for
$\alpha_B$ as determined from cosmological cluster
simulations\cite{1999A&A...348..351D, 2001A&A...378..777D} and Faraday rotation
measurements\cite{2005A&A...434...67V}. We note, however, that these techniques
are not capable of resolving or measuring small-scale turbulent dynamo processes
in the cluster outskirts and potentially miss the field amplification by these
processes (as long as the resulting field strength is small compared to that in
the cluster centre). Hence the true plasma beta in the outskirts of galaxy
clusters could be in the range $10^2\aplt \beta \aplt 10^3$ as suggested by
recent theoretical work\cite{2006PhPl...13e6501S}.

Resolution constraints, however, mean that we aren't completely free to choose
arbitrary $\beta$.  In the draping region, along the stagnation line, we expect
from earlier analytic kinematic work\cite{1980Ge&Ae..19..671B,
  lyutikovdraping,2008ApJ...677..993D}
\begin{equation}
B \simeq B_0 \left [1 - \left ( \frac{R}{R+s} \right )^3 \right ]^{-1/2},
\end{equation}
where $s$ is the distance along the stagnation line from the leading
edge of the (assumed spherical) projectile; for $s/R \ll 1$,
\begin{equation}
B(s) \simeq B_0 \sqrt{\frac{R}{3 s}}.
\label{eq:stagnation-line-b}
\end{equation}
In reality, the field growth is truncated where the back reaction becomes
significant, $\alpha \rho_\rmn{icm} \vel_\rmn{gal}^2 \approx B^2/(8 \pi)$,
and so the drape thickness should be 
\begin{equation}
  \label{eq:l_drape}
  l_\rmn{drape} \simeq \frac{1}{6 \alpha {\cal{M}}_A^2}\, R \simeq
  \frac{1}{3 \alpha \beta \gamma {\cal M}^2}\, R 
\end{equation}
where ${\cal M} \approx 1$ for our case, and in our previous work
with spherical projectiles, $\alpha \approx 2$.  Imposing the constraint $l_\rmn{drape}
\apgt \Delta x$ for an object of size $R = L\,\sqrt{2}/4$ where
$L$ is the size of our box, we derive a resolution-based restriction on $\beta_\rmn{sim}$:
\begin{equation}
\beta_\rmn{sim} \aplt \frac{\sqrt 2}{12 \alpha \gamma {\cal M}^2} N
\end{equation}
where $N$ is the number of points in the domain.  Computational requirements
lead us to resolution of $(N_x, N_y, N_z) = (640, 768, 640)$, meaning that the
natural $\beta_\rmn{sim}$ to choose is $\beta_\rmn{sim} \approx 40$; smaller
plasma beta parameters lead to a thicker layer.  As a compromise between the
resolution requirements and fidelity to the physical conditions of the
intra-cluster medium (ICM), we choose $\beta_\rmn{sim} = 100$.

From the discussion above, then, using the parameter $\alpha \approx 2$ from our
simulations of spherical projectiles, we would expect $l_\rmn{drape} \approx
R/600 = 33\,\mathrm{pc}$, or $\Delta x \approx 2.7 l_\rmn{drape}$.  However,
$\alpha$ is a geometric parameter; the thickness of the layer is set by the
requirement that at the top of the layer, flow and excess magnetic pressure do
enough work on the tension of the magnetic field lines to push them around the
obstacle.  Thus, $\alpha$ depends sensitively on the geometry of both the flow
and the magnetic field. In our highly flattened geometry representing a galaxy,
the layer thickness is greater than over a sphere for zero inclination. We note
that the layer becomes thinner for increasing inclinations of the galaxy with
respect to the direction of motion\cite{D&P} which increases the requirements
for the numerical resolution that is need to resolve the layer.

\begin{figure}
\begin{center}
\includegraphics[width=3in]{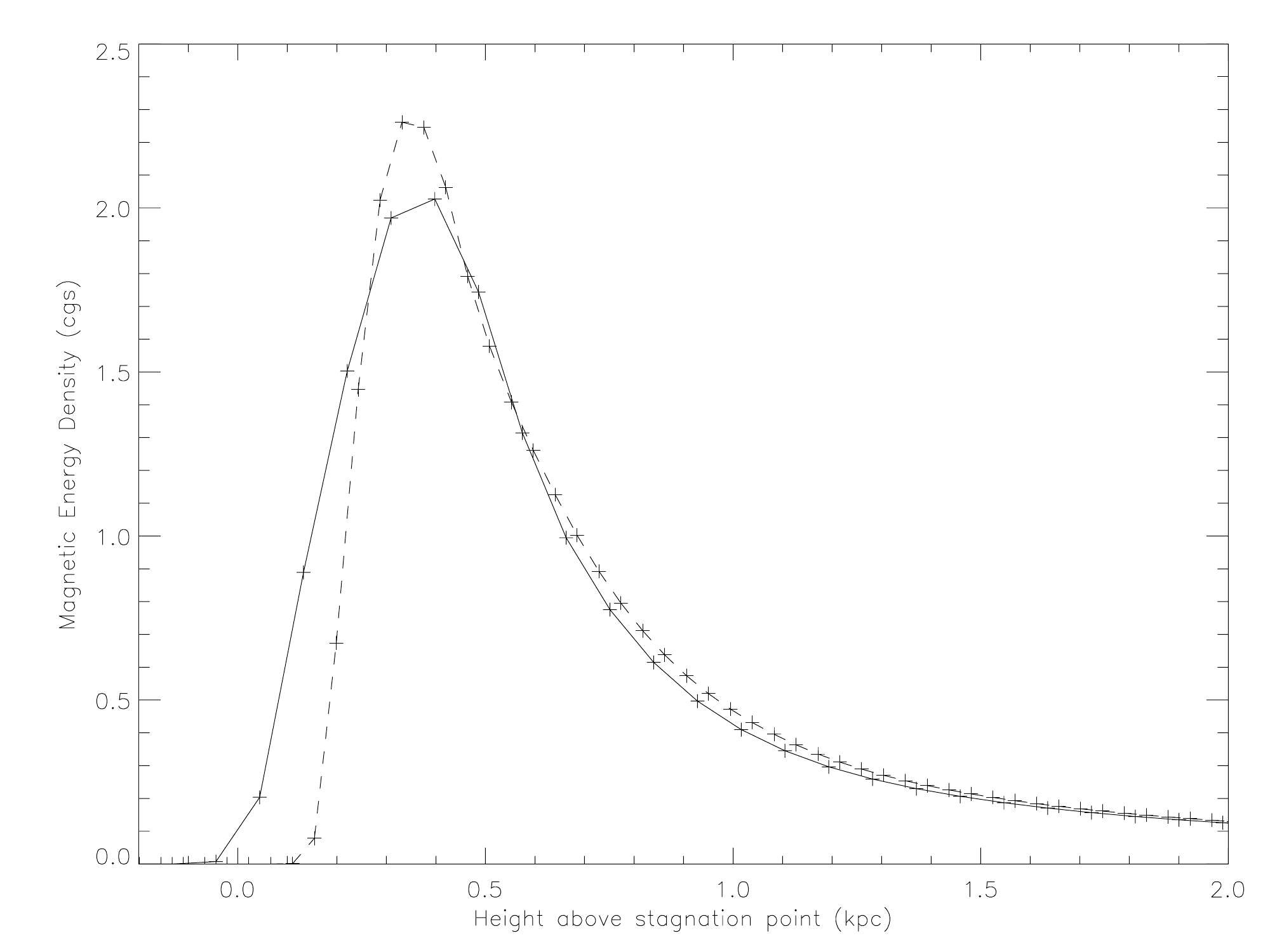} 
\end{center}
\caption{  
\label{fig:resolutionStudy}
{\bf Resolution study.} Plotted is the magnetic energy density (in
$10^{-12}\,\mathrm{erg}\,\mathrm{cm}^{-3}$) along the stagnation line for one
of our fiducial models (shown in the left hand side of Fig.~\ref{Fig2} of the main
body), as points connected by a solid line, and the results for the same
elapsed simulation time for a simulation with twice the number of resolution
elements in each direction -- $(N_x, N_y, N_z) = (1280, 1536, 1280)$ shown with
a dashed line and points.  The high resolution simulation contained 2.5 billion
points, required 300,000 CPU hours, and generated 1.2~TB of data.
We see that the peak of the magnetic energy density
- the draping layer - seems to be of order 100-150pc thick, and is resolved by
two grid cells even in the lower resolution simulation.  The main difference
between the two is a somewhat reduced magnetic field strength in the
lower-resolution case, possibly due to the increased numerical magnetic
diffusion at the lower end of the layer.  } \end{figure}

\begin{figure}
\begin{center}
\includegraphics[width=3in]{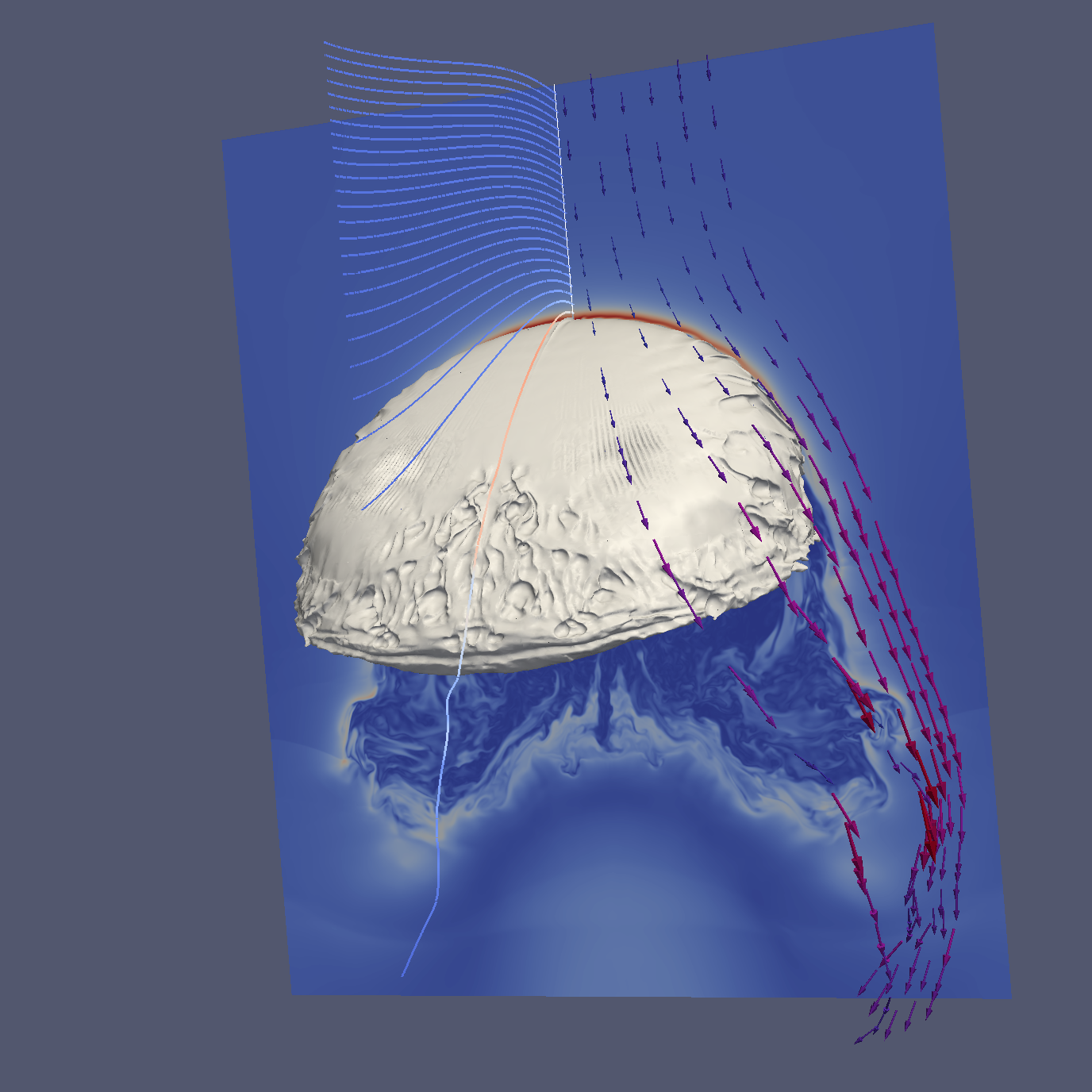} 
\end{center}
\caption{  
\label{fig:highresrendering}
{\bf Rendering of the high resolution run.} Plotted are field lines along the stagnation line, and velocity
vectors coloured and sized by magnitude of velocity.  This is the high resolution case that is used for
comparison; just the rendering of this data required 256 processors on SciNet's GPC cluster.}
\end{figure}

The calculation of $\alpha$ for varying geometries is beyond the scope of this
work, but for our geometry, we can measure the thickness of the drape layer in
our simulations, and run resolution tests to ensure we correctly resolve the
layer.  This is done in Fig.~\ref{fig:resolutionStudy}, where we compare our
fiducial simulations with an inclination of 45 degrees (solid line) to a
simulation of double the resolution.  A rendering of the high-resolution
simulation is shown in Fig.~\ref{fig:highresrendering}.  We see that even in our
fiducial resolution case we resolve the layer thickness by two cells, although
the field is somewhat reduced from its true value due to increased numerical
diffusion.  We also note that the stagnation line is the point of maximum field
line compression and places the most stringent condition on resolution.  Well
off the stagnation line, the magnetic drape is clearly well resolved by four or
more grid cells.  However, it is the stagnation line, where the magnetic field
is greatest and in the thinnest layer, which is most relevant for the
synchrotron signal from the draped layer.

\subsection{Magnetic field orientations}

The orientation of our galaxy model with respect to the direction of
motion and the ambient magnetic field are parameters we are rather free
to choose.  

For the 'inclination' of the galaxy (the angle between the normal to the
galactic plane and the velocity of the galaxy), a face-on orientation
(inclination of zero) is somewhat trivial due to the high degree of symmetry and
closely follows our previous work\cite{2008ApJ...677..993D}, so is unnecessary
to reproduce here.  At the other extreme, an edge-on orientation (inclination of
90 degrees) is somewhat less symmetric, but has a small phase-space for
occurrence, and is too sensitive to the differences between our galaxy model and
a realistic galaxy.  Thus the simulations performed for this work typically had
intermediate inclinations. The simulations in Figs.~1 and 2 had an inclination
of 45 degrees as a representative orientation, breaking enough symmetries to
give non-obvious results.

The 'tilt' between the magnetic field orientation and the plane defined by the
galaxy velocity vector and its normal is taken to be constant in our models of a
uniform magnetic field, from zero to ninety degrees.  In
Sects.~\ref{sec:reorientation} and \ref{sec:tilt} we discuss some intermediate
cases in the context of determining how accurately one can infer magnetic field
orientation.  We also consider some cases of a time-varying orientation of
magnetic field.

In addition to a homogeneous magnetic field in our initial conditions, we want
to explore a field with a characteristic scale in order to address the question
on how to measure the magnetic correlation length with our proposed effect.
Analytic force-free magnetic field configurations with an isotropic
3D-correlation scale do not exist. There is a number of non-isotropic 3D force
free solutions. Since we are most interested in variations along the direction
of motion as we want to study how these imprint themselves as changes in the
expected polarisation signal. Hence, we decided to maintain the homogeneous
field in the plane perpendicular to the direction of motion, but to allow for
variations of it along the direction of motion (that we take to be the
$y$-axis). Employing the force-free conditions, namely
\begin{eqnarray}
  \label{eq:force-free}
  \mathbf{\nabla}\cdot\vecbf{B} & = & \vecbf{0}, \\
  (\mathbf{\nabla} \times \vecbf{B})\times\vecbf{B} & = & \vecbf{0},
\end{eqnarray}
we arrive at one solution $\vecbf{B} = B_0 (\cos(ky), 0, \sin(ky))^T$ that
satisfies these equations. It represents a uniform field with an orientation
that rotates along the $y$-direction (our assumed direction of motion) thereby
forms a helical structure.

\section{Physics of magnetic draping}

Here, we are summarising the most important results regarding the physics of
magnetic draping over moving objects while we refer the reader to previous
work\cite{2008ApJ...677..993D} for a more detailed study.


\subsection{Magnetic energy of the draping layer}

We just saw that the magnetic energy density in the draping layer, $\eps_B\simeq
\alpha\rho\vel^2$, is solely given by the ram pressure wind and {\em completely}
independent of the strength of the ambient cluster fields. However, the total
energy in the draping sheath is proportional to the energy density of the
ambient cluster field. We demonstrate this analytically for the simple example
of a sphere with radius $R$ and volume $V_\rmn{sph}$, while we assume a constant
thickness of the drape $l_\rmn{drape}$ (see Eqn.~\ref{eq:l_drape}). The total
energy in the drape covering the half sphere with an area $A=2\pi\,R^2$ is given
by
\begin{eqnarray}
  \label{eq:E_drape}
  E_{B,\,\rmn{drape}} 
  &=& \frac{B_\rmn{drape}^2}{8\pi}\, A\,l_\rmn{drape} = 
  \frac{B_\rmn{drape}^2}{8\pi}\,\frac{A\,R}{6\alpha\,\M_A^2} \nonumber\\
  &=&  \alpha\rho\vel_\rmn{gal}^2\,\frac{A\,R}{6\alpha}\,
  \frac{B_\rmn{icm}^2}{4\pi\,\rho\vel_\rmn{gal}^2}=
  \frac{1}{2}\,\eps_{B,\,\rmn{icm}}\,V_\rmn{sph}.
\end{eqnarray}
In our case of draping over a galaxy, we expect geometrical factors to enter so
that our assumptions will be modified and we need to numerically simulate
and study this problem in more detail\cite{D&P}.


\subsection{Developing the draping layer}

As suggested by Eqn.~\ref{eq:E_drape}, the time necessary to sweep up the
magnetic drape is much shorter than the orbital crossing time (the time
it takes for a galaxy to cross the entire cluster; $\tau_\rmn{cross} = 2\,
R_\rmn{cluster}/\vel_\rmn{gal}$) of the galaxies in the cluster. We have
$E_{B,\,\rmn{drape}} \simeq 0.5\, \eps_{B,\,\rmn{icm}} V_\rmn{gal} \simeq 0.5\,
E_{B,\,\rmn{icm}}(< V_\rmn{gal})$.  Energy conservation tells us that we need at
least half a galactic crossing time (the time it takes for a galaxy to travel
its own size; $\tau_\rmn{c} = 2\, R_\rmn{gal}/\vel_\rmn{gal}$) to get the
magnetic energy in the drape - and possibly a factor of a few more as not all
the magnetic energy represented by field lines in the sweep-up region ahead
of the galaxy ends up in the drape as the circulation flow around the galaxy
establishes itself and magnetic back reaction starts to build up causing longer
residency timescales of the field lines in the drape. Once the magnetised
boundary layer around the galaxy is set up, the galaxy is very `sticky'; field
lines that enter the boundary layer stay there for on order ten galactic
crossing times.

\begin{figure}
\includegraphics[width=3in]{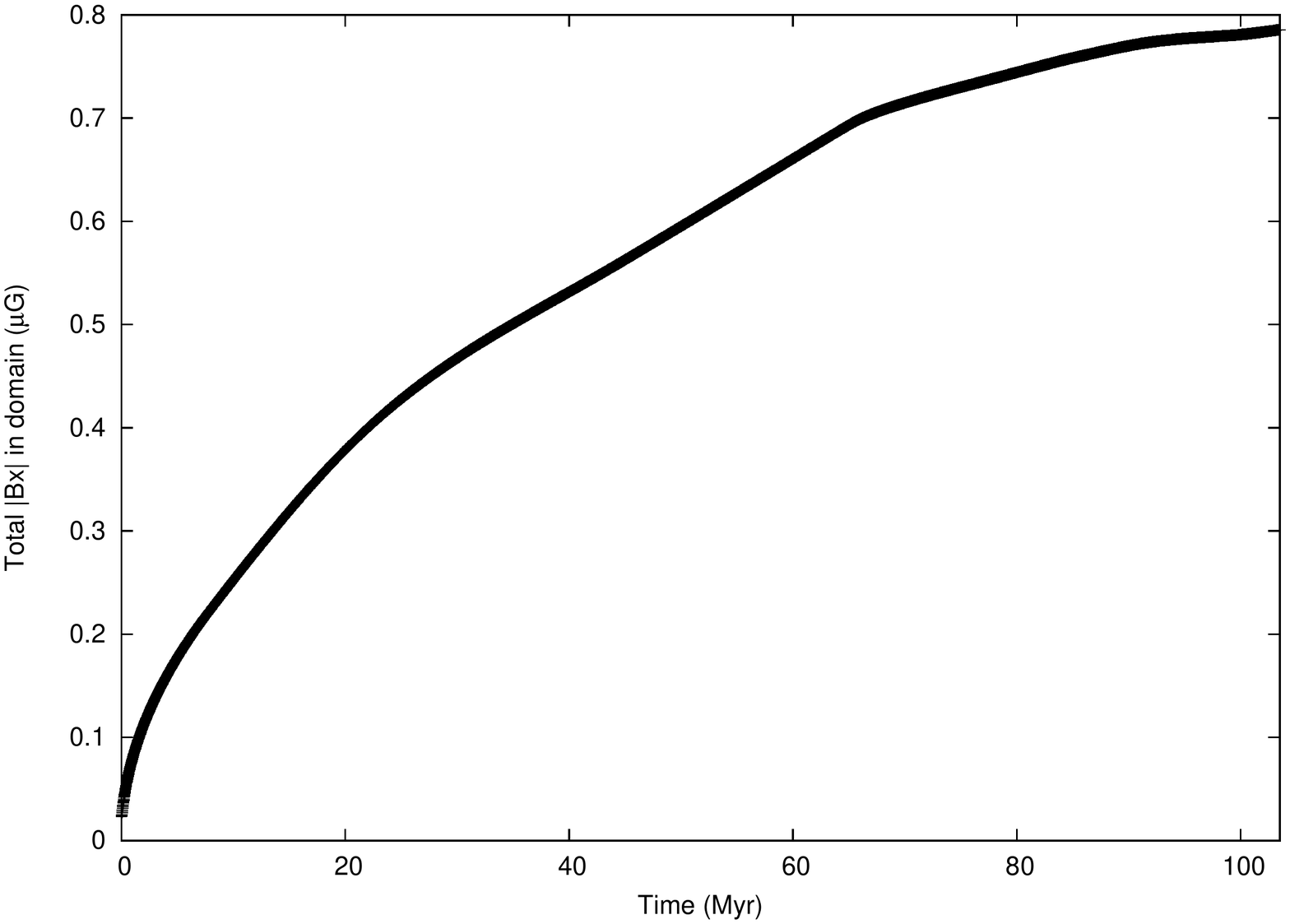} 
\caption{ {\bf Time evolution of field strength.}  Shown here is the
  contribution to magnetic energy from the $x$ component of the magnetic field
  integrated through the domain for one of our fiducial simulations with the
  ambient magnetic field oriented in the $x$ direction.  It is seen to level off
  by $t \approx 82.8\,\mathrm{Myr}$; most of our analysis in this work is based
  on snapshots of the simulation at time $t = 85.25\,\mathrm{Myr}$.  Simulations
  were run to $t = 103.5\,\mathrm{Myr}$; when there are differences these are
  noted.  }
\label{fig:drapeVsTime}
\end{figure}

Exactly how long the boundary layer takes to become established, however, is a
subtle question, and one we don't address here.  We note instead that
empirically here and in previous work\cite{2008ApJ...677..993D} we find that
within five or so galactic crossing times the layer seems to have established a steady
state.  Plotted in Fig.~\ref{fig:drapeVsTime} is the $x$-contribution to the
magnetic energy in the simulation domain in a fiducial simulation where the
field was oriented in the $x$ direction.  A steady state begins to be reached by
about $t = 82.8\,\mathrm{Myr}$; our simulations run to $103.5\,\mathrm{Myr}$,
and most of our analysis is done on snapshots from $t = 85.25\,\mathrm{Myr}$.

\begin{figure}
\includegraphics[width=3in]{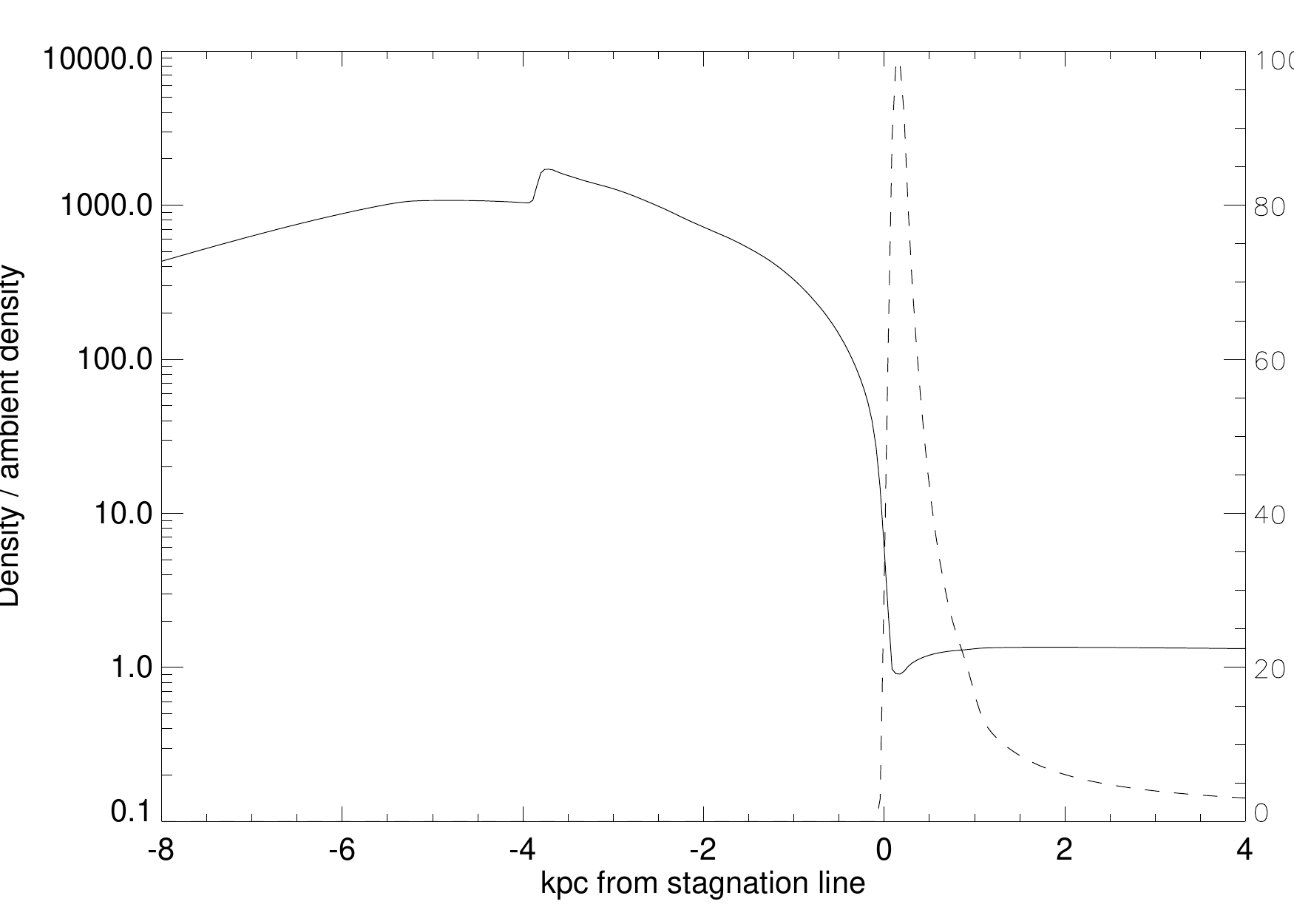} 
\caption{  
\label{fig:magdens-stagnationline}
{\bf Density and magnetic field along the stagnation line}.  The density
(in units of ambient density) along the stagnation line of a very
high-resolution draping run plotted with a solid line and using the
logarithmic scale on the left, overplotted with the magnetic energy
density (in units of the ambient magnetic energy density) plotted with
a dashed line and using the linear scale on the right.  Data was taken
from a snapshot in the simulation when the magnetic drape is still
being established.   Note that the peak of the magnetic energy density
occurs where there is actually a small {\em drop} in local density, because
of the influence of the now dynamically-important magnetic pressure.  
Crucially, the magnetic field pileup is {\em not} a compression effect.
}
\end{figure}

It is important to note that the buildup of the magnetic draping layer is a
boundary-layer effect, and {\em not} due to compression.  The flows considered
here are roughly sonic, and any resulting adiabatic compressions of fluid
elements (and of the field frozen into them) is much too small to result in such
a large magnetic field enhancement.  Indeed, as
Fig.~\ref{fig:magdens-stagnationline} shows, the peak of magnetic field energy
actually corresponds to a local {\em drop} of density.  The data from this plot
was taken from the high-resolution simulation shown in
Fig~\ref{fig:highresrendering}.

\subsection{Dissecting the draping process -- draping of turbulent fields}

Here we will argue that our choice of homogeneous fields in the plane
perpendicular to the direction of motion is in fact representative and captures
the essence of the observable draping signal in the polarised synchrotron
emission even if we were to consider turbulent fields. To this end, we need to
understand which region in the upstream ends up the the magnetic draping
boundary layer. In other words, we want to know which scale {\em perpendicular
  to the direction of motion} is mapped into the drape. In our previous work, we
demonstrated that outside the draping sheath in the upper hemisphere along the
direction of motion, the smooth flow pattern resembles that of an almost perfect
potential flow solution\cite{2008ApJ...677..993D}. The thickness of the magnetic
boundary layer $l_\rmn{drape}$ is given by Eqn.~\ref{eq:l_drape}. Hence, we want
to derive the impact parameter of the streamline to the stagnation line, that is
tangent to the outermost boundary layer.  Fluid further away from the stagnation
line than this critical impact parameter is deflected away from the galaxy and
will never become part of the draping layer. Using for simplicity the spherical
approximation, we previously derived in Appendix A.2 an approximate equation for
the impact parameter $p$ in the region close to the
sphere\cite{2008ApJ...677..993D},
\begin{equation}
  \label{eq:impact}
  p = \sqrt{3 s R}\,\sin \theta.
\end{equation}
Here $s=r-R$ is a radial variable measured from the contact of the sphere with
radius $R$, and $\theta$ denotes the the usual azimuthal angle measured from the
stagnation line. Taking $s=l_\rmn{drape}$, we derive a critical impact parameter
\begin{equation}
  \label{eq:impact_cr}
  p_\rmn{cr} = \frac{1}{\sqrt{\alpha\beta\gamma}}\frac{R\,\sin \theta}{\M}
  \simeq\frac{R\,\sin \theta}{\sqrt{2\beta}}
  \simeq\frac{R}{15}.
\end{equation}
In our previous work with spherical projectiles\cite{2008ApJ...677..993D},
$\alpha \approx 2$ and we adopted numerical values used in our simulation, i.e.
${\cal M} = 1 /\sqrt{\gamma}$ and $\beta=100$. In the last step we adopted the
condition that the streamline never intercepts the draping region,
i.e. $\theta=\pi/2$. We note that most of the polarised intensity comes from a
region $\theta<\pi/6$ which implies a critical impact parameter of $R/30$ for
streamlines to intercept the synchrotron emitting region.  The theoretically
expected value of Eqn.~\ref{eq:impact_cr}, $p_\rmn{cr}\simeq R/15$, agrees well
with that found in our simulations (Fig.~\ref{Fig3}), where $p_\rmn{cr}\simeq
R/10\simeq 2$~kpc. The differences can easily be accounted by the differing
geometry that might modify the effective curvature radius in the vicinity of the
stagnation line (and possibly cause a different value of $\alpha$) and/or
magnetic back-reaction in the draping sheath that could also offset the
streamlines with respect to the simplified case of a potential flow.

These considerations suggest a criterion that the magnetic coherence scale has
to meet for magnetic fields to be able to drape and modify the dynamics of the
draped object. This draping criterion reads
\begin{equation}
  \label{eq:lambda_B}
  \frac{\lambda_B}{2\,R} \apgt \frac{1}{\sqrt{\alpha\beta\gamma\, \M^2}}.
\end{equation}
Here $R$ denotes the curvature radius of the working surface at the stagnation
line (which can be identified as its initial simulation value while it might be
modified for cases of marginal stability due to hydrodynamic instabilities),
$\alpha \approx 2$ (as measured for spherical objects in our earlier
work\cite{2008ApJ...677..993D} and can also be slightly modified of order unity
for non-spherical cases), $\gamma=5/3$, and $\M$ is the sonic Mach number.  We
can test our formula with magnetohydrodynamical simulations of fossil AGN
bubbles that rise buoyantly in a cluster atmosphere with tangled magnetic fields
of various coherence scale\cite{2007MNRAS.378..662R}. These simulations have a
plasma beta parameter of $\beta=20$ and the bubbles have a sonic Mach number of
$\M\simeq0.25$\cite{Ruszkowski2010}. The condition for bubble stability due to
draped magnetic fields reads
\begin{equation}
  \label{eq:lambda_B2}
  \frac{\lambda_B}{R} \apgt \frac{2}{\sqrt{2\times20\times\frac{5}{3}\times 0.25^2}}
  \simeq 1.
\end{equation}
These simulations explored two archetypal cases. (1) The `draping case' had a
magnetic coherence scale was much larger than the bubble radius,
$\lambda_B/R\simeq 4.8$. It showed that draping occurs under these circumstances
(just as our formula of Eqn.~\ref{eq:lambda_B} suggested) and showed the ability
of the draped magnetic fields to suppress Kelvin-Helmholtz instabilities. (2)
The `random case' had a magnetic coherence scale that was of order the initial
bubble radius, $\lambda_B/R\simeq 0.96$ -- hence it was at marginal stability
according to Eqn.~\ref{eq:lambda_B}. In such a case, hydrodynamical
instabilities start to flatten the contact surface which increases the curvature
radius at the stagnation line; an effect that violates the condition for draping
of Eqn.~\ref{eq:lambda_B}. Hence, in the absence of draping the interface cannot
be longer stabilised against the shear so that a smoke ring develops as these
simulations show! We caution the reader that the quantitative predictions stated
here need to be thoroughly tested in future simulations that will help to
consolidate the picture put forward.

\begin{figure}[t]
\includegraphics[width=3in]{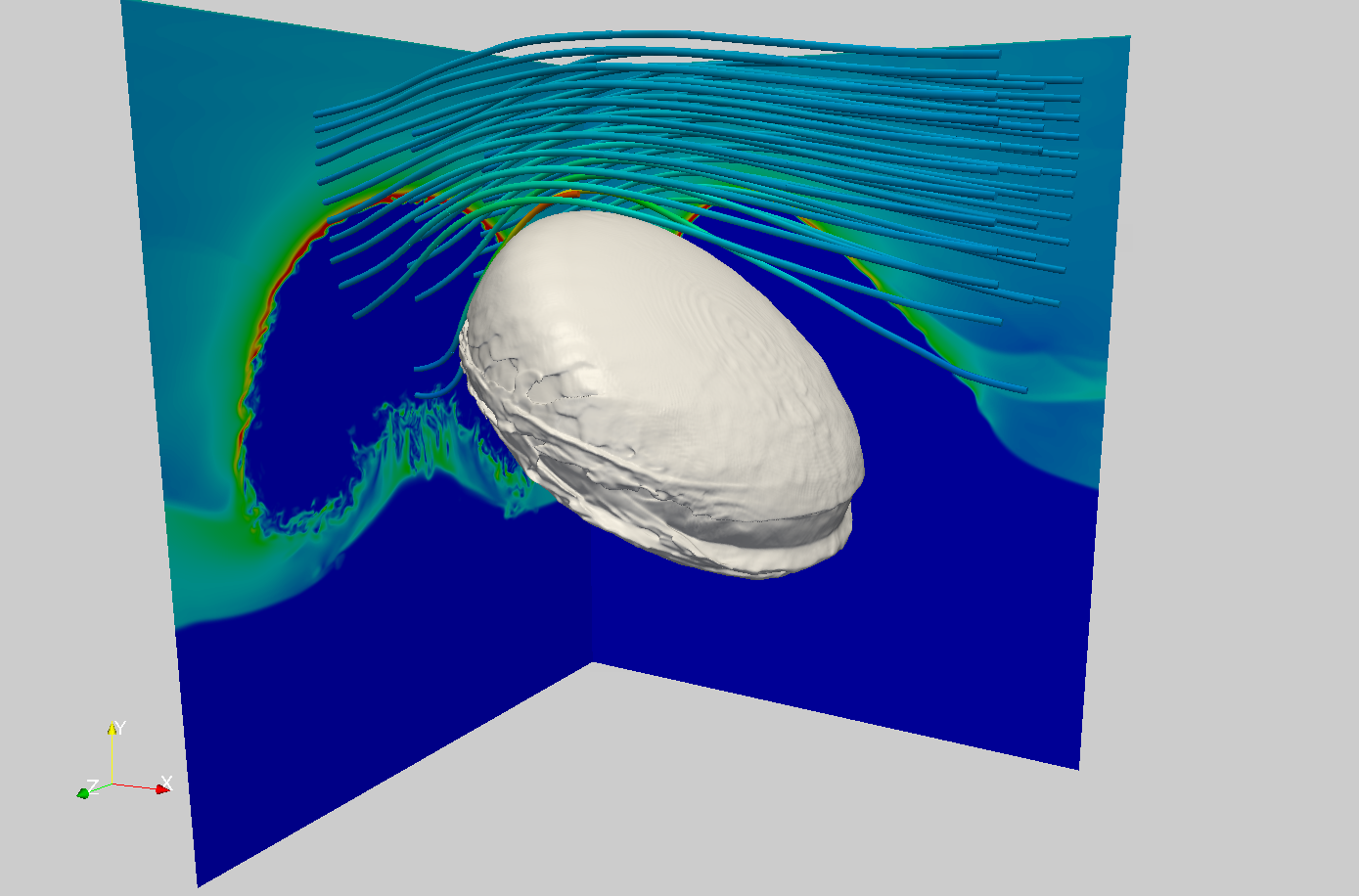}\\
\includegraphics[width=3in]{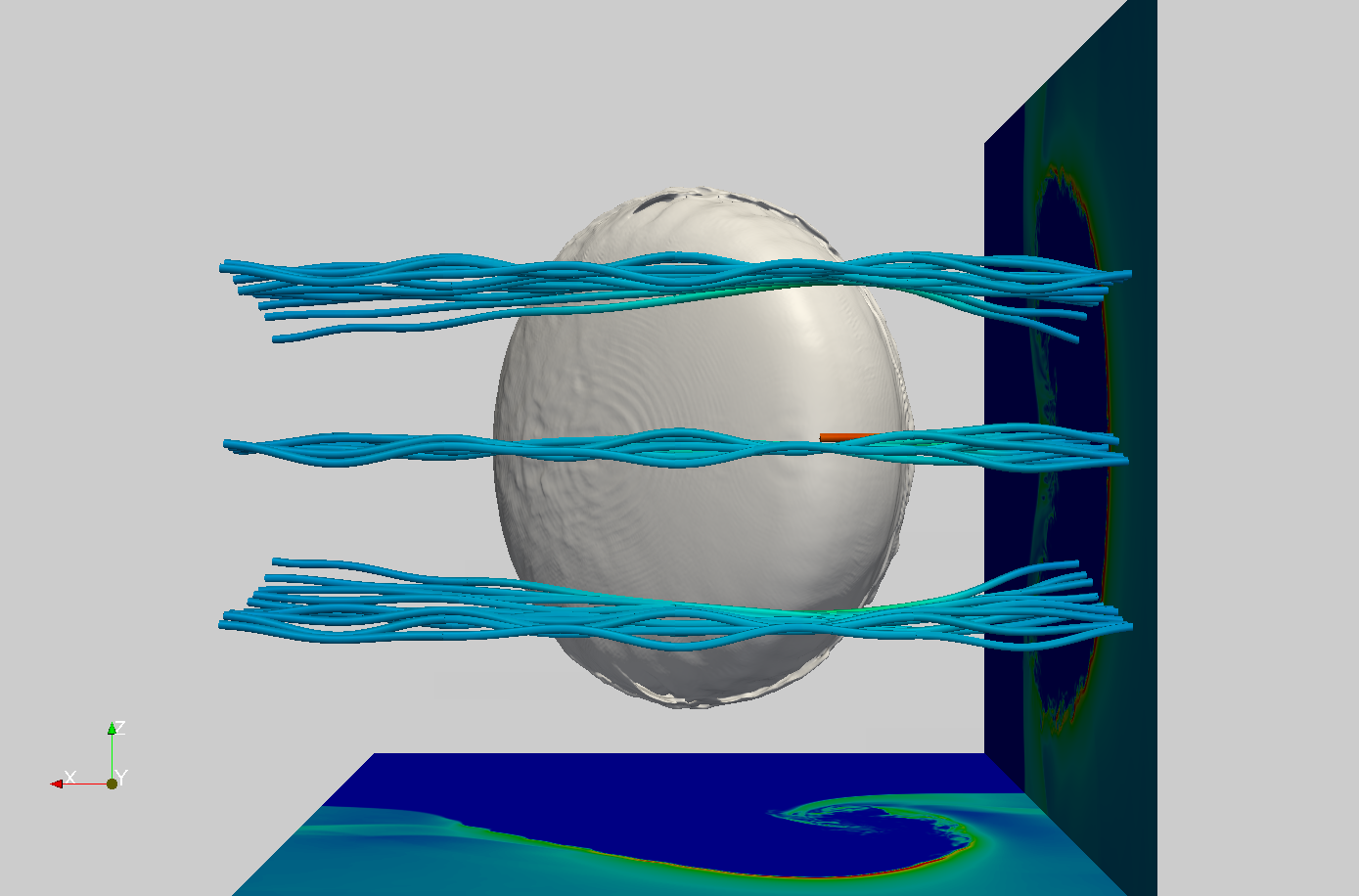}
\caption{{\bf Simulating magnetic draping of perturbed homogeneous fields:}
  Shown are views from the side and top of a simulation with homogeneous guide
  fields (similar to those in Fig.~\ref{Fig2}A). However, we added a sinusoidal
  perturbation that was pointing perpendicular to the guide field and direction
  of motion that we also varied along the direction of motion. The energy
  density in the perturbation was 1/10 of that in the guide field. Note that we
  still form a dynamically important draping sheath.}
\label{fig:perturbations} 
\end{figure}

In Figure~\ref{Fig4} we demonstrated that the magnetic coherence length {\em
  parallel to the direction of motion} needs to be at least of order the
galaxy's size for polarised emission to be significant. Otherwise, the rapid
change in field orientation leads to depolarisation of the emission. For the
sake of the following argument, we assume that the ICM were turbulent and
statistically isotropic. Assuming a Kolmogorov spectrum with a coherence scale
that is larger than the galaxy's diameter, the ratio of magnetic energy density
of that large scale to the small scale that is mapped onto the polarised emission
signal is given by 
\begin{equation}
  \label{eq:Kolmogorov}
  \frac{\delta B)^2}{(\delta B_0)^2} = \left(\frac{k}{k_0}\right)^{-2/3} \simeq 30^{2/3} \simeq 10.
\end{equation}
We simulated such a case with a homogeneous guide field with a sinusoidal
perturbation with an energy density of 1/10 of that in the guide field. As a
result we find still draping that is virtually indistinguishable from that of
the pure homogeneous case. We conclude that the draping process is robust with
respect to small perturbations and that our assumptions of homogeneous cluster
fields on the scale of a galaxy are consistent with observations and sufficient
to capture the essence of polarised signal due to draping. We note, that we are
able to place a lower limit to the coherence scale of ${\sim100}~\rmn{kpc}$ at
the position of NGC~4501 from the non-observation of the polarisation twist in
its polarised ridge. This behaviour is in fact expected if the MTI is ordering
fields on such large scales which are comparable to that of the temperature
gradient in a galaxy cluster (see also Sect.\ref{sec:MTI}).

\begin{figure}[t]
\includegraphics[width=3in]{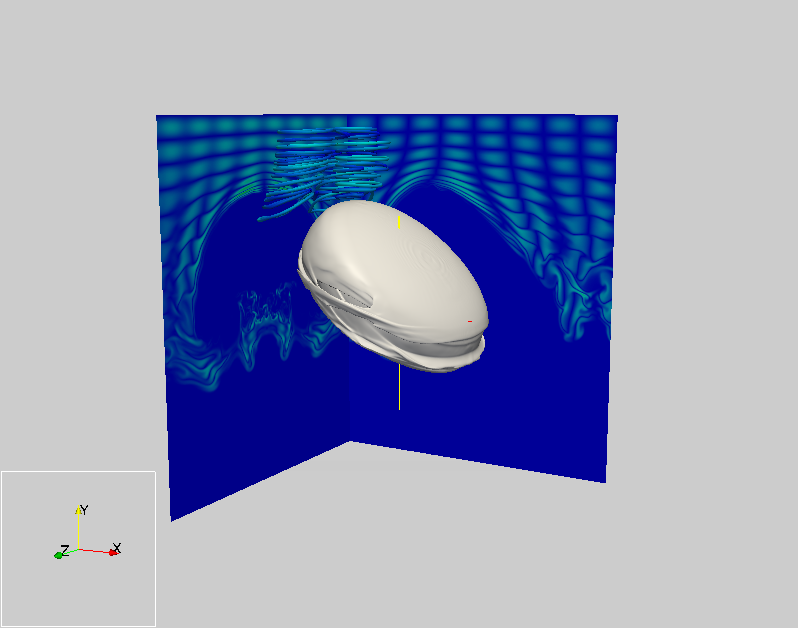}
\caption{{\bf Simulating magnetic draping of field loops:}
  Shown is a side view of a draping simulation where the field consisted of
  loops of size $\lambda = 14~\mathrm{kpc}$.
  None of the loops contain the stagnation line, so they are simply swept individually
  around the galaxy, experiencing only modest stretch; no significant draping occurs
  (field lines and cut planes are coloured by magnetic energy density using the same
  colour scale as in Fig~\ref{fig:perturbations}.  }
\label{fig:loops} 
\end{figure}

It is worth noting that not every field configuration that formally meets the
criterion of Eqn.~\ref{eq:lambda_B} forms a coherent drape. The field also needs
to meet certain morphological requirements.  In Fig~\ref{fig:perturbations}, we
show our canonical galaxy model moving through a field of `loops' of size
$14\,\mathrm{kpc}$; the field configuration coming in through the top boundary
is given by $\vecbf{B} = B_0 \cos(2 \pi y/\lambda) \left ( \sin(2 \pi
  z/\lambda), 0, \sin(2 \pi x/\lambda)\right)$ where $B_0$ is chosen to give us
$\beta = 100$.  Because the loops do not contain the stagnation line, they
simply pass over the galaxy experiencing only modest shear and thus there is
very little amplification and no draping layer is formed.


\subsection{Does the draped field orientation reflect the ambient field?}
\label{sec:reorientation}

To be able to make the claim that the draped field reflects the
orientation (in that plane) of the ambient field, we must ensure that the
field is not greatly changed by the process of draping, or at least is not
reoriented by significantly more than existing uncertainties.   This can
be tested by examining our simulations, as well as in observations of
draping in other environments.

In considering draping of the Solar wind over the Earth, recent
work\cite{2005AnGeo..23..885C} comparing the instantaneous ambient solar wind
field to the draped field over the earth found `perfect draping', that is a
draped field within $10$ degrees of the solar wind field, approximately 30\% of
the time; the remaining time the draped field orientation remained within $30$
degrees of the ambient field.  This is already promising, as the Earth's
magnetospheric system is much more complicated and rapidly changing than the
system we are considering; the Earth's tilted dipole rotates once per day, the
solar wind field changes on still shorter timescales, and imperfect ionisation
of the medium means that reconnection between these two very dynamic sources of
field plays an important role, particularly at the bottom boundary of the drape.
The upper boundary, however, we would expect to remain unchanged.

While we don't expect such dynamics in the physical system considered
here, there is however an added hydrodynamic effect; because of the
reduced symmetry compared to draping over a sphere, the flow is not
symmetric.  At the stagnation point, flow lines separate into
lines going down along the diameter of the galaxy and into lines going
along the side, dragging field lines with them.  This can tend to drag
intermediate field lines towards alignment with the plane containing
the galaxy normal and the direction of motion of the galaxy.

\begin{figure}[t]
\includegraphics[width=1.5in]{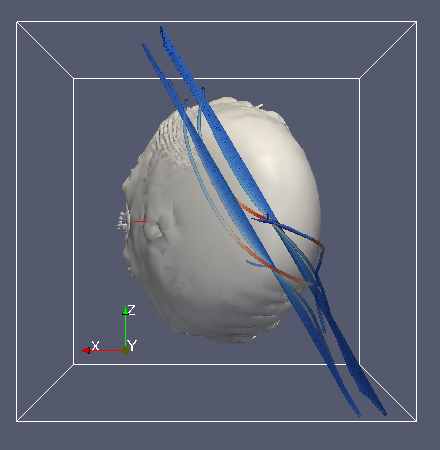} 
\includegraphics[width=1.5in]{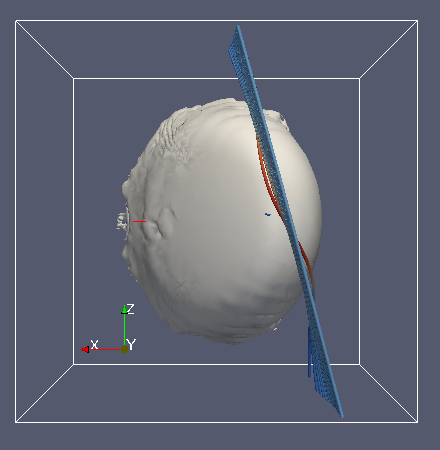} 
\caption{  
  \label{fig:varyFieldAngle} {\bf Intermediate field `tilts' and reorientation}:
  Shown are views from the top (that is, from the direction of the wind felt by
  the galaxy) of two simulations run with field tilts intermediate to those of
  the field configurations shown in Fig.~\ref{Fig2} of the main body.  As shown in the
  streamlines of the 3D renderings in Fig.~\ref{Fig4} of the main body, the flow over the
  front of the galaxy is pulled over the diameter of the galaxy; this leads to
  some modest field reorientation, shown here.  }
\end{figure}

As shown in Fig.~\ref{fig:varyFieldAngle}, for field twists of 45 and 60
degrees, we see only modest (at most $15$ degrees of rotation) effects; and most
crucially, it is quite monotonic, and therefore predictable and can be corrected
for.  The difference is largest for intermediate angles; for fields nearly
aligned with the direction of stretch, the effect is not important, and for
field lines nearly orthogonal to the effect, the effect is extremely small.
Further, because the flow follows a known pattern, the bending of the field
lines has a characteristic signature with its magnitude growing then decreasing
across the face of the galaxy.  This pattern is transverse to the flow, so
cannot be confused with a polarisation twist due to field changes in the
direction of motion of the galaxy.

We also note that this field adjustment varies with galaxy orientation
relative to its direction of motion.  For galaxies moving face-on through
the ICM, symmetry prevents any reorientation.   For galaxies moving
edge-on, the effect would be maximal, but measuring the magnitude of
the effect will require simulations with more realistic galaxy models.

On the basis of these measurements, then, we can see that field lines
nearly orthogonal to the plane connecting the galaxy normal and
direction of motion (in the coordinate system used here, this means
field lines near the $z$-coordinate direction) do not get
significantly reoriented; neither do field lines draped over a nearly
face-on galaxy.  Field lines near a 45-degree angle to the plane
connecting the normal to the direction of motion can be reoriented by
about $15$ degrees in a predictable (and thus correctable) way;
however in our current sample of galaxies we do not have any of these
cases.


\subsection{Magnetic field components in the direction of motion}
\label{sec:DOMA}

\begin{figure}
\includegraphics[width=3in]{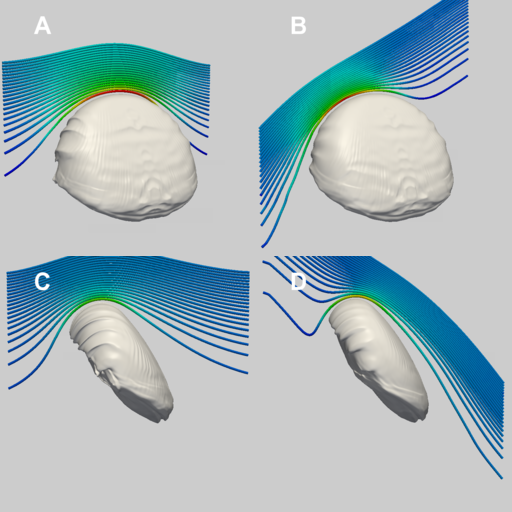} 
\caption{  
\label{fig:tiltfield}
{\bf Field component in direction of motion}:  Shown are views from
four simulations of a galaxy with a 60 degree inclination towards the
direction of motion moving into a uniform magnetic field.  In panel
{\bf A} the field is oriented in the $z$-direction, and in panel {\bf
B}, the same magnitude field is given an equal component along the $y$
direction, the direction of motion.  Similarly, in panel {\bf C}, the
field is oriented in the $z$ direction, whereas in {\bf D} that same
field now also has a component in the direction of motion.   In both cases
{\bf B} and {\bf D}, the position of the peak magnetic field strength
in the drape is shifted away from the hydrodynamic stagnation point;
the magnetic field `hot spot' will now not be co-located with the HI
ram-pressure induced hot spot.   This direction-of-motion asymmetry
is a sign that the field the galaxy is moving into has a significant component
along the direction of motion.
}
\end{figure}

Does the location of the maximum value of the magnetic energy density in the
drape, $\vecbf{x}(B_\rmn{max,~drape})$, always coincide with the stagnation
point? In contrast to $\vecbf{x}(B_\rmn{max,~drape})$, the stagnation point is a
direct observable as the ram pressure causes the neutral gas -- as traced by the
HI emission -- to be moderately enhanced at the stagnation point which results
in an `HI hot spot'. Hence this question addresses a potential bias of the
morphology of the polarised intensity with respect to the direction of motion.

Let's first look at the two extreme cases: (1) a homogeneous magnetic field
perpendicular to the galaxy's direction of motion and (2) a field that is
aligned with the galaxy's proper motion. In the first case,
$\vecbf{x}(B_\rmn{max,~drape})$ coincides with the stagnation point as this is
the point of first contact with an undistorted field line. In the latter case,
draping is absent as the galaxy smoothly distorts the field lines and causes a
minor enhancement of the magnetic energy density around the equator with respect
to the vector of proper motion.  For any arbitrary field angle in between these
two cases (with a field component in direction of motion), we expect
$\vecbf{x}(B_\rmn{max,~drape})$ to be perpendicular to the local orientation of
the magnetic field in the upstream as this is the point of first contact with an
undistorted field line. This `direction-of-motion asymmetry' is clearly
demonstrated in Fig.~\ref{fig:tiltfield} and has to be corrected for when
estimating the upstream orientation of the magnetic field.


\subsection{Kelvin-Helmholtz instability}
\label{sec:KHI}

We expect the contact boundary layer between these galaxies and the ICM to
become unstable to shearing motions on time scales less than the orbital
crossing time of these galaxies $\tau_\rmn{cross}\simeq 2\,
R_\rmn{cluster}/\vel_\rmn{gal}\simeq2$~Gyr.  In the absence of gravity, the
growth rate of this Kelvin-Helmholtz instability is given by\cite{chandra}
\begin{equation}
  \label{eq:KH}
  \omega_\rmn{KH} =  
  k\,\Delta \vel\, \frac{\sqrt{\rho_\rmn{ism}\,\rho_\rmn{icm}}}
  {\rho_\rmn{ism}+\rho_\rmn{icm}} \simeq 
  \frac{3\pi\,\vel_\rmn{gal}}{\lambda}\,\sqrt{\frac{\rho_\rmn{icm}}{\rho_\rmn{ism}}}\simeq
  \frac{3\pi\,c_\rmn{ism}}{\lambda}.
\end{equation}
Here we used the maximum shear velocity of the potential flow solution in
the absence of a magnetic field\cite{2008ApJ...677..993D}, $\Delta\vel=
1.5\,\vel_\rmn{gal}$, and energy conservation at the interface,
$\rho_\rmn{ism}^{}\,c_\rmn{ism}^2 = \rho_\rmn{icm}^{}\, \vel_\rmn{gal}^2$. It turns
out that the self-gravity of the disk has a negligible effect on stabilising the
interface.  Hence the interface becomes unstable to the shear on a timescale
\begin{equation}
  \label{eq:KH2}
  \tau_\rmn{KH}\simeq \frac{2}{3}\,\frac{\lambda}{c_\rmn{ism}} \simeq 
  0.5~\rmn{Gyr}\,\frac{\lambda}{10\mbox{ kpc}}.
\end{equation}
Even though the multi-phase structure of the ISM could provide some stabilising
viscosity, we still expect the small scales to become unstable on timescales
much less than an orbital crossing time of the cluster in the absence of a
stabilising agent such as magnetic tension in the magnetic draping layer in the
direction of the field\cite{2007ApJ...670..221D}.

\section{Modelling  cosmic ray electrons}

Our MHD simulations of a projectile moving through the intracluster magnetic
field self-consistently models the magnetic field distribution in the contact
boundary layer between the galaxy and the ICM.  The observable synchrotron
radiation, however, requires modelling a second ingredient -- high energy
electrons or cosmic ray electrons (CRes) which are accelerated by the field, thus
emitting radiation at a characteristic frequency\cite{RybickiLightman}
\begin{equation}
  \label{eq:nu_s}
  \nu_\rmn{sync} = \frac{3 e B}{2\pi\, m_\rmn{e} c}\,\gamma^2 \simeq 
  5 \mbox{ GHz}\, \frac{B}{7\,\mu\mbox{G}}\, 
  \left(\frac{\gamma}{10^4}\right)^2,
\end{equation}
where $e$ denotes the elementary charge, $c$ the speed of light, $m_\e$ the
electron mass, and the particle kinetic energy $E / (m_\e c^2)= \gamma-1$ is defined
in terms of the Lorentz factor $\gamma$. In the draping layer, these electrons
cool on a synchrotron timescale of
\begin{equation}
  \label{eq:tau_syn}
  \tau_\rmn{sync} = \frac{E}{\dot{E}} = 
  \frac{6 \pi m_\e c}{\sigma_\rmn{T} B^2 \gamma} = 
  5\times10^7 \,\rmn{yr}\,\left(\frac{\gamma}{10^4}\right)^{-1}
  \left(\frac{B}{7\,\mu\rmn{G}}\right)^{-2}
\end{equation}
where $\sigma_\rmn{T}$ is the Thomson cross section. In the absence of a
magnetic field, we can derive an upper limit on the cooling time by considering
inverse Compton interactions of these CRes with the photons of the cosmic
microwave background, yielding
\begin{equation}
  \label{eq:tau_IC}
  \tau_\rmn{ic} = 
  \frac{6 \pi m_\e c}{\sigma_\rmn{T} B_\rmn{cmb}^2 \gamma} = 
  2\times10^8 \,\rmn{yr}\,\left(\frac{\gamma}{10^4}\right)^{-1},
\end{equation}
where we use the equivalent magnetic field strength of the energy density of the
CMB at redshift $z=0$, $B_\rmn{cmb} = 3.24\,\mu\mbox{G}$.


\subsection{Origin of cosmic ray electrons}

We now turn to the question about the source of the CRes. There are two
possibilities -- they could either come from the ICM, or from within the
interstellar medium of the galaxy itself. We will show that the first
possibility is inconsistent with the observations by discussing three detailed
models, namely injection by a population of multiple active galactic nuclei,
shock acceleration ahead of the galaxies, and turbulent re-acceleration of an
aged population of CRes.  1)~Since these electrons have very short cooling
timescales they would have to be injected at a distance $L < \tau_\rmn{ic}
\vel_\rmn{gal} \simeq 240$~kpc ahead of the galaxy using $\vel_\rmn{gal}\simeq
1000\,\rmn{km/s}$, and possibly much closer when considering non-zero magnetic
fields within the ICM. However we neither observe the radio lobes of the AGN or
starburst driven winds nor strong intra-cluster shocks that would be responsible
for the injection of CRes. Also the published values of the radio spectral index
at the stagnation point of the polarisation ridge\cite{2004AJ....127.3375V} with
$\alpha_\nu=0.7$ are consistent with a freshly injected CRe population similar
to that observed in our Galaxy. This would be difficult to arrange with a
causally unconnected process in the ICM which had to be in the immediate
vicinity of all these galaxies. 2) Since these galaxies orbit trans-sonically,
we can safely assume that their velocities never exceed twice the local sound
speed yielding a Mach number constraint of $\M<2$. Adopting the standard theory
of diffusive shock acceleration in the test particle
limit\cite{1987PhR...154....1B}, we can relate the spectral index $p$ of the
accelerated CR distribution to the density compression factor $R$ at the shock
through $p=(R+2)/(R-1)$. Adopting the well known Rankine-Hugoniot jump
conditions and using $\Gamma=5/3$, we arrive at\cite{2007A&A...473...41E}
\begin{equation}
  \label{eq:p}
  p=\frac{4+\M^2(3\Gamma-1)}{2(\M^2-1)} > 3.3,\mbox{ for }\M<2.
\end{equation}
Hence the synchrotron spectral index $\alpha_\nu=(p-1)/2>1.15$ of such a
shock-accelerated population of CRes would be inconsistent with the
observations\cite{2004AJ....127.3375V}. 3) Re-acceleration processes of `mildly'
relativistic electrons ($\gamma\simeq 100-300$) that are being injected over
cosmological timescales into the ICM by sources like radio galaxies, structure
formation shocks, or galactic winds can provide an efficient supply of
highly-energetic CR electrons.  Owing to their long lifetimes of a few times
$10^9$ years these `mildly' relativistic electrons can accumulate within the
ICM\cite{1999ApJ...520..529S}, until they experience continuous in-situ
acceleration either via interactions with magneto-hydrodynamic waves, or through
turbulent spectra\cite{1977ApJ...212....1J, 1987A&A...182...21S,
  2001MNRAS.320..365B, 2002ApJ...577..658O, 2004MNRAS.350.1174B,
  2007MNRAS.378..245B}. This acceleration process produces generically a running
spectral index with a convex curvature\cite{1987A&A...182...21S,
  2001MNRAS.320..365B} which is not observed in these polarisation
ridges. Secondly, this process would predict an ubiquitous CRe distribution that
should light up the entire magnetic drape which is not observed. In fact, the
non-observation of this emission limits the efficiency of this process!
Finally, the universality of the ratio of total-to-polarised synchrotron
intensity\cite{Vollmer2009} $TI/PI\simeq 10$ in these galaxies suggests a
galactic source of these CRes and would again require a fine-tuning of the
available turbulent energy density everywhere in the ICM -- in contrast with
theoretical expectations\cite{2006PhPl...13e6501S}.


Having argued against the ICM as a source of these CRes, we now consider the
possible sources of CRes within the ISM. We know that the sources of CRes in our
Milky Way are supernova remnant shocks\cite{1999ApJ...525..357S,
  2006ApJ...648L..33V} as well as hadronic reactions of CR protons with ambient
gas protons\cite{2006Natur.439..695A}. Are the CRes that are injected deeply
inside the gas disk able to diffuse through the galaxy and populate the entire
magnetic drape? As we saw in Fig.~\ref{Fig2}, the process of magnetic draping causes
the field lines to become aligned with the contact discontinuity. This implies
that the CRe that enter the draping sheath from inside the galaxy gyrate the
innermost field lines with a Larmor radius of
\begin{equation}
  \label{eq:Larmor}
  r_\rmn{L} = \frac{p_\bot c}{e B} \simeq 10^{-6} \,\rmn{pc}\, \frac{\gamma}{10^4} 
  \left(\frac{B}{7\,\mu\rmn{G}}\right)^{-1}.
\end{equation}
The diffusion process in such a low-collisionality plasma perpendicular to the
field lines is suppressed by a factor of order $10^{12}$ relative to the Spitzer
value\cite{2003A&A...399..409E} and makes it impossible for these CRes to
efficiently populate the drape. 

As explained in the text, there is a mechanism that accelerate CRes directly
within the draping layer: The ram pressure that the galaxy experiences in the
ICM displaces and strips some of the outermost layers of gas in the galaxy, but
the stars are largely unaffected.  Thus the stars lead the galactic gas at the
leading edge of the galaxy at which the magnetic draping layer forms. As these
stars become supernovae, they drive powerful shock waves into the ambient medium
that accelerates electrons to relativistic energies.  Assuming a conservative
supernova rate of one per century, the number of supernova events above one
galactic scale height within $\tau_\rmn{syn}$ is given by
\begin{equation}
  \label{eq:N_SN}
  N_\rmn{sn} = \frac{\tau_\rmn{syn}}{\tau_\rmn{sn}} = \frac{5 \times
    10^7\, \rmn{yr}}{2\, e\, 100\,  \rmn{yr}} \simeq 10^5.
\end{equation}
Consider the supernova that dumps its energy $E_\rmn{sn} \simeq 10^{51}$~erg
explosively into the surrounding gas. After an initial phase of free expansion,
there will be a second phase when the nuclear energy of the explosion drives a
self-similar strong shock wave into the ambient medium (the so-called Sedov
solution). We are interested in the maximum radius of such a remnant shock as it
determines the homogeneity of the resulting CRe distribution when considering an
ensemble of supernova explosions within a synchrotron cooling time. The
condition for the shock phase is such that the energy deposited by the supernova
has to be larger than the sum of internal energy and ram pressure times the
shock volume of the ambient medium (neglecting radiative losses throughout this
process), or
\begin{equation}
  \label{eq:Sedov}
  E_\rmn{sn} \stackrel{!}{>} \frac{4}{3}\pi R_\rmn{sn}^3 
  (\eps_\rmn{icm}^{} + \rho_\rmn{icm}^{} \vel_\rmn{gal}^2).
\end{equation}
Using a typical velocity for the galaxy of $\vel_\rmn{gal}\simeq
1000\,\rmn{km/s}$, and a gas of primordial element abundance in Virgo with
$n_\rmn{icm} \simeq 10^{-4}\,\rmn{cm}^{-3}$ and $kT\simeq 6.7$~keV, we compute a
maximum shock radius of $R_\rmn{sn} = 150$~pc. It turns out that the maximum
radius for supernovae exploding inside the galaxy, but in the immediate vicinity
of the contact will be exactly the same as the ISM reacts to the ram pressure
and slightly increases its density.  It is now straight forward to calculate the
filling factor of supernova remnant shocks in the region where the stellar disk
intersects the magnetic draping layer, yielding
\begin{equation}
  \label{eq:filling}
  f_\rmn{sn} = \frac{N_\rmn{sn} \pi R_\rmn{sn}^2}{\pi R_\rmn{gal}^2} \simeq 6,
\end{equation}
where we adopted $R_\rmn{gal}\simeq 20$~kpc. This implies that multiple supernova
remnants intersect each other which leads to a smooth distribution of CRes,
consistent with what we observe in our Galactic synchrotron
emission\cite{2009ApJS..180..265G}. It follows that the CRes can maintain a
smooth distribution that follows that of the star light. 


\subsection{Spatial distribution}

As these CRes are injected in a homogeneous manner, their steady state
distribution within the draping layer is governed by transport and cooling
processes. Since these CRe are largely constrained to stay on any given line,
and this field line is `flux-frozen' into the medium, it is advected with the
ambient flow over the galaxy leading to an advection length scale of
$L_\rmn{adv}\simeq 10\,\rmn{kpc}$ as described in the text. The CRes interact
with Alfv\'en waves that cause them to random walk along that field line. Hence
the distribution along these field lines can be described by a diffusive process
with a characteristic length scale from the injection region,
\begin{equation}
  \label{eq:diff}
  L_\rmn{diff} = \sqrt{2 \kappa \tau_\rmn{sync}} = 
  6\,\rmn{kpc}\,\left(\frac{\kappa}{10^{29}\,\rmn{cm}^2\,\rmn{s}^{-1}}\right)^{1/2}.
\end{equation}

These considerations detail the spatial extent of the CRe distribution that
emits high-frequency (4.85~GHz) radio synchrotron emission. How do we normalise
their number density? Simulating all-sky maps in total intensity, linear
polarisation, and Faraday rotation measure for the Milky Way, the CRe
distribution has been derived to be exponential both in cylindrical radius from the
centre and height above the mid-plane of the galaxy\cite{2008A&A...477..573S,
  2009A&A...495..697W}
\begin{equation}
  \label{eq:n_CRe}
n_\rmn{cre} = C_0\, e^{-(R-R_\odot)/h_R} e^{-|z|/h_z}
\end{equation}
with the radial scale `height' $h_R \simeq 8\,\mathrm{kpc}$, the $z$ scale
height corresponding to the thick disk height $h_z \simeq 1\,\mathrm{kpc}$, and
$C_0$ a normalisation chosen to give representative CRe densities at the Solar
position, $C_0 \simeq 10^{-4}\,\mathrm{cm}^{-3}$. In our Galaxy, all three
components, gas, magnetic fields, and CR protons are in equipartition with an
energy density of about 1 eV per cm$^3$. The energy density of CRes is a
factor of 100 below that due to properties of electromagnetic acceleration
processes and the fact that CRe become relativistic at much smaller energies
compared to protons\cite{2002cra..book.....S}.

It is a non-trivial task to map this CRe distribution and its normalisation from
our Galaxy to Virgo cluster spirals. To this end we employ the far-infrared
(FIR)-radio correlation of these galaxies and compare it to typical parameters
for our Galaxy in order to determine whether we would need to rescale the CRe
normalisation.  As already mentioned earlier, the total synchrotron intensity
dominates over the polarised intensity which strongly suggests that star forming
regions are the primary source of the total synchrotron emission since the
turbulence -- perhaps induced by supernova explosions -- destroys the large
scale interstellar magnetic field and causes depolarisation of the intrinsically
polarised emission.  Hence the star formation rate is also expected to strongly
correlate with the overall spatial distribution of CRes in these galaxies. We
use the surface density of gas, $\Sigma_\rmn{gas}$, as a proxy of the star
formation rate\cite{1998ApJ...498..541K} and find $\Sigma_\rmn{gas} \simeq
2.5\times 10^{-3} \rmn{g\,cm}^{-2}$ in five galaxies\footnote{The subsample that
  we use for the FIR-radio correlation study consists of NGC 4402, NGC 4501, NGC
  4535, NGC 4548, and NGC 4654\cite{2006ApJ...645..186T}.} out of our sample of
eight where we had data available for this comparison. This agrees exactly with
$\Sigma_\rmn{gas}$ at the solar circle of our Galaxy\cite{2006ApJ...645..186T}.
Since the radio luminosity depends on the uncertain emission volume, we decided
to use the minimum energy magnetic field strength\cite{1994hea..book.....L} as a
robust measure of the energy density in magnetic fields and CRes. Except for
NGC~4548, we find values of $B_\rmn{min} \simeq (5\ldots8)\,\mu$G which again
agrees very well with the Galaxy's value at the solar circle of $B_\rmn{min}
\simeq 6\,\mu$G\cite{2001SSRv...99..243B}. In fact, the low value of NGC~4548,
$B_\rmn{min} \simeq 2.3\,\mu$G could indicate a lower normalisation of the CRe
distribution which would be a natural explanation of the weakness of the observed
polarised emission.

For this work, we decide to use a two-step phenomenological model in the
post-processing step and postpone a dynamical model of the transport of CRes to
future work\cite{D&P}. 1) We use the CRe distribution of an undisturbed galaxy of
Eqn.~\ref{eq:n_CRe} `inside' our model galaxy.  The centre of mass of our galaxy
is found, and the orientation of the galaxy at time zero is used to determine
the mid-plane.  (The much more massive galactic gas is never measurably
re-oriented during the course of this simulation.)  Because our model galaxy is
thicker for numerical reasons than the aspect ratio implied with the scale
heights above (our aspect ratio for the galaxy is 10:1 rather than the 20:1
given by $R:h_z$) we used $h_z = 2\,\mathrm{kpc}$; thus the cosmic ray number
density has fallen off by one factor of $e$ by the top of the model galaxy.

2) In a second step, the exponential profiles are truncated at the contact
discontinuity between the galaxy and the ICM -- \textit{e.g.}, corresponding to
the initial unperturbed cosmic-ray containing gas in the outer regions of the
galaxy being efficiently stripped away. Instead, we attach another exponential
distribution of CRes perpendicular to the 3D surface of the contact outside the
galaxy while requiring a continuous transition of the CRe distribution from
within the galaxy. This setup mimics the CRe acceleration by supernovae in the
draping region where the stars lead the ram pressure displaced interstellar gas
as well as subsequent advective and diffusive transport processes of the CRes
around the galaxy. We choose the scale height to be of the characteristic size
of a supernova remnant (Eqn.~\ref{eq:Sedov}) and leave the specific value
$h_\bot\simeq (150 \ldots 250)$~pc of this exponential as a free parameter that
will be explored in Sect.~\ref{sec:param-study}.

In detail, the contact discontinuity between the galaxy and the ICM is
determined by a density and magnetic field cutoff; the magnetic field is zero
within the galaxy and the density is low outside of the galaxy, i.e.{\ }$\rho <
\rho_c$ or $|B| > B_c$, where $\rho_c$ and $B_c$ are cutoff values chosen to
reproduce the contact discontinuity.  Any suitably small value of the magnetic
cutoff was seen to correctly separate the upper part of the galaxy from the ICM.
The implied location of the contact discontinuity was much more sensitive to the
density cutoff; after trial and error, this value was chosen to be approximately
40\% above the ambient density, $\rho_c = 1.4$.  The ICM density is seen to drop
to below the ambient value in the draping layer, increasing the contrast; and no
shocks occur above the galaxy, meaning there is no compressive density
enhancement to values above this to confuse galactic material with
shock-compressed ICM material.  A further cutoff is placed so that no cosmic
rays exist at $3~\mathrm{kpc}$ below the galactic mid-plane or beyond; this
ensures that no stripped galactic material in the wake is allowed to `light up'.
Dense stripped material travels quite slowly (approximately 1/10th or less the
speed of the ICM `wind') and the flow lines around the galaxy increase the flow
time of the stripped material to $\tau > 5 \times 10^7\,\mathrm{yr}$ to be
advected that far below the galaxy; at this time, the cosmic rays would have
already cooled to the point that they would not emit significant amounts of
synchrotron radiation at high radio frequencies.

\section{Modelling the synchrotron emission}

\subsection{Formulae and algorithms}

Here we summarise the equations for calculating the maps of the polarised
synchrotron emission. The synchrotron emissivity (power per unit volume, per
frequency, and per solid angle) is partially linearly polarised. In the limit of
ultra-high energy CRes with $\gamma \gg 1$, the synchrotron polarisation and intensity
depends on the transverse component of the magnetic field, $B_\rmn{t}=
B_\rmn{t}(\vecbf{x})$ (perpendicular to the line-of-sight), as well as on the
spatial and spectral distribution of CRes. The emissivities $j_{\bot,\|}= \dd^4
E_{\bot,\|} / \dd t\, \dd\omega\, \dd\Omega\, \dd V$, perpendicular and parallel
to $B_\rmn{t}$ are given by\cite{RybickiLightman}
\begin{eqnarray}
j_{\bot,\|} & = & \frac{1}{4 \pi} \frac{\sqrt{3} \,e^3}{8 \pi m_\rmn{e} c^2} 
\left ( \frac{2 m_\rmn{e} c \omega}{3 e} \right )^\frac{1-p}{2} C B_\rmn{t}^\frac{p+1}{2} 
\Gamma \left ( \frac{p}{4} - \frac{1}{12} \right ) \nonumber\\
&\times & \left [ \frac{2^{\frac{p+1}{2}}}{p+1} 
\Gamma \left ( \frac{p}{4} + \frac{19}{12} \right ) \pm 
2^\frac{p-3}{2} \Gamma \left ( \frac{p}{4} + \frac{7}{12} \right ) \right ].
\end{eqnarray}
Here $C=C(\vecbf{x})$ denotes the normalisation of the CRe distribution and is
defined by $N(\gamma)\dd \gamma = C \gamma^{-p} \dd\gamma$, $N(\gamma)$ the
number density of CRes $\in [\gamma,\gamma+\dd \gamma]$, $p = 2\alpha_\nu+1 =
2.4$ is the spectral index that we fix by considering the observed synchrotron
spectral index $\alpha_\nu=0.7$ at the stagnation
point\cite{2004AJ....127.3375V}, and $\omega=2\pi\nu$, where $\nu$ is the
observational frequency.  Denoting the angular position on the sky  by a unit
vector $\hat{\vecbf{n}}$ on the sphere, the polarised specific intensity
$P=P(\omega,\hat{\vecbf{n}})$ can be expressed as a complex variable (cf.{\
}reference\cite{1966MNRAS.133...67B}),
\begin{equation}
\label{eq:PI}
P(\omega,\hat{\vecbf{n}}) = 
\int_0^L \dd r \left [ j_\bot (\omega,\hat{\vecbf{n}}) - 
j_\| (\omega,\hat{\vecbf{n}}) \right ] 
\e^{2 i \chi(r,\hat{\vecbf{n}})}.
\end{equation}
A process called Faraday rotation arises owing to the birefringent property of
magnetised plasma causing the plane of polarisation to rotate for a nonzero
magnetic field component along the propagation direction of the photons. The
observed polarisation angle $\chi$ is related to the source intrinsic
polarisation angle $\chi_0$ (orthogonal to the magnetic field angle in the plane
of the sky) through
\begin{eqnarray}
  \label{eq:RM}
  \chi(r,\hat{\vecbf{n}}) &=& RM(\hat{\vecbf{n}})\,\lambda^2 + \chi_0(\hat{\vecbf{n}}) \\
  RM(\hat{\vecbf{n}}) &=& a_0
  \int_0^L B_r(\vecbf{x})\, n_{\mathrm{e}}(\vecbf{x})\, \dd r 
  \label{eq:RM2}\\
  &\simeq& 812\, \frac{\rmn{rad}}{\rmn{m}^2}\, \frac{B}{\mu G} \,
  \frac{n_{\mathrm{e}}}{10^{-3}\mbox{ cm}}\, \frac{L}{\mbox{Mpc}},
  \label{eq:RM3}
\end{eqnarray}
where $ a_0=e^3/(2\pi m_\e^2c^4)$ and $n_{\mathrm{e}}$ is the number density of
electrons. In the last step, we assumed constant values and a homogeneous
magnetic field along the line-of-sight to give an order of magnitude estimate
for $RM$ values. The Stokes $Q$ and $U$ parameters correspond to the real and
imaginary components of the polarised intensity and are the integrals over solid
angle,
\begin{equation}
\label{eq:Stokes}
Q+i U = \int \dd \Omega \,P.
\end{equation}
We discretise the equations above, rotate the simulation volume to the desired
viewing angle, and sum up the individual Stokes $Q$ and $U$ parameters of each
computational cell to obtain the line-of-sight projection. From the integrated
Stokes parameters we obtain the magnitude of the polarisation $|P| =
\sqrt{Q^2+U^2}$ and the polarisation angle $\chi = 0.5\,\arctan(U/Q)$. This
enables us to infer the orientation of the magnetic field in the draping layer
which is identical to the direction of the polarisation vectors rotated by 90
degrees and uncorrected for Faraday rotation (hereafter referred to as 
`B-vectors').

\subsection{Faraday rotation}

It turns out that Faraday rotation from within the drape can be safely ignored,
since an order of magnitude estimate yields a negligible rotation of the plane of
polarisation with
$$
  \Delta \chi \simeq 0.01 \deg \,\left(\frac{\lambda}{\rmn{6\,cm}}\right)^2\, 
  \frac{B_\rmn{drape}}{7\, \mu \rmn{G}} \,
  \frac{n_{\mathrm{e}}}{10^{-4}\mbox{ cm}^{-3}}\, 
  \frac{L_\rmn{drape}}{100\mbox{ pc}}.
$$
How does the Faraday rotation of the ICM bias the inferred magnetic field
direction?  Since the gradient of the ICM density and magnetic field strength
across the galaxies' solid angle is negligible, we expect a coherent rotation of
all polarisation angles.  Using typical values from deprojected Virgo X-ray
profiles (see Fig.~\ref{fig:profiles}), we estimate the magnitude of the overall
rotation for a line-of-sight that does not intersect the Virgo core region to
$$
  \Delta \chi \simeq 2 \deg \,\left(\frac{\lambda}{\rmn{6\,cm}}\right)^2\, 
  \frac{B_\rmn{icm}}{0.1\, \mu \rmn{G}} \,
  \frac{n_{\mathrm{e}}}{10^{-4}\mbox{ cm}^{-3}}\, 
  \frac{L_\rmn{cluster}}{1\mbox{ Mpc}}.
$$
This is small enough not to bias our inferred magnetic field direction. The
adopted parameters for the ICM outside the dense cluster core regions imply an
$RM\simeq 8~\rmn{rad/m}^2$ which is also much smaller than the measured $RM$
values towards central radio lobes. Hence this explains why the radial field
structure that we find in this work has not yet been measured in the $RM$
statistics from radio lobes. Using polarised sources with a large angular extent
such as radio relics\cite{2009MNRAS.393.1073B}, future low-frequency radio
interferometers ({\em GMRT, LOFAR, MWA, LWA}\footnote{{\bf G}iant {\bf
    M}eterwave {\bf R}adio {\bf T}elescope (GMRT), {\bf LO}w {\bf F}requency {\bf
    AR}ray (LOFAR), {\bf M}ileura {\bf W}idefield {\bf A}rray (MWA), {\bf L}ong {\bf
    W}avelength {\bf A}rray (LWA)}) should be able to detect this field geometry.

\subsection{Parameter study}
\label{sec:param-study}

\begin{figure}
\includegraphics[width=1.5in]{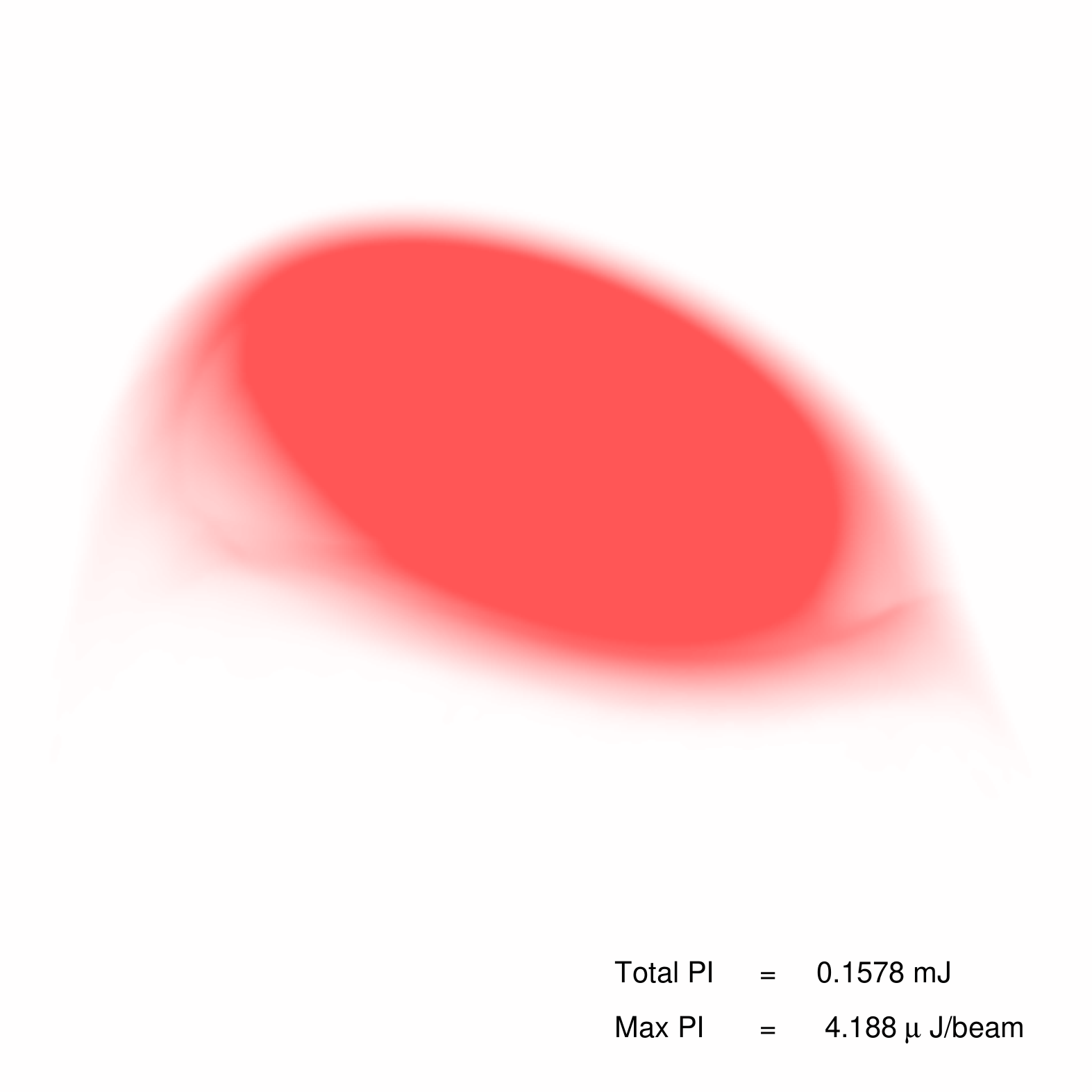} 
\includegraphics[width=1.5in]{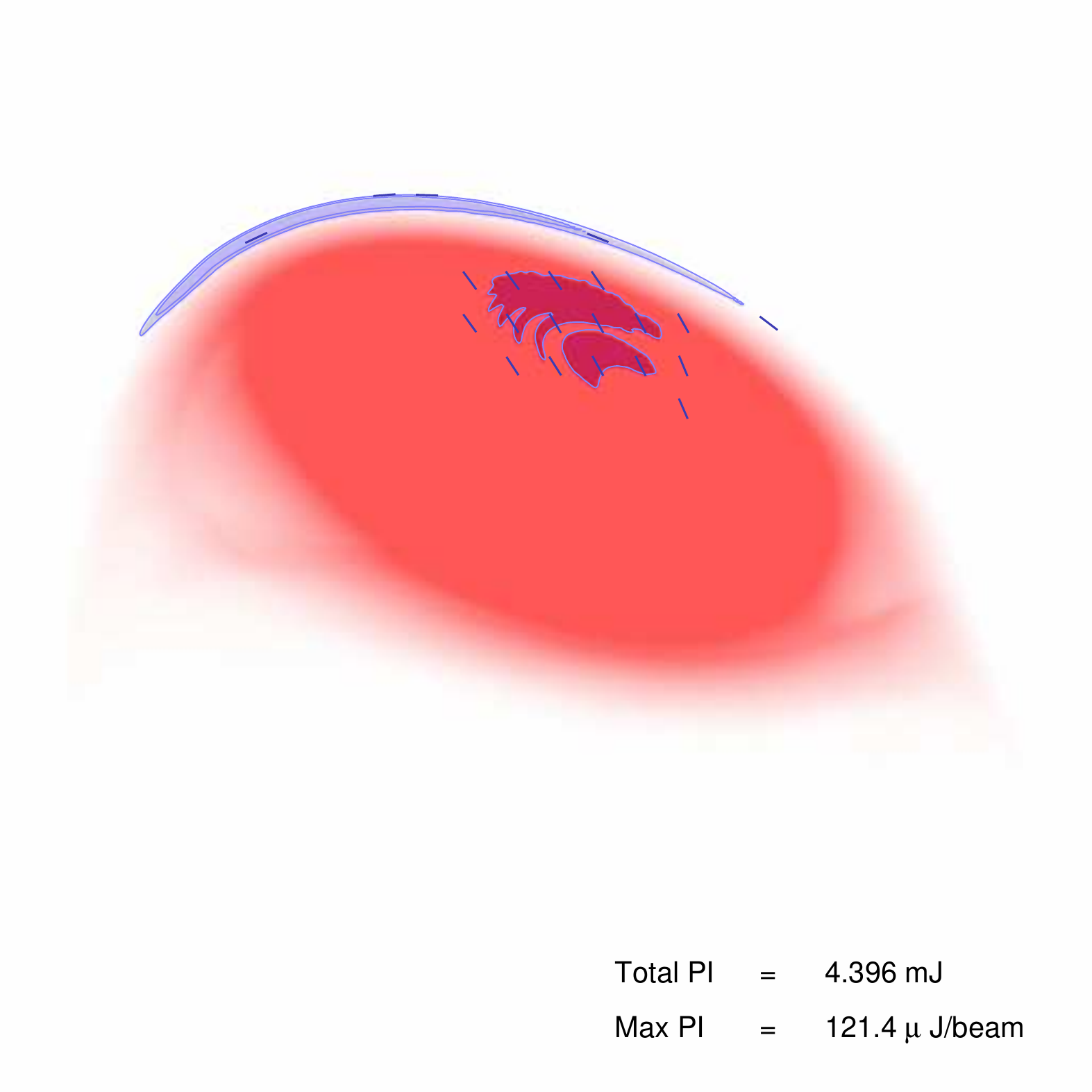} \\
\includegraphics[width=1.5in]{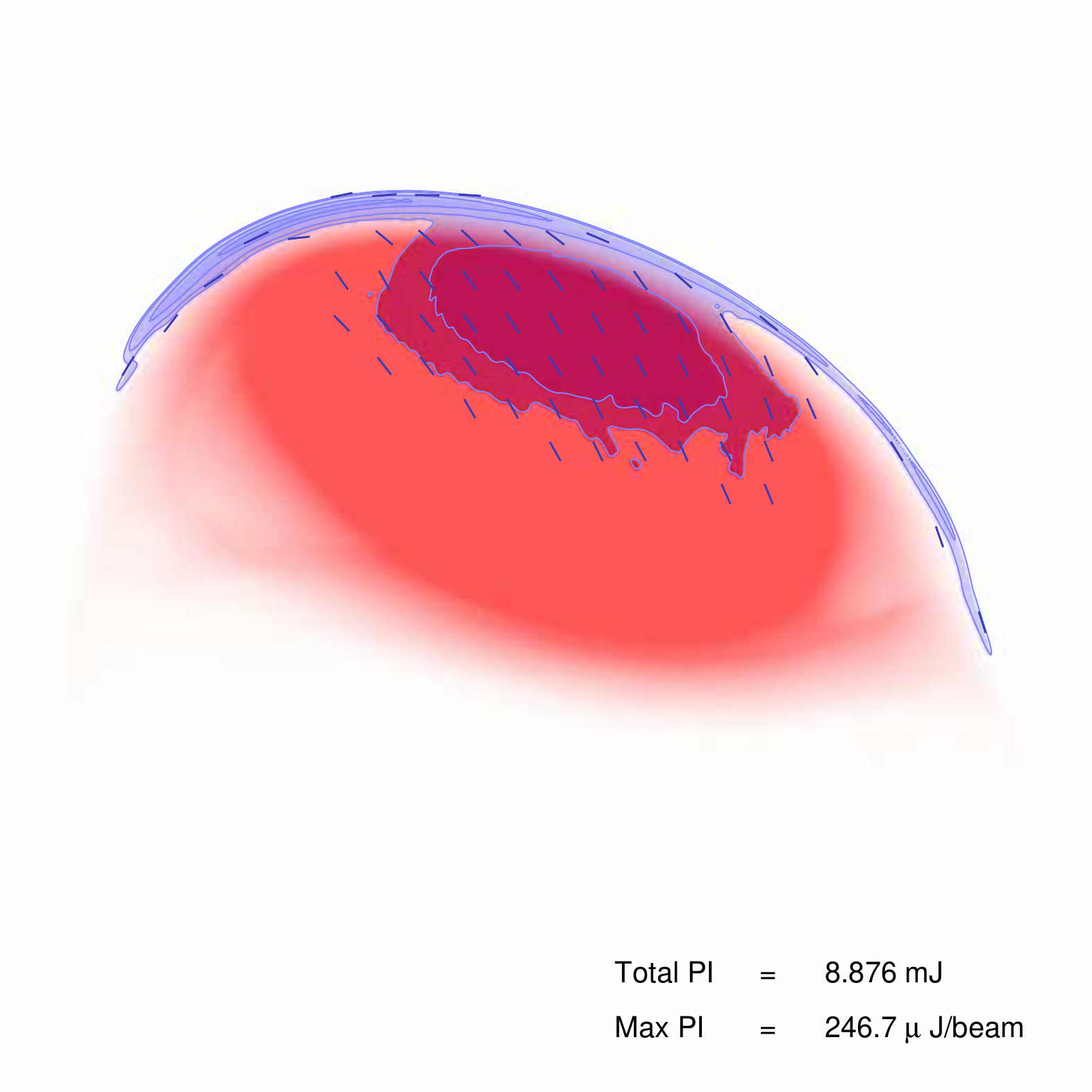} 
\includegraphics[width=1.5in]{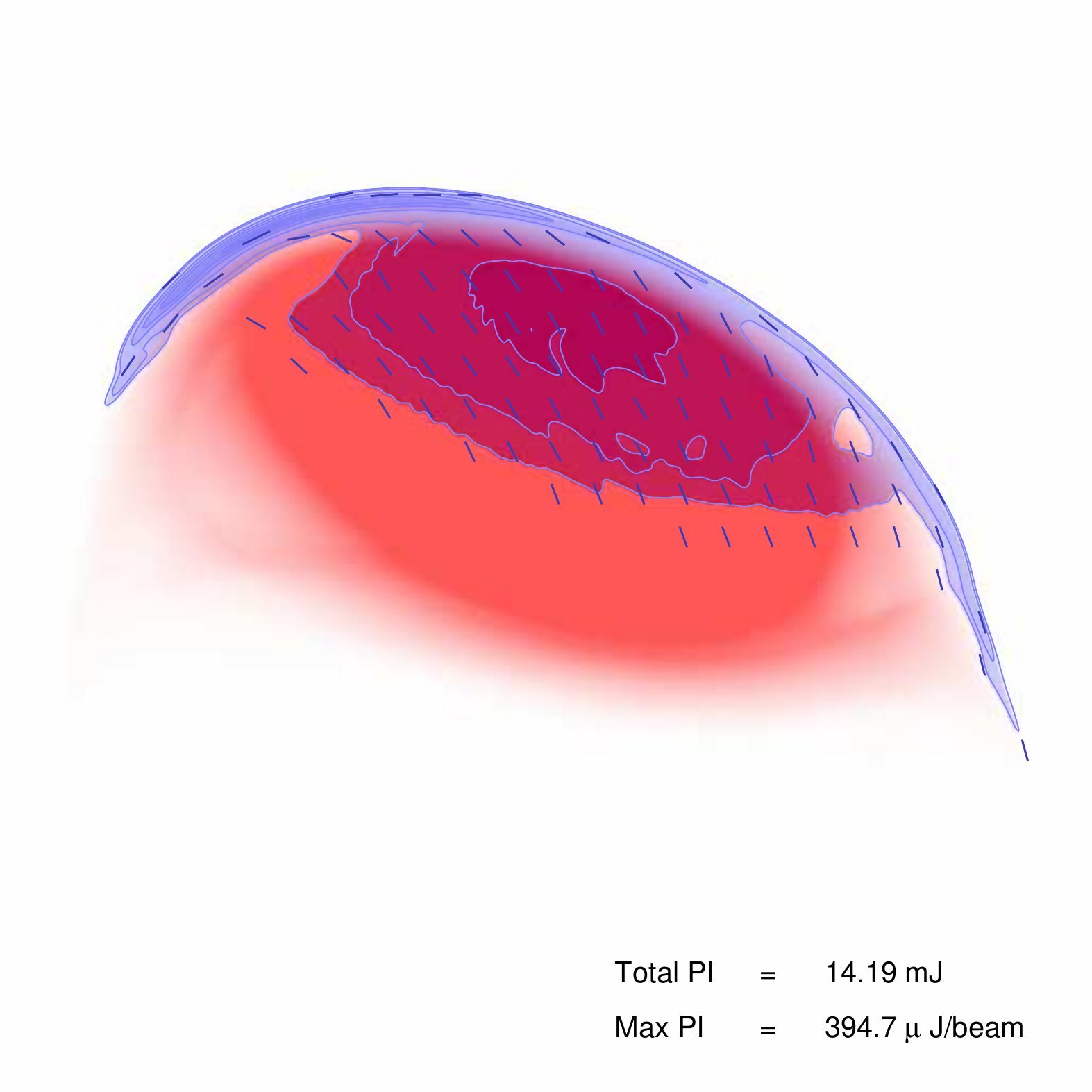} 
\caption{  
\label{fig:varyDiffusionLen}
{\bf Parameter study of cosmic ray distribution scale height.}
Parameter study on varying the cosmic ray scale height out of the galaxy.  Shown
is a projection through the simulation volume for our fiducial $\vecbf{B} = B_o
\hat{\vecbf{x}}$ case of density (red), synchrotron polarised intensity (blue
contours), and inferred magnetic field angle (lines) Compare are the cases with
a tiny scale height (scale height = $0.1\,\mathrm{pc}$), top left; one with
somewhat smaller height than used in this work (scale height =
$150\,\mathrm{pc}$), top right; with our default value (scale height =
$200\,\mathrm{pc}$, bottom left); and with a larger one (scale height =
$250\,\mathrm{pc}$, bottom right.  Contours are fixed to to those used in recent
observational work on Virgo cluster spirals\cite{Vollmer2008}.  The
extra-galactic scale height of the exponential distribution of CRes effects the
overall normalisation of the observed polarised intensity but does not alter the
morphology seen.  }
\end{figure}

In Fig~\ref{fig:varyDiffusionLen} we see the effects for one of our runs (the
same simulation and orientation as shown on the right hand side of
Fig.~\ref{Fig2} in the main body) as we vary our distribution of CRe outside of
the galaxy.  The relevant scale for these cosmic rays from supernovae should be
something of the order 300~pc, corresponding to a 150~pc radius for a typical
supernova remnant under these conditions as described above.  We plot the
resulting polarised synchrotron radiation for cases with a cosmic ray
distribution with scale heights 0.1, 150, 200, and 250~pc.  Because the
synchrotron intensity varies linearly with the CRe number density but roughly
quadratically as the magnetic field, and the magnetic field energy density is
extremely sharply peaked ($B^2 ~ s^{-1}$, {\em
  cf}. Eqn.~\ref{eq:stagnation-line-b}) at the contact discontinuity, it is the
magnetic field which is most important in setting the morphology of the region
of synchrotron emission, and adjusting the scale height of the cosmic ray
distribution primarily changes the overall normalisation.  Further, since we are
modelling our distribution as an exponential tail times the nearest galactic
value, we very quickly hit diminishing returns; once the draped region is fully
populated ({\em e.g.}, the scale height is several times the drape thickness,
88~pc in our fiducial simulation case) then further increases in scale height do
little to increase the synchrotron emission.

We choose 200~pc as a fiducial scale height for modelling our cosmic ray
distribution through the draped region, and use it through the body of this
work.  It is entirely possible that the true cosmic ray distribution will fall off
much more slowly than this, particularly at the leading edge of the galaxy where
are large stellar population exists beyond the ICM/ISM boundary.  With larger
values, however, our approach of adding the distribution everywhere over the
galaxy is too simplistic to fully account for the realistic modelling of the
transport of these cosmic rays through the draped layer by advection and
diffusion. It may even produce unphysical results by {\em e.g.}, implying
transport over distances by which the cosmic rays would have cooled down to
invisibility in the 6~cm radio observations.  On the other hand, sizes much
smaller ( {\em e.g.}, less than 150~pc --- the radius of a single supernova
remnant, a characteristic length scale for injection {\em
  cf.}~Eqn.~\ref{eq:Sedov}) starts to become difficult to justify.

To get more realistic spatial distribution of polarisation intensities, 
future work must include both realistic gas/star separation and proper
cosmic ray transport; however, we emphasise again that the magnetic field
geometry sets the geometry and the region where one could expect synchrotron
emission, the most intense emission must be near the stagnation point on
the leading end, and the remaining physics simply determines how quickly
the emission falls off throughout the rest of the drape.

\subsection{Spectral ageing effects}
\label{sec:aging}

As the CRes are injected by supernovae in the draping region where the stars
lead the ram pressure displaced interstellar gas, they are subsequently
transported advectively around the galaxy alongside the magnetic field and their
distribution is smoothed out along the field lines through diffusion. Hence we
expect the freshly accelerated CRes to show a spectral index of $p\simeq 2.3$, similar
to what we observe at supernova remnants in our Galaxy. As they cool
radiatively, their spectrum steepens so that we expect a larger spectral index
as we move towards the edges of the polarised synchrotron ridge (to be modelled
in future work\cite{D&P}). There we expect to probe either a simply ageing CRe
population or observe a superposition of an aged CRe population and freshly
injected CRes with a decreasing number density there as the ram pressure component
normal to the gas decreases for a given inclination of the galactic disk (due to
a smaller region of displacement of stars and the disk). This exact behaviour is
in fact observed along the polarised ridge in NGC~4522\cite{2004AJ....127.3375V,
  2009ApJ...694.1435M}. This is a universal prediction of our model!

\section{Bringing it together: magnetic draping and synchrotron polarisation}
\label{sec:cartoons}

In order establish an intuition of the physics of magnetic draping at
galaxies as well as the resulting radio polarisation we contrast 3D
volume renderings of our simulations and the mock polarisation maps in
the following. We will vary the galactic inclination, the tilt of the
cluster magnetic field, and the viewing angle separately.

\subsection{Varying the galactic inclination}
\label{sec:inclination}

\begin{figure*}
\begin{center}
\includegraphics[width=6.5in]{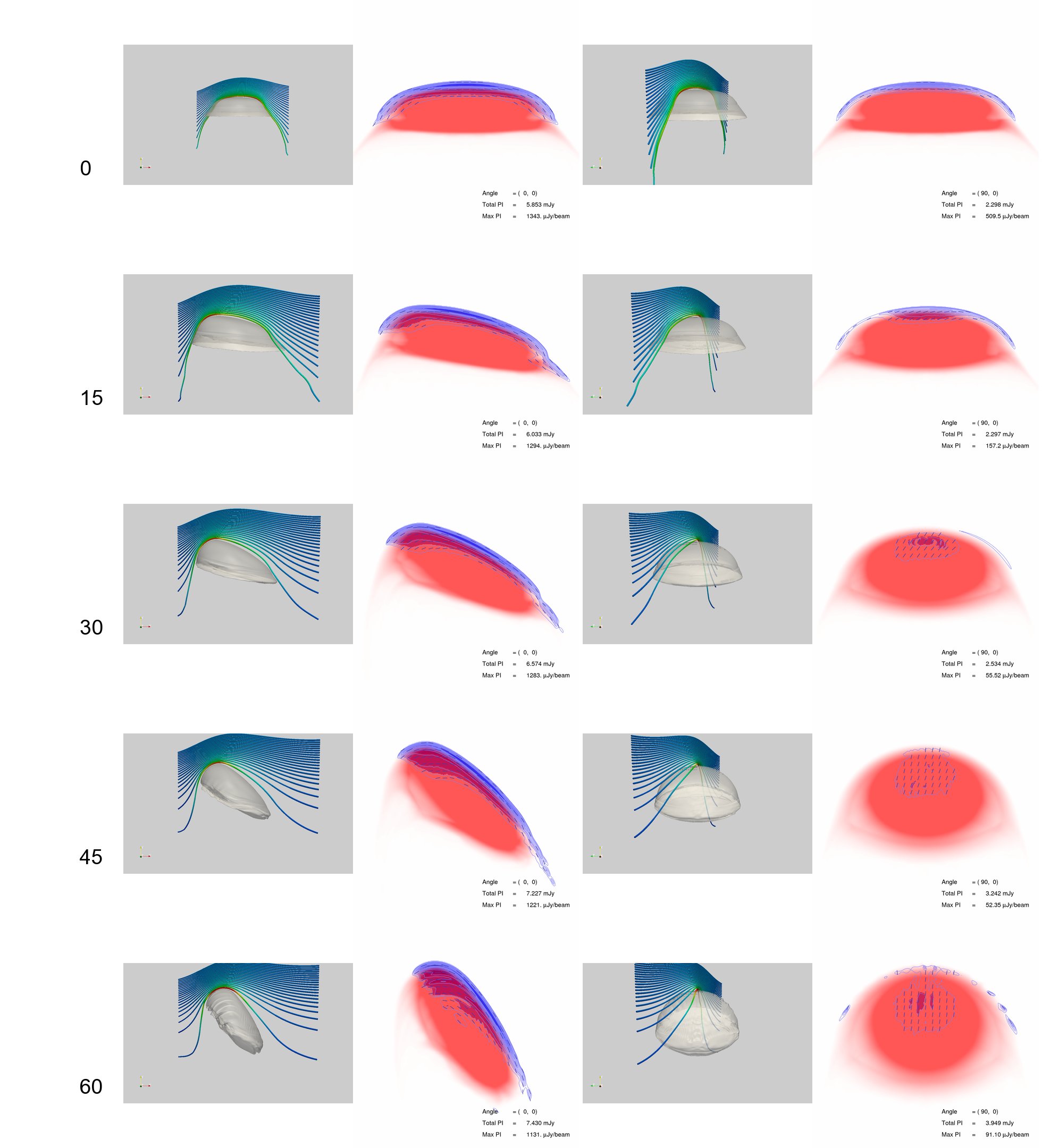}
\end{center}
\caption{
  \label{fig:InclinationsPanel} {\bf{3D Renderings and polarised synchrotron
      radiation for models of different inclination}}.  For five different
  simulations of galaxies with different inclinations moving upwards into a
  homogeneous magnetic field of a fixed orientation (30 degrees from the
  $z$-direction towards the $x$-direction), we see 3D renderings and simulated
  polarised synchrotron emission views from two points of view -- along the $z$
  axis (left) and $x$ axis (right).  Note that in this case peak polarised
  intensity is much greater viewed `edge on' than `face on'. The reason for this
  is that one observes a much greater column density of polarised synchrotron
  emitting material as one is seeing along the drape width, of order the galaxy
  size; face on, one integrates only through the drape thickness at the rim.
  Also note that peak magnetic energy density is higher in the
  higher-inclination cases; as the field gets draped over a region of smaller
  curvature radius, it is harder to move the field off of the stagnation point,
  and so the steady-state field strength is greater.}
\end{figure*}

In Figure \ref{fig:InclinationsPanel} we show two views (edge-on, face-on) of
multiple simulations with varying the galactic inclination.  For the edge-on
views on the left, we observe an increase of asymmetry of the polarised
intensity (PI) for increasing inclination. The face-on views on the right show a
decrease in PI an the stagnation point. This is due to a combination of two
effects: 1) for an increase in inclination, magnetic field lines get more
effectively reoriented as they are advected with the flow over the galaxy as
discussed in detail in Sect.~\ref{sec:reorientation}. 2) This leads to a large
line-of-sight component of the magnetic field if the galaxy is viewed
face-on. As the polarised synchrotron emission only maps out the transverse
magnetic field component $B_\rmn{t}$ (perpendicular to the line-of-sight), this
effect biases the polarised emission low relative to the total magnetic energy
density in the draping sheath (causing a so-called `geometric bias').  Within a
factor of three, the magnitude of the total PI is very similar for the face-on
and edge-on view. This suggests that the geometric bias only affects the
polarised intensity at the rim, but not in projection across the galaxy. This is
supported by the direction of the B-vectors across the galaxy that resemble the
orientation of the upstream magnetic field.  

Interestingly, the overall morphology of galaxy changes due to ram pressure, in
particular the contact of the galaxy and the ICM becomes more susceptible to
Kelvin-Helmholtz instabilities for higher inclinations.  The magnetic draping
layer becomes thinner for increasing inclinations of the galaxy with respect to
the direction of motion\cite{D&P} which increases the requirements for the
numerical resolution that is need to resolve the layer.  Our resolution study in
Fig.~\ref{fig:resolutionStudy} demonstrates that our simulations with an
inclination of 45 degrees were just able to resolve the draping layer with our
fiducial resolution discussed in Sect.~\ref{sec:sim-B}. Most likely that is to
lesser extend the case for higher inclination cases causing the Kelvin-Helmholtz
instability to modify the interface of the galaxy with the ICM. The enhanced
instability for high-inclination cases might also partly be an artifact of our
numerical modelling of the galaxy that neglects the detailed physics (multiphase
ISM including its clumpiness, self-gravity) of the interface, at least for 
time scales of order $10^8$~years that we based our analysis on. We note that
these enhanced Kelvin-Helmholtz troughs that are visible in the iso-density
contours also facilitate the reorientation effect of the magnetic field as flux
tubes can be trapped more easily within them. The accompanying filamentary
behaviour of the PI also result from this artificially enhanced effect.

\subsection{Varying the tilt of the magnetic field}
\label{sec:tilt}

\begin{figure*}
\begin{center}
\vspace{-0.5in}
\includegraphics[width=5.5in]{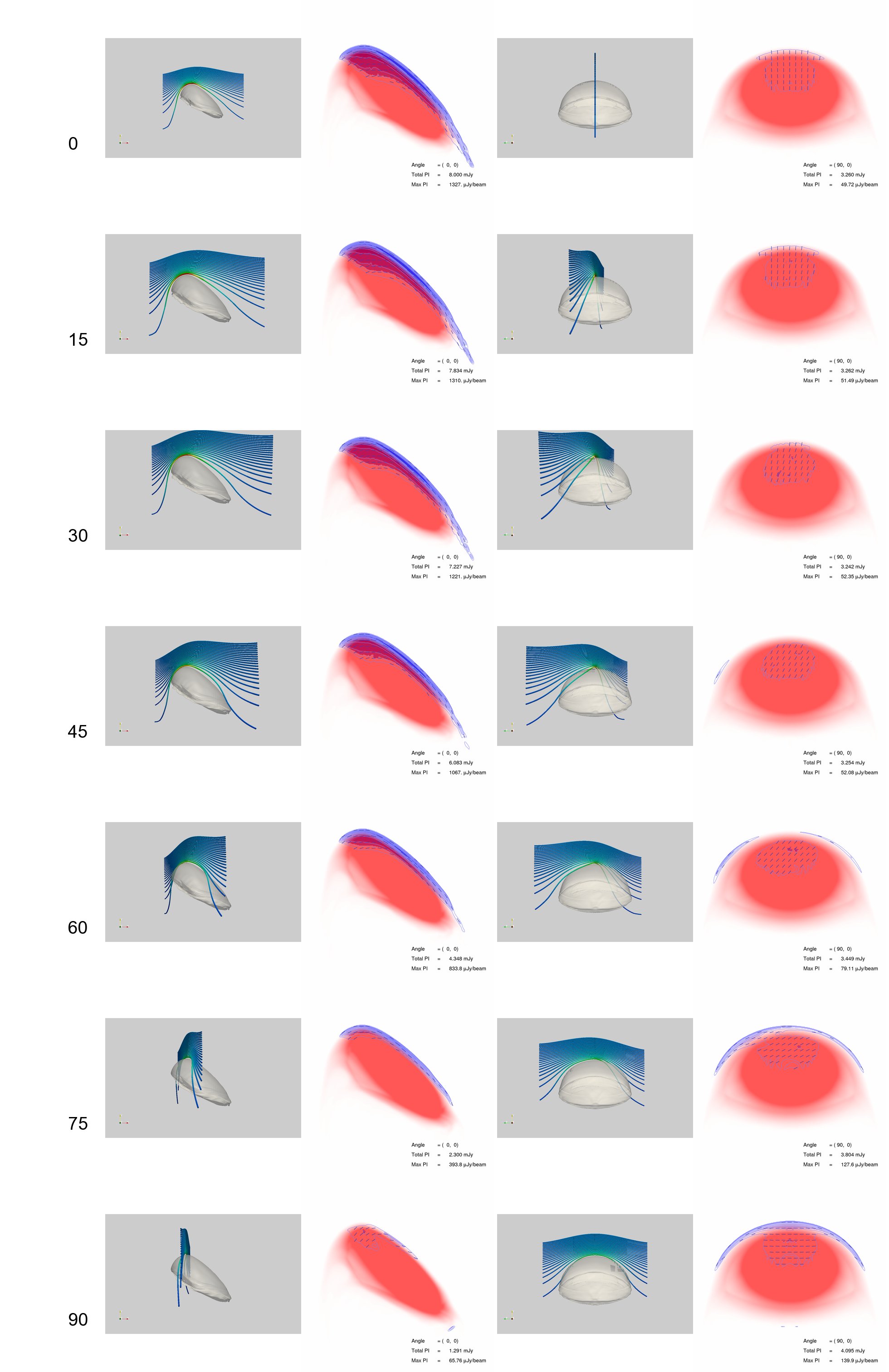}
\end{center}
\caption{
  \label{fig:TiltsPanel} {\bf{3D Renderings and polarised synchrotron radiation
      for models of different tilt}}.  Shown are seven different simulations of
  galaxies with a fixed 45 degree inclination to the direction of motion, moving
  into fields with orientations ranging from along the $x$-axis (`tilt' of zero,
  top) to along the $z$ axis (`tilt' of ninety, bottom).  On the left two
  columns are 3d renderings and simulated polarised synchrotron emission viewed
  from along the $z$ as is, and on the right, from the $x$ axis.  We see the
  progression of polarised intensity mostly being observable when viewed from
  the $z$ axis to mostly being viewable when viewed from the $x$ axis.  }
\end{figure*}

\begin{figure*}
\begin{center}
\vspace{-0.15in}
\includegraphics[width=6.5in]{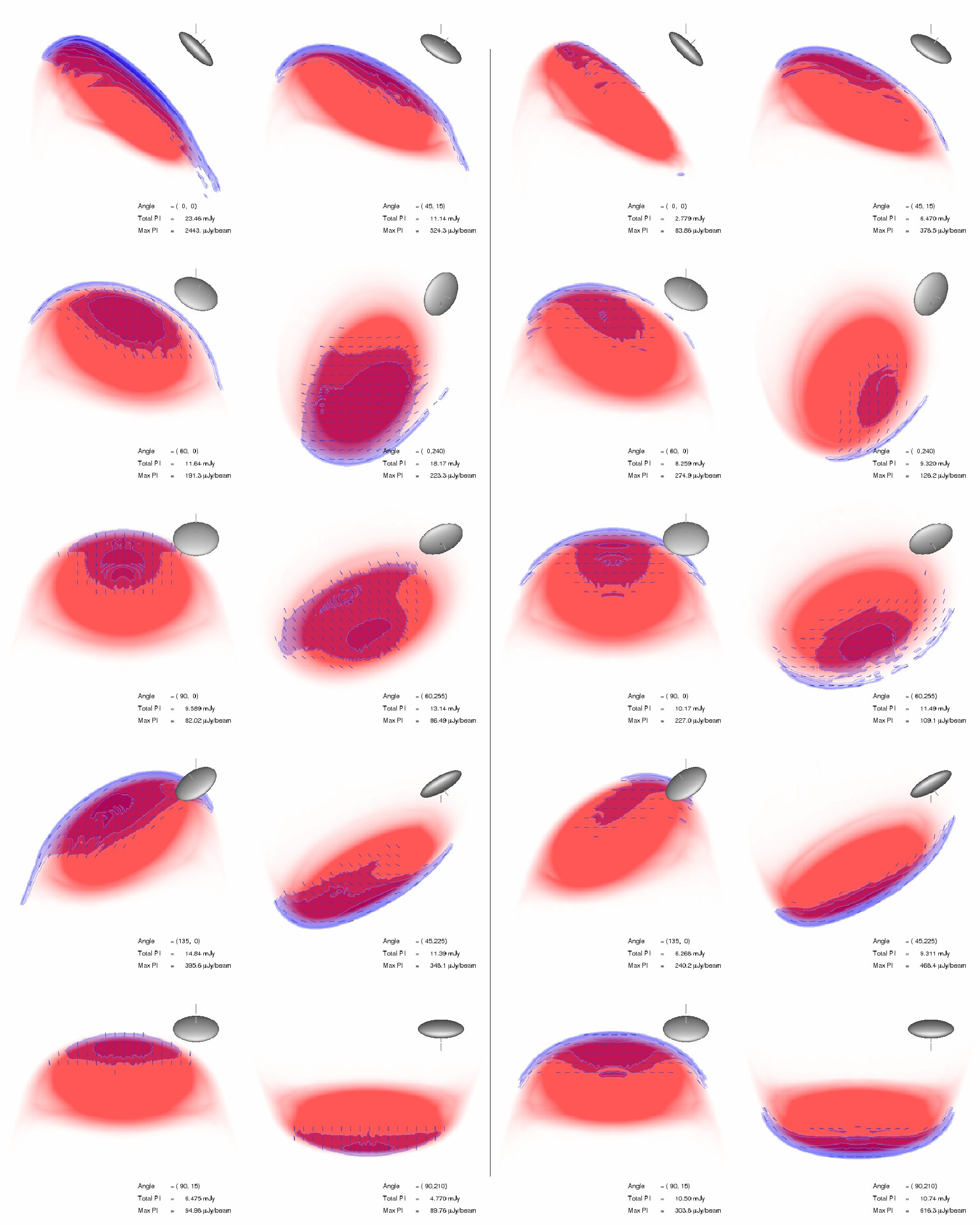}
\end{center}
\caption{
  \label{fig:ViewingAnglesPanel} {\bf{Polarised synchrotron radiation of the two
      models shown in Fig~1}} in the main paper, observed from ten different
  viewing angles.  On the left are shown different viewpoints of the model with
  a homogeneous magnetic field in the $x$-direction, as in Fig~1A and 1C; on the
  right are the same viewpoints for the model with the magnetic field in the $z$
  direction, as shown in Fig~1B and 1D.  Even for a fixed model, different
  points of view result in a rich range of synchrotron ridge morphologies.
  Nonetheless, for any given viewing angle, the difference between these two
  very different field orientations is clearly discernable.}
\end{figure*}

Figure \ref{fig:TiltsPanel} shows two views (edge-on, face-on) of multiple
simulations with varying the magnetic tilt from 0 to 90 degrees. Along this
sequence, the PI decreases for the edge-on view (left) and increases for the
face-on view (right). The reason is again the geometric bias mentioned
above. For the face-on view, this effect is accompanied with a moderate
reorientation effect of intermediate tilted magnetic fields. However, the total
PI remains almost unchanged - apparently it mostly impacts through the maximum
value for the PI and this face-on view (which has a very small configuration
space density) amplifies this effect due the high symmetry of this
configuration.  This makes it possible to infer ambient field orientation due to
the dynamics of draping even in cases where it would be impossible to see
synchrotron polarisation of an upstream B-field that is e.g., aligned with the
line-of-sight.  Note that there is a larger amplification of the magnetic field
as well as the PI in the edge-on view due to a combination of two effects.  1)
The emission region intersected by the line-of-sight is larger for the edge-on
case than it is for the face-on view where it only picks up the emission from
the rim.  2) The peak magnetic energy density is higher for the case of small
tilts compared to the perpendicular case with a tilting angle of 90 degrees. As
the field gets draped over a region of smaller curvature radius, the magnetic
tension in the field is increased. More work is needed to reach steady state
which results also in a higher magnetic energy density or similarly synchrotron
emission.  However, we caution the reader that the configuration space density
is very small in both cases and particularly chosen for their high degree of
symmetry that facilitates the understanding of the underlying draping physics
and the resulting PI maps.

\subsection{Varying the viewing angles}

Figure \ref{fig:ViewingAnglesPanel} shows multiple views of two simulations of
galaxies encountering a homogeneous magnetic field. It demonstrates that even in
this very simple case, there is a richness of synchrotron ridge morphologies
visible that clearly allows to discriminate different field geometries.  In most
cases, the total PI only varies within a factor of two barring very symmetric
cases (that however have a small configuration space density).

\section{Observational evidence and model comparison}
\label{sec:obs}

The surprising discovery of these polarised radio synchrotron ridges poses two
big questions. 1) Why does the polarised radio emission from spirals look so
different from what we observe in field spiral galaxies where the polarisation
is generally relatively symmetric and strongest in the inter-arm
regions\cite{2001SSRv...99..243B}? 2) What is the origin of these
polarisation ridges and why are they so persistent?

The answer to the first question is straightforward as we observe the large
scale magnetic field in between the dense star forming regions that constitute
the spiral arms -- probably caused by the strong shear in the disk.
High-resolution 3D hydrodynamical simulations of the ram pressure stripping of
galaxies that model the multiphase structure of the ISM self-consistently show
that the lower density gas in between spiral arms is quickly stripped
irrespective of its radius within the disk, while it takes more time for the
higher density gas to be ablated\cite{2009ApJ...694..789T}. Being flux-frozen
into this dilute plasma, the regular large scale field will also be stripped,
leaving behind the small scale field in the star forming regions. Beam
depolarisation effects and superposition of causally unconnected star forming
patches along the line-of-sight cause the resulting radio synchrotron emission
to be effectively unpolarised.  

While this work proposes a novel model to explain the second question, it is
worthwhile to consider all the observational evidence\cite{2004AJ....127.3375V,
  2007A&A...464L..37V, 2007A&A...471...93W, 2009ApJ...694.1435M,
  2010arXiv1001.3597V} that any model has to explain (we note that not all the
characteristics in the following list have yet been studied for every galaxy in
our sample).
\begin{enumerate}
\item The polarised radio morphology shows strongly asymmetric distributions of
  polarised intensity with elongated ridges located in the outer galactic
  disk. All edge-on galaxies with polarisation data (NGC 4388, NGC 4396, NGC
  4402, NGC 4522) show extraplanar polarised emission which extends further than
  the HI emission.
\item The polarisation vectors are coherently aligned across the polarised ridge
  that often extends over entire leading edge of the galaxy over scales of $\sim
  30$~kpc.
\item The distribution of stars lead the polarised emission
  ridge\cite{2007A&A...464L..37V} which itself leads the HI emitting galactic
  gas at the leading edge of the galaxy. This is the case for all inclined
  galaxies (NGC 4501, NGC 4522, NGC 4654) in our sample (see Fig.~\ref{Fig1}).
\item In a few cases (most prominently at NGC 4654), we observe radio emission
  not just at the leading edge (as inferred from the HI morphology) but
  additionally at the side of the galaxy.
\item The ram pressure causes the neutral gas as traced by the HI emission to be
  only moderately enhanced at the stagnation point by less than a factor of
  two. This `HI hot spot' is very localised spatially and much smaller in extent
  compared to the elongated polarised ridge emission.
\item The region of the `HI hot spot' is almost coincident with a flat spectral
  index ($\alpha_\nu=0.7$), while $\alpha_\nu$ steepens towards the edges of the
  polarised ridge where the degree of polarisation is very large and approaching
  values of 40\%. Most prominently this is seen in NGC 4388, NGC 4396, NGC 4402,
  NGC 4654\cite{2010arXiv1001.3597V} and NGC 4522\cite{2004AJ....127.3375V}.
\item Recent work\cite{2009ApJ...694.1435M} finds a radio-deficit region which
  lies just exterior to a region of high radio polarisation and flat radio
  spectral index, although the total 20 cm radio continuum in this region does
  not appear strongly enhanced. The strength of the radio deficit is inversely
  correlated with the time since peak pressure as inferred from stellar
  population studies and gas-stripping simulations, consistent with the strength
  of the radio deficit being a good indicator of the strength of the current ram
  pressure. This work also finds that galaxies having local radio deficits
  appear to have enhanced global radio fluxes.  
\item The ratio of total-to-polarised synchrotron intensity $TI/PI\simeq 10$ in
  these galaxies seems to be a universal property with a maximum deviation of
  only 30\%\cite{Vollmer2009}. 
\item At the contact boundary layer between these galaxies and the ICM, there
  are no Kelvin-Helmholtz instabilities observed even though these interfaces
  should become unstable to shearing motions on time scales less than the
  orbital crossing time of these galaxies (Eqn.~\ref{eq:KH2}).
\end{enumerate}

Our model of draping weak cluster magnetic fields over these galaxies is
powerful enough to naturally explain all of the mentioned properties. In fact,
the morphology of the polarised and total radio intensity, and the resulting
degree of polarisation support the picture that the total intensity traces the
star forming regions whereas the polarised intensity arises from magnetic
draping of cluster fields.  In particular, magnetic draping causes a thin sheath
of strong fields to develop that suppresses Kelvin-Helmholtz instabilities in
the direction of the field\cite{2007ApJ...670..221D} and causes a coherent
synchrotron polarisation at the leading edge of the orbiting galaxy: how exactly
our model addresses points 1-5 has been explained in the main body of the paper,
the issue of spectral ageing (point 6) in Sect.~\ref{sec:aging}, and the issue
of the Kelvin-Helmholtz shearing instability (point 9) in Sect.~\ref{sec:KHI}.

The flat radio spectral index indicates a freshly injected CRe population into
the drape that originates from supernovae that went of just outside the drape and
we do not expect to see synchrotron emission there as the magnetic field is only
starting to ramp up towards the drape -- hence the {\em radio deficit} relative
to the stellar IR emission (point 7). The global enhanced radio flux can be
understood from the additional contribution of the synchrotron emission from the
magnetic drape. The universal ratio of total-to-polarised synchrotron intensity
$TI/PI$ indicates that draping cluster fields over orbiting galaxies is a
universal process (point 8). It is important to point out that our model also
predicts galaxies with radial (with respect to the galaxies' centre) B-field
configurations. However, for galaxies on circular or low eccentric cluster
orbits and a preferentially radially aligned magnetic field (with respect to the
cluster centre), the toroidal field configuration where the B-field is aligned
with the disk, and which is observed in most of our cases, is more likely.

In contrast, there are models that attempt to explain the highly polarised radio
ridge emission as a result of compression of the interstellar magnetic
field\cite{2003A&A...402..879O, 2006A&A...458..727S, Vollmer2008} or as the
result of shear as the ram pressure wind stretches the magnetic field.  These
models however have several theoretical and observational shortcomings.  In
particular, the observed extra planar emission (point 1), the coherence of the
polarisation across the entire galaxy (2), the fact that the polarisation leads
the galactic gas (3), and the polarised ridges at the side of some galaxies, far
from the stagnation point (4) is very difficult to achieve.  Another challenge
are the observed small-extended `HI hot spots' indicating only a very moderate
compression. These would suggest a very localised (at best) polarised emission
due to field compression or shearing.  The fact that the total 20 cm radio
continuum is not enhanced also argues against these class of models as the small
scale field is expected to be also increased by these processes.  In particular,
models\cite{2003A&A...402..879O, 2006A&A...458..727S,Vollmer2008} that suggest
ram pressure modification of the magnetic field in the disk could play an
important role are likely flawed due to their approximation of the ICM/ISM
hydrodynamic interaction with `sticky particles', which contains no pressure
term and thus no back reaction, instead modelling all hydrodynamic interactions
as inelastic collisions and omitting explicit modelling of the ICM entirely.
Finally, these models do not explain the absence of Kelvin-Helmholtz
instabilities that are expected to act on scales $< 10$~kpc (see
Eqn.~\ref{eq:KH2}).

\section{Measuring the magnetic orientation}

\subsection{General considerations}

We start with general remarks that were inferred from our numerical simulations
and observations. (1) The distribution of the ram-pressure stripped gas in the
sky points in the opposite direction of the galaxy's proper velocity component
in the sky. Additionally, the maximum of the HI emission (the `HI hot spot')
that is located at the opposite direction of the galaxy disk points towards the
maximum value of the ram pressure and hence is an estimate of the 3D direction
of motion of the galaxy. (2) The maximum value of the magnetic energy density in
the drape $B_\rmn{max,~drape}$ is the point of first contact with an undistorted
field line -- hence it is perpendicular to the local orientation of the magnetic
field in the upstream (the `direction-of-motion asymmetry',
cf.~Sec.~\ref{sec:DOMA}). As the magnetic field orientation approaches the
direction of motion, draping becomes weaker until it ceases to operate for an
exactly parallel magnetic field: in this case the field lines cannot pile up
anymore at the contact of the ICM to the galaxy and the field is not amplified.
(3) The polarised synchrotron emission only maps out the transverse magnetic
field component $B_\rmn{t}$ (perpendicular to the line-of-sight). Hence, the
location of the maximum polarised intensity (PI) potentially biases the location
of $B_\rmn{max,~drape}$ towards that location in the drape where the draping
field which has a large component in the plane of the sky (the so-called
`geometric bias').

Based on these considerations, we conclude on different possibilities regarding
the alignments of the hot spots of the HI emission and polarised intensity. A
misalignment of these two hot spots implies either that there is a component of
the magnetic field along the proper motion of the galaxy $\bupsilon_\rmn{gal}$
or that the draped magnetic field at the position of the HI hot spot has only a
line-of-sight component (geometric bias). Contrary, if the hot spots of the HI
emission and polarised intensity are aligned, this implies either that there is
no magnetic field component along $\bupsilon_\rmn{gal}$ or that there is a
cancellation of the effect that the magnetic field has a component along
$\bupsilon_\rmn{gal}$ and the mentioned geometric bias.

We now establish different archetypal cases that we base our detailed galaxy
study on. (1) The proper motion of the galaxy is predominantly along the
line-of-sight (LOS). The non-draping case has a magnetic field along the
LOS. Any magnetic field configuration in the plane of the sky is easily
distinguishable through the polarisation vectors. If the magnetic field has
additionally a component along the LOS, geometric bias would cause the HI and PI
hot spots to approach. Future studies are needed to characterise the asymmetry
that accompanies this behaviour of geometric bias in order to correct for it. We
note that this configuration would tend to underestimate the LOS component of
the magnetic field due to geometric bias. (2) The proper motion of the galaxy is
mainly in the plane of the sky. As a first case, we study the case of a magnetic
field perpendicular to the proper motion where we do not expect a misalignment
of the HI and PI hot spots along the leading edge of the galaxy. A transverse
magnetic field shows a clear polarisation signature of the draping sheath.  For
a magnetic field along the LOS, there won't be much of a PI signature at the
leading edge due to geometric bias. However, as field lines are bend by the flow
in the process of draping, the PI hot spot will be shifted slightly inwards and
allow us to infer the upstream orientation of the magnetic field. Considering
now the case of a component of the magnetic field along the proper motion of the
galaxy. The case with a transverse magnetic field is easily distinguishable due
to the misalignment of the HI and PI hot spots. In the case where the magnetic
field shares components both in the direction of motion and the LOS, the LOS
component is likely to be underestimated: two competing effects such as
geometric bias and the direction-of-motion asymmetry can be responsible for the
shift of the maximum of the polarised intensity onto the projected image of the
galaxy.

\subsection{Detailed discussion of  our galaxy sample}

We now detail our procedure of measuring the field orientations from the
direction of the B-vectors (polarisation vectors rotated by 90 degrees). We
obtain the LOS velocities from the `Third Reference Catalogue of Bright
Galaxies' (RC3). They agree fairly well with those measurements from the VIVA
survey\cite{2009AJ....138.1741C}.  In general, we use the morphology of the HI
and (if available) the total synchrotron emission to obtain an estimate of the
galaxy's velocity component on the sky.  If the galaxy is well resolved in HI
and not completely edge-on, then we determine the projected stagnation point by
localising the `HI hot spot' and dropping a perpendicular to the edge of the
galaxy. In case this is not possible (because the galaxy is {\em e.g.,} edge-on)
we additionally use the location and morphology of the polarised radio emission
as an independent estimate while keeping in mind the potential biases that are
associated with it.  We then determine the orientation of the B-vectors in a
region around this stagnation point and identify this with the orientation of
the ambient field before it got swept up.

In the following, we provide important information on our measurements of the
magnetic field orientation at individual galaxies while we refer the reader to
Fig.~\ref{Fig1} in the main body of the paper for visual impression of the
polarisation ridges. HI data is based on the VIVA
survey\cite{2009AJ....138.1741C} while we take the data for the polarisation
ridges for the majority of our galaxies from the quoted
references\cite{2007A&A...464L..37V, 2010arXiv1001.3597V} except for
NGC~4522\cite{2004AJ....127.3375V} and NGC~4548\cite{2007A&A...471...93W}.  For
convenience, we identify their location in the Virgo map (Fig.~\ref{Fig6})
relative to the cluster centre (M87), use the velocity colour coding of
Fig.~\ref{Fig6}, and order them by catalogue number. In addition, we provide a
measurement of the transverse component of the magnetic field, $B_\rmn{t}$, and
quote its deviation with respect to the projected radial cluster direction
$\theta$ which is measured with respect to a radial line connecting M87 and the
respective HI hot spot. We emphasise that $\theta$ is defined to be the
difference between two projected angles on the sky.
\begin{description}
\item[NGC 4388] Small, red galaxy at 3 o'clock, nearly edge-on orientation
  ($\vel = 2538\,\rmn{km/s}$). There is no clear tail of stripped gas visible in
  HI which suggests a large LOS component of the velocity which is also
  consistent with the large red-shift. This galaxy shows two HI hot spots of
  nearly equal strength, one towards the central-eastern part, the other towards
  the central-western part of the disk ($\alpha_\rmn{e}=12:25:48$,
  $\delta_\rmn{e}=12:39:44$, $\alpha_\rmn{w}=12:25:45$,
  $\delta_\rmn{w}=12:39:45$). Each of these spots seem to show its own draping
  sheath around it which appears as a closed ridge of polarised emission with a
  B-field that is mostly aligned with the disk, but there are perpendicular
  components above and below the central regions of the disk visible. The
  vertical B-vectors in the north are most likely associated with a central
  nuclear outflow that is visible in H$\alpha$\cite{1999ApJ...520..111V}. It has
  been suggested that NGC~4388 hosts a Seyfert 2 nucleus that powers a
  $15''$-jet to the north of the galactic disk\cite{1991A&A...249...43H}.
  Judging from the morphology of OIII and H$\alpha$, the nuclear outflow is
  probably not responsible for the extraplanar emission towards the south(east)
  and the northwest which could be caused by a reoriented or even ram-pressure
  stripped field of the drape in the downstream (along the LOS).  We estimate a
  radial deviation of the transverse field component to $\theta = 35\deg$.
\item[NGC 4396] Small, blue galaxy at 1 o'clock, edge-on orientation ($\vel =
  -133\,\rmn{km/s}$).  The location of the HI hot spot coincides with that of
  the PI and is centred on the galaxy in projection ($\alpha=12:25:58$,
  $\delta=15:40:27$). There is a polarised radio continuum ridge in the
  northwestern part of the outer disk where the B-field mostly aligned with the
  disk. In the southeastern part we observe weak extraplanar polarised emission
  with field vectors pointing perpendicular to the disk. In our interpretation,
  the galaxy is moving out of the plane towards southeast and draping the radial
  cluster field which has a non-negligible LOS component around it while the
  planar and extraplanar emission directly traces the magnetic drape. This
  picture is also consistent with the long one-sided HI-tail that is extending
  to the northwest of the galaxy\cite{2007ApJ...659L.115C}.  We estimate a
  radial deviation of the transverse magnetic field component to $\theta =
  20\deg$.
\item[NGC 4402] Small, blue galaxy at 2 o'clock, edge-on orientation ($\vel =
  190\,\rmn{km/s}$). There is no stripped gas visible in HI which suggests a
  large LOS component of the velocity which is also consistent with the large
  blue-shift.  The polarised intensity shows a strong emission with a maximum in
  the western part of the disk which coincides the the HI hot spot that ranges
  from the centre to the western disk ($\alpha=12:26:06-07.5$,
  $\delta=13:06:45$). There is additional extraplanar polarised emission above
  the disk plane in the northeast of the galactic disk. This is a clean draping
  case with the field vectors aligned with the disk. The hydrodynamic drag that
  the field in the drape experiences as it is stretched and transported over the
  galaxy causes it to slightly bend over the galaxy. This can be seen as slight
  change of B-vectors of the extraplanar polarised emission which is a smoking
  gun for the ICM origin of magnetic fields that give rise to the polarised
  emission.  We estimate a radial deviation of the transverse field component to
  $\theta = 26\deg$.
\item[NGC 4501] Medium, size red galaxy at 12 o'clock, inclined orientation ($\vel
  = 2120\,\rmn{km/s}$). The HI morphology shows stripped material towards the
  north-eastern part of the disk which suggests a velocity component in the
  plane of the sky towards the south-west. The redshift as well as the fact that
  the HI hot spot is not exactly at the rim but in projection slightly inwards
  located implies an additional LOS component into the image plane.  The galaxy
  shows a strongly asymmetric distribution of polarised intensity with an
  extended global maximum in the outer southwestern part of the disk, which
  almost coincides with a region of high column density HI ($\alpha=12:31:57$,
  $\delta=14:24:55$), but not perfectly, which is one of our model
  predictions. The symmetric distribution of PI along the ridge with respect to
  the maximum of the distribution implies a negligible component of the magnetic
  field along the proper motion of the galaxy which yields us to the magnetic
  field configuration in Fig.~\ref{Fig2}B.  We believe that this very beautiful
  case allows us to unambiguously determine the ambient field orientation as
  shown in Fig.~\ref{Fig5}. We observed no sign of a `polarisation tilt' across
  the polarised emission region which allows us to obtain a lower limit on the
  magnetic coherence scale of $\lambda_B > 100$~kpc. This also suggests a
  negligible LOS component of the magnetic field.  A very symmetric local
  maximum of polarisation intensity at the very centre of the galaxy might be
  due to a core emission component. Since it does not spatially coincide with
  the region of the HI hot spot, it is probably unrelated to the ram pressure
  wind that the disk experiences.  We estimate a radial deviation of the
  transverse field component to $\theta = 39\deg$.
\item[NGC 4522] Small, red galaxy at 6 o'clock, edge-on/sightly inclined
  orientation ($\vel = 2332\,\rmn{km/s}$).  The HI morphology shows stripped
  material towards the north and western part of the disk which reminds at
  Kelvin-Helmholtz instabilities (KHI) and suggests a velocity component in the
  plane of the sky towards the south-east. The strong redshift as well as the
  fact that the HI hot spot is not at the rim of the HI emission but more
  centred imply an additional LOS component into the image plane. The HI and PI
  hot spots almost coincide ($\alpha=12:33:44$, $\delta=09:10:26$). There is a
  strong polarised intensity towards the north-eastern part of the galaxy with
  the field vectors aligned with the disk. We observe extraplanar polarised
  emission above the disk plane towards the northwest with the field vectors
  bent in the direction of extraplanar emission as expected from our draping
  model. The upstream magnetic field could be tilted along the LOS as this
  component would not be visible due to the geometric bias effect. If the
  stripped material shows indeed KHI this would imply a non-negligible magnetic
  field component along the LOS. A draped magnetic field lying purely in the
  plane would suppress KHI\cite{2007ApJ...670..221D}.  We see a steepening of
  the spectral index map\cite{2004AJ....127.3375V} as we move away from the hot
  spot which is consistent with synchrotron cooling CRes that are advectively
  and diffusively transported with the flux frozen field lines.  We estimate a
  radial deviation of them transverse field component to $\theta = -28\deg$.
\item[NGC 4535] Large, yellow galaxy at 6 o'clock, face-on orientation ($\vel =
  1973\,\rmn{km/s}$). The HI morphology shows stripped material towards the
  south and eastern part of the disk which suggests a velocity component in the
  plane of the sky towards the north-west. The modest redshift implies an
  additional LOS component into the image plane. This picture is consistent with
  an HI hot spot that is visible in the face-on view towards the north-west
  ($\alpha=12:34:13-14$, $\delta=08:13:21-31$). The other HI hot spot towards
  the south might indicate that this galaxy moves close to face-on.  The galaxy
  shows a low emission of polarised radio continuum emission in the centre and
  the west of the outer galactic disk which extends further south. Notably, the
  maximum of PI is slightly shifted southwards compared to the HI hot
  spot. There is no polarisation twist visible which would hint at a
  considerable LOS component of the magnetic field. We are looking at a
  reasonably clean case of magnetic draping where the magnetic field is mostly
  in the plane of the sky and the HI/PI misalignment is due to the
  `direction-of-motion asymmetry'.  We estimate a radial deviation of the
  transverse field component to $\theta = -23\deg$.
\item[NGC 4548] Medium size, blue galaxy at 11 o'clock, face-on orientation
  ($\vel = 498\,\rmn{km/s}$).  The large blue-shift implies a substantial LOS
  component out of the image plane and the hole in the central HI distribution
  of the galaxy suggests an almost face-on proper motion that is responsible for
  this impressive case of ram-pressure stripping.  The galaxy has an overall low
  emission of polarised radio continuum emission with a maximum PI that
  corresponds to that hole in the HI distribution ($\alpha=12:35:26$,
  $\delta=14:30:05$). This again would be very difficult to reconcile with any
  ISM-based magnetic field model responsible for the PI. We note however that
  the detection of the polarised emission suffers from the large beam size
  relative to the scale of the galaxy which makes the determination of the field
  orientation of the draped magnetic field somewhat uncertain. However, the
  weakness of the observed polarised emission can be explained by the low value
  of the minimum energy magnetic field strength of $B_\rmn{min} \simeq
  2.3\,\mu$G that is inferred from the total synchrotron
  emission\cite{2006ApJ...645..186T}. This possibly indicates a lower
  normalisation of the CRe distribution.  We estimate a radial deviation of the
  transverse field component to $\theta = 45\deg$.
\item[NGC 4654] Medium size, green galaxy at 9 o'clock, inclined orientation
  ($\vel = 1035\,\rmn{km/s}$). This galaxy has a large velocity component in the
  plane of the sky as its redshift corresponds to the cluster mean. The location
  of the `HI hot spot' in the northwest and the ram pressure stripped tail
  pointing in the opposite direction makes this galaxy a beautiful case for
  studying magnetic draping. The fact that the HI hot spot is not exactly at the
  rim but in projection slightly inwards located suggests an additional small
  LOS velocity component.  Almost coincident with the HI hot spot
  ($\alpha=12:43:52$, $\delta=13:08:18$), we observe a ridge of polarised
  emission at the northwestern outer rim of the galactic disk with a maximum
  that is slightly shifted north due to the `direction-of-motion
  asymmetry'. This rim even shows a nose extending in the direction of motion
  outside of the HI disk where the exponential stellar disk got stripped of its
  gas and injects CRes into the magnetic drape. We note that this configuration
  is not sensitive to a possible B-component along the LOS due to geometric
  bias.  Additionally, we observe another extended polarisation ridge in the
  southern part of the outer disk and there is a hint of correlated polarisation
  noise along the southeastern extended HI tail\cite{1995ApJ...453..154P,
    2007ApJ...659L.115C}.  

  As shown in Fig.~\ref{Fig5} of the main body, our best-fit model galaxy is
  inclined by 68 deg and encounters a homogeneous magnetic field that is tilted
  by 30 deg. Our choice of viewing angle results in an inclination of 25 deg
  with respect to edge-on configuration which reproduces the overall morphology
  of the PI as well as the separation of the two main polarisation ridges
  surprisingly well. However, numerical limitations force us to use a disk
  height-to-radius ratio of ten (which is at least a factor of ten below
  realistic ratios) as well as to neglect the detailed physics (multiphase ISM
  including its clumpiness, self-gravity) which makes a very detailed comparison
  difficult. Potentially, our model underestimates the inclination of the disk
  with respect to the line-of-sight and hence underestimates the amount of PI in
  the south-eastern polarisation ridge.  The filamentary structure at the
  leading radio ridge is also an artifact of the enhanced Kelvin-Helmholtz
  interface instabilities for this highly inclined case that trap magnetic flux
  tubes and hence suppress their bending in toroidal direction in the draping
  process as discussed in Sec.~\ref{sec:inclination}.  These instabilities,
  worsened by our affordable numerical resolution, grow stronger with
  inclination, contaminating simulations run with larger inclinations.  Models
  with a larger inclination would have a magnetic field tilted counterclockwise
  as seen in Fig.~\ref{Fig5}; thus we slightly adjust the field line to
  compensate for limitations of the numerical models used in this work. We note
  however that we preserve the field line mapping from our simulated PI map to
  the upstream orientation of the field in our simulation.  Hence, we estimate a
  radial deviation of the transverse field component to $\theta = 21\deg$.
\end{description}

To summarise, we find a preferentially radial field configuration relative to the
centre of Virgo, M87 (for the statistical significance of this finding, please
refer to Sect.~\ref{sec:stats}). In addition, we note that all three
galaxy-pairs that are close-by in the sky show very correlated local
B-orientations with the following relative angles in the plane of the sky:
NGC~4402 and NGC~4388 have $\Delta \theta = 1\deg$, NGC~4535 and NGC~4522 have
$\Delta \theta = 0\deg$, and NGC~4548 and NGC~4501 have $\Delta \theta =
19\deg$.  At the position of NGC~4501, we have evidence for a magnetic coherence
scale of $\lambda_B > 100$~kpc.

\section{Statistical significance of radial fields}
\label{sec:stats}

\subsection{Uniform sum distribution}

In order to assess the statistical significance of our finding we consider the
probability of finding the angles so close to radially aligned in the case where
the angles were uniformly distributed on the sky.  The distribution of the sum
of $n$ uniform variates on the interval $[0,1]$ is given by the uniform sum
distribution,
\begin{eqnarray}
  \label{eq:USD}
  \lefteqn{P_{X_1+X_2+\ldots+X_n}(u)} \nonumber\\
  &=& \int\idotsint \delta(x_1+x_2+\ldots+x_n-u)\,
  \dd x_1 \dd x_2\ldots\dd x_n \nonumber\\
  &=& \frac{1}{2 (n-1)!}\,\sum_{k=0}^{n}(-1)^k\,\binom{n}{k}\,(u-k)^{n-1}\,\rmn{sgn}(u-k).
\end{eqnarray}
We do a Monte-Carlo sampling of this distribution; by considering 250,000
realisations of 8 uniformly-oriented angles and summing their absolute deviation
from radial alignment we find a cumulative probability that
$u\leq\sum_{k=1}^8|\theta_i|/90=205.5/90=2.28$ of 1.68\%.  Hence we can reject
the null hypothesis of the projected field being uniformly distributed in angle
on the sky.

This result could be understood by assuming we were in the regime of the central
limit theorem, where the distribution of the sum of the $n$ random variables
$|\theta|$ of mean $\mu=45$ and variance $\sigma^2=675$ would then be a Gaussian
distribution, with mean $n\mu$ and variance $n \sigma^2$.  We derive that the
alignment of the field in the plane of the sky is significantly more radial than
expected from random chance, with a significance level of
\begin{equation}
  \label{eq:significance} \frac{\sum_{k=1}^8|\theta_i|-n\mu}{\sqrt{n
  \sigma_{|\theta|}^2}} = 2.1.
\end{equation} This corresponds to a probability of
$(1-\rmn{erf}(\sigma_{|\theta|}/\sqrt{2}))/2 = 0.0168$, which is the
same result as obtained above.

We point out a major caveat as the statistical analysis presented here does not
include systematic uncertainties. In particular, line-of-sight effects could
introduce a larger systematic scatter which is however impossible to address
which such a small observational sample at hand.

\subsection{Interpretation}

We conclude with a few important remarks on how to interpret this finding. (1)
If the galaxy that probes the magnetic field orientation is offset from the
plane perpendicular to the LOS that contains the centre of the Virgo cluster,
M87, then even a purely radial transverse magnetic field would attain an
azimuthal component. On the other side, if there is a LOS-component of the
magnetic field, that could counteract this effect.  In this respect, our
estimates of deviations from a radial field should be taken as lower limits to
the true deviation from 3D radial structure. (2) Using 3D simulations of the
magneto-thermal instability (MTI) that fix the temperature at the boundary of
the unstable layer, the magnetic field is reoriented as a result of the MTI and
saturates at an average deviation from a radial field of $\theta\simeq 35 \deg$
(see reference\cite{2007ApJ...664..135P} as well as the discussion in
Sect.~\ref{sec:MTI} for more detail). (3) All but two estimates for $\theta$
using our galaxy sample are smaller than the average deviation from a radial
field as inferred from idealised MHD simulations of a stratified atmosphere in a
box. This in principle allows some room for the LOS effect mentioned under (1)
to explain for the mismatch of our average value and that of the simulations. On
the other hand, the distribution of field orientations in a galaxy cluster that
evolves within a cosmological framework is not known to date. It could be a
function of time since the last major merger or other evolutionary parameters
including the cumulative effect of non-gravitational energy injection, {\em
  e.g.}  by AGN. We note that our findings are consistent with our current
theoretical understanding of this effect.

\section{Magneto-thermal instability and conduction}
\label{sec:MTI}

\subsection{Physics of the instability}

For a decreasing temperature profile with increasing radius, the ICM is subject
to the magneto-thermal instability (MTI) as described in the text. Since the
growth time of the MTI for typical cluster conditions amounts to 0.9
Gyr\cite{2008ApJ...688..905P}, we have tens of growth times during a Hubble time
which potentially allows significant rearrangements of the magnetic field
structure as well as the atmosphere.  Depending on the boundary conditions at
the maximum of the temperature profile (around $r\simeq (0.1-0.2) R_{200}$) as
well as at the cluster boundary that is set by accretion shocks, the asymptotic
saturated state of the MTI is different. (1) For adiabatic boundary conditions,
the temperature gradient as the source of free energy of the instability relaxes
in order to yield an isothermal profile\cite{2007ApJ...664..135P}. The timescale
over which this happens seems to depend on the geometry of the magnetic field in
the initial conditions. The relaxation occurs over times scales of 3.5 to 7 Gyr,
depending whether the field was initially turbulent or purely
azimuthal\cite{2008ApJ...688..905P}. (2) For fixed boundary conditions, the MTI
grows a predominantly radial component with an average deviation from the radial
direction $\theta \simeq 35 \deg$. This configuration counteracts the source of
instability and allows efficient heat conduction by conductively transporting
the heat rather than through buoyant motions\cite{2007ApJ...664..135P}.

The immediate question which arises here is which of these two cases is indeed
realised in Nature? A sample of 13 nearby, relaxed galaxy clusters and groups
observed by the X-ray satellite Chandra shows a declining temperature profile in
the outer parts of all of these clusters\cite{2005ApJ...628..655V}. This
strongly suggests that the second possibility of a temperature profile with
approximately fixed boundary conditions is at work. MHD simulations with
anisotropic heat conduction of a toy cluster that is not growing within a
cosmological framework demonstrates that the temperature profile becomes
eventually isothermal\cite{2008ApJ...688..905P} -- in conflict with above
mentioned observations. Hence we need a heating source that preferentially heats
the cluster centre so that this heat can be conductively transported outwards by
means of radial fields that result from the MTI.

\subsection{Consequences for cluster physics -- conduction}

\begin{figure}
\begin{center}
\hspace{0.2in}\includegraphics[width=2.8in]{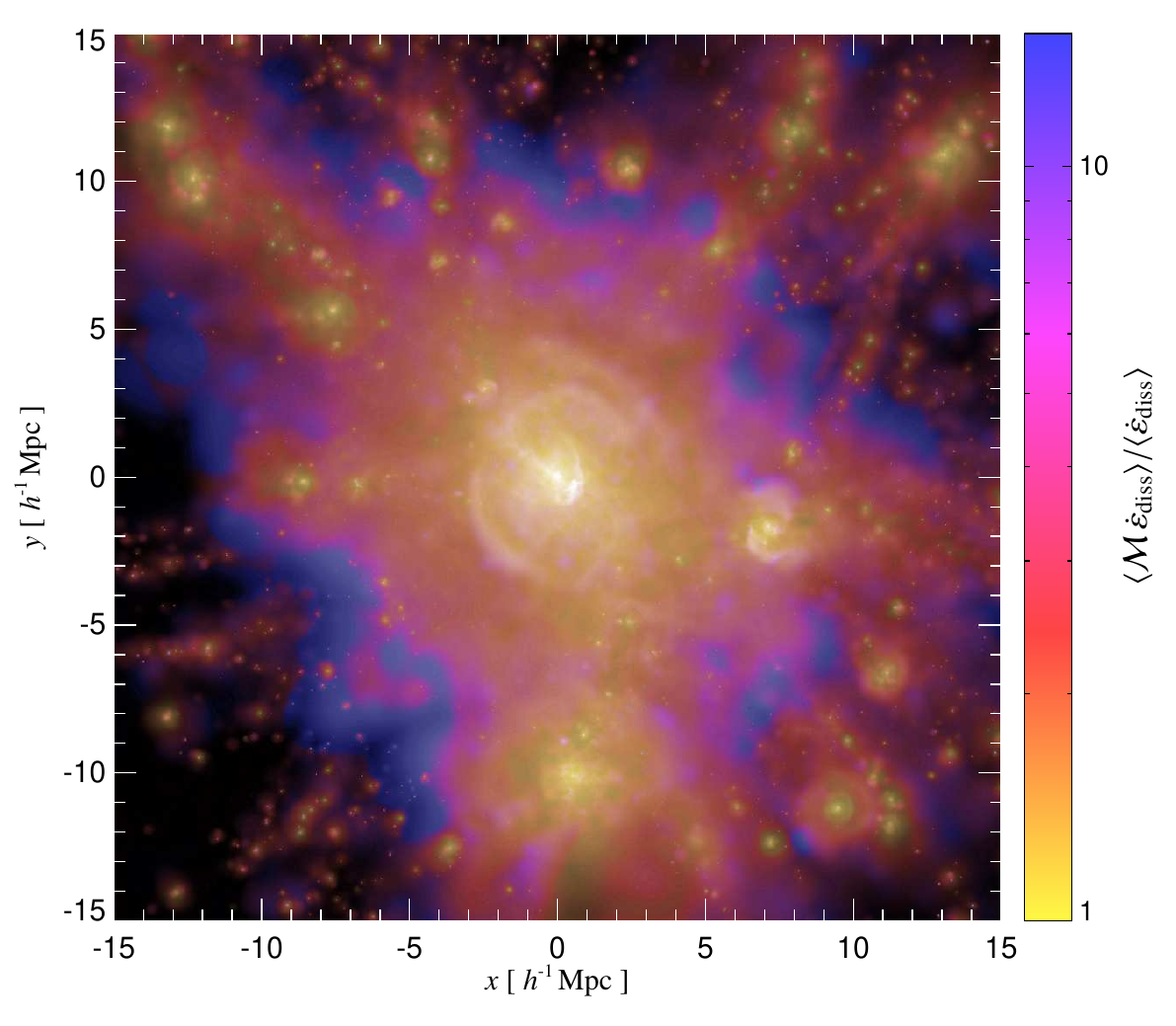}\\
\includegraphics[width=2.8in]{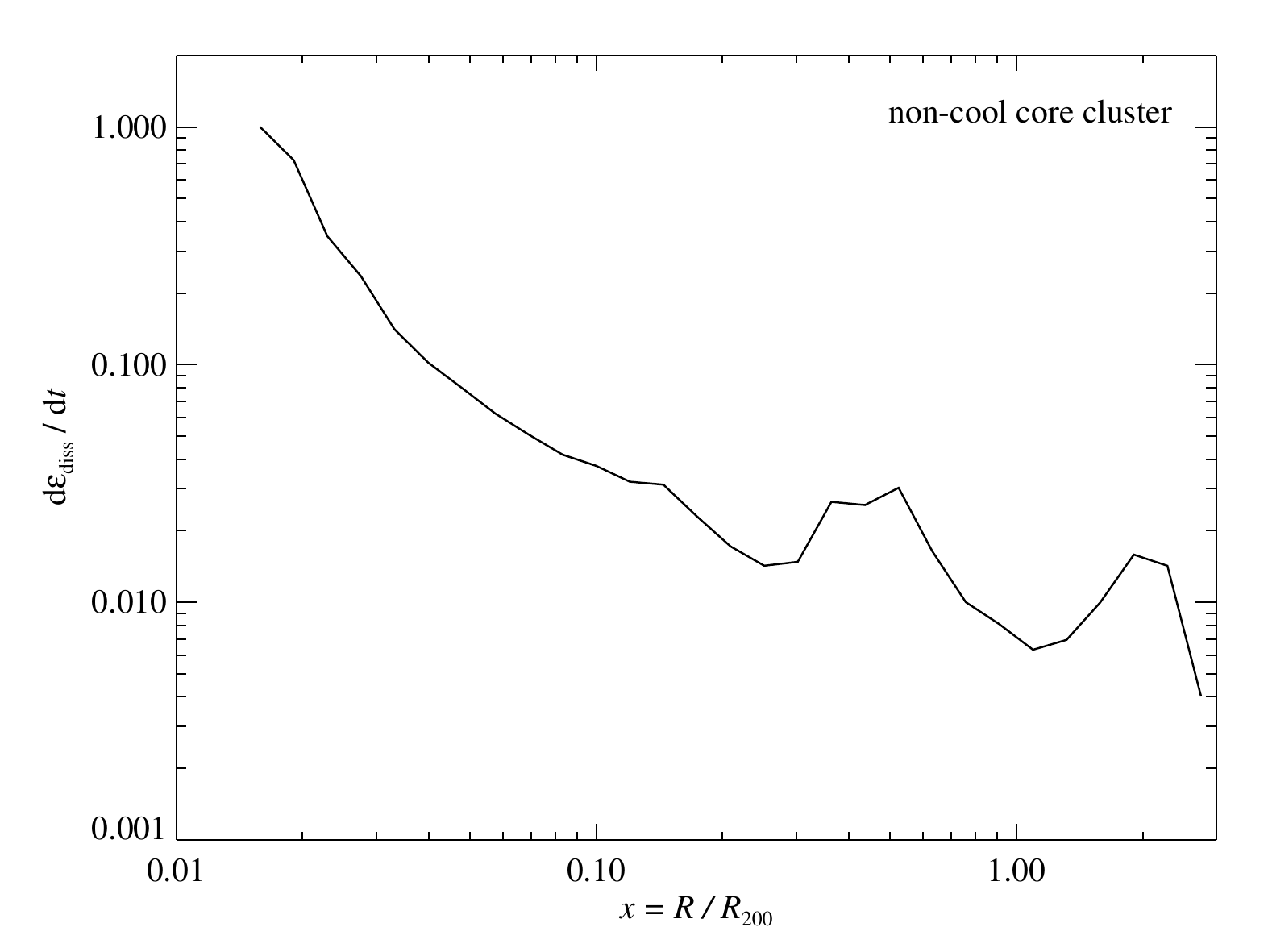}\hspace{0.2in}
\end{center}
\caption{{\bf Energy dissipation rate at shocks in a simulated galaxy cluster
    (non-cool core, $M=1.5\times 10^{15}\,\rmn{M}_\odot$).}  {\em Top:} the Mach
  number of structure formation shocks weighted by the energy dissipation rate
  in colour (while the brightness displays the logarithm of the dissipation
  rate)\cite{2007MNRAS.378..385P}. Note that most of the energy density is
  dissipated in weak shocks inside groups and clusters. {\em Bottom:} normalised
  energy density dissipation rate as a function of cluster-centric radius. The
  profile peaks in the centre and declines towards the periphery. On top of this
  average profile, there is a merging shock wave visible at
  $r=0.6\,R_{200}=1.4\,\rmn{Mpc}$ as well as the accretion shock at
  $r=2\,R_{200}$; both of which have counterparts in the projected image at the
  top.}
\label{fig:mach}
\end{figure}

A very interesting piece of evidence arises from studies of the evolution of the
temperature profile in cosmological cluster simulations that employ isotropic
conduction at $1/3$ of the classical Spitzer value. They show that even hot
clusters, that have very efficient heat conduction, are not able to establish an
isothermal profile\cite{2004MNRAS.351..423J, 2004ApJ...606L..97D}. This suggests
that shocks driven by gravitational infall can act as a selective heating
source.  It is very instructive to consider the energy dissipation at structure
formation shocks in cosmological simulations\cite{2006MNRAS.367..113P,
  2007MNRAS.378..385P}. Figure~\ref{fig:mach} shows a high-resolution
hydrodynamic simulation of a galaxy cluster using the `zoomed initial
conditions' technique within a cosmological
framework\cite{2008MNRAS.385.1211P}. Using the distributed-memory parallel
TreeSPH code GADGET-2, the simulations followed radiative cooling of the gas and
star formation. They also identified the Mach numbers of cosmological formation
shocks\cite{2006MNRAS.367..113P} that dissipate gravitational energy in the
process of hierarchical structure growth. The important point to take away is
that most of the energy of these shocks is dissipated towards the cluster
centres in weak flow shocks.  This can be understood as the energy flux through
the shock surface, $\dot{E}_\rmn{diss}/R^2 \sim \rho \vel^3$, that will be
dissipated at these formation shocks, is maximised in the dense cluster
cores. To order of magnitude, we expect the heating to take place on a dynamical
time scale of $\tau_\rmn{dyn} \simeq 1$~Gyr. This time scale is shorter or
comparable to the thermal conduction time over significant cluster scales,
$\tau_\rmn{cond}=\lambda^2/ \chi_\rmn{C}\simeq 2.3
\,\rmn{Gyr}\,(\lambda/1\,\rmn{Mpc})^2$, where $\chi_\rmn{C} = 8\times
10^{31}(T/10\,\rmn{keV})^{5/2}(n/5\times10^{-3}\,\rmn{cm}^{-3})^{-1} \,
\rmn{cm}^2\,\rmn{s}^{-1}$ is the Spitzer thermal diffusivity. Hence one would
expect a stationary temperature profile that should always show a moderate
temperature gradient which would be needed to power the MTI and keep the
magnetic field lines preferentially radial. 

Since the thermal conduction time scales with the square of the length scale in
question, it will always win over the dynamical time or sound crossing time on
small enough length scales and thermally stabilise these regions -- provided we
have a mechanism that ensures efficient conduction close to the Spitzer value.
Clearly, we need detailed MHD simulations of cluster formation in a cosmological
framework that include anisotropic conduction to confirm this picture.  However,
our result of a preferentially radial field orientation in Virgo strongly
suggests that gravitational heating by accreted substructures seems to maintain
the temperature gradients even in the presence of the MTI. This indirectly
confirms a prediction of the hierarchical structure formation scenario in which
a halo constantly undergoes mergers and accretes smaller mass objects that
gravitationally heat the ICM through shock wave heating.

A note here is in order as completely randomly-oriented magnetic fields could
also sustain effective conduction at $1/3$ of the Spitzer value. This however
assumes that field line-wandering is in fact volume filling, i.e. a set of
initially close-by field lines is not constrained to a two dimensional hyper
surface. One could image idealised models that would not predict conduction at
$1/3$ of Spitzer; one realisation of such a model for the growth of clusters
that is able to explain the observed entropy profile in clusters is the model of
smooth accretion\cite{2005RMP...77...207V}.  Suppose that mass accretes in a
series of concentric shells that have negligible pressure and entropy and are
initially comoving with the Hubble flow. Once these shells decelerate and reach
zero velocity at the turnaround radius, they fall back under the influence of
gravity of the previously collapsed shells and encounter the accretion shock at
about the virial radius. There will be a higher entropy generated for later
accreting shells leading to an overall stable entropy
profile\cite{2005RMP...77...207V}. Presumably each of these shells carries their
flux-frozen magnetic field that in this simple picture does not intersect with
the field in the other shells. Once established, the stably stratified entropy
profile suppresses convective motions and prevents radial mixing (assuming for a
moment that the MTI is suppressed by some mechanism).  Of course within
hierarchical structure formation, these idealised hyper surfaces will never be
perfectly separated and there will be some degree of mixing. However, a small
degree of mixing should still suppress efficient conduction and calls for an
effect that actively builds up radial field components -- one prime candidate
would be the MTI.

\section{The Virgo cluster}

\begin{figure}
\begin{center}
\includegraphics[width=2in, angle=-90]{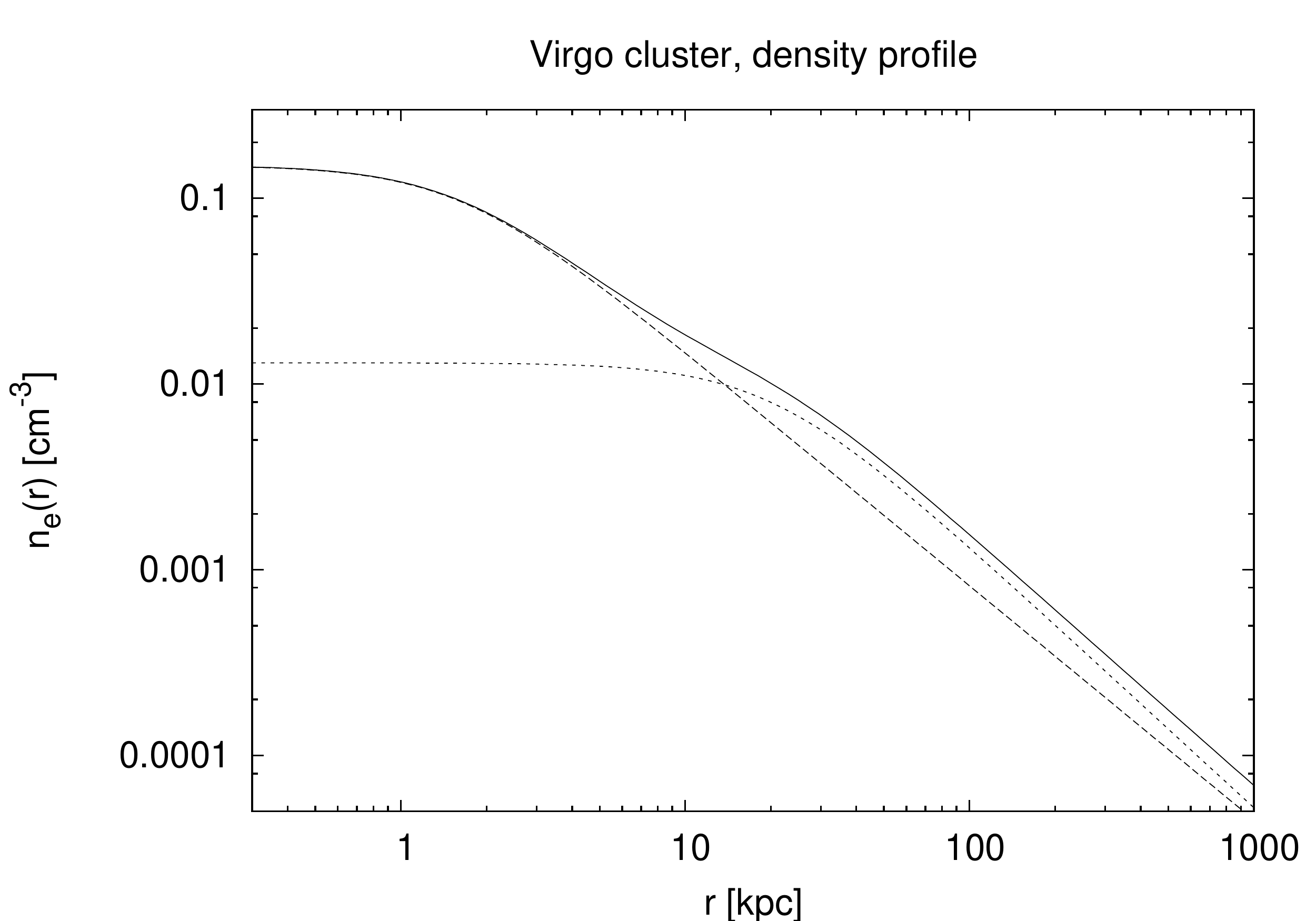}\\
\hspace{0.1in}\includegraphics[width=1.9in, angle=-90]{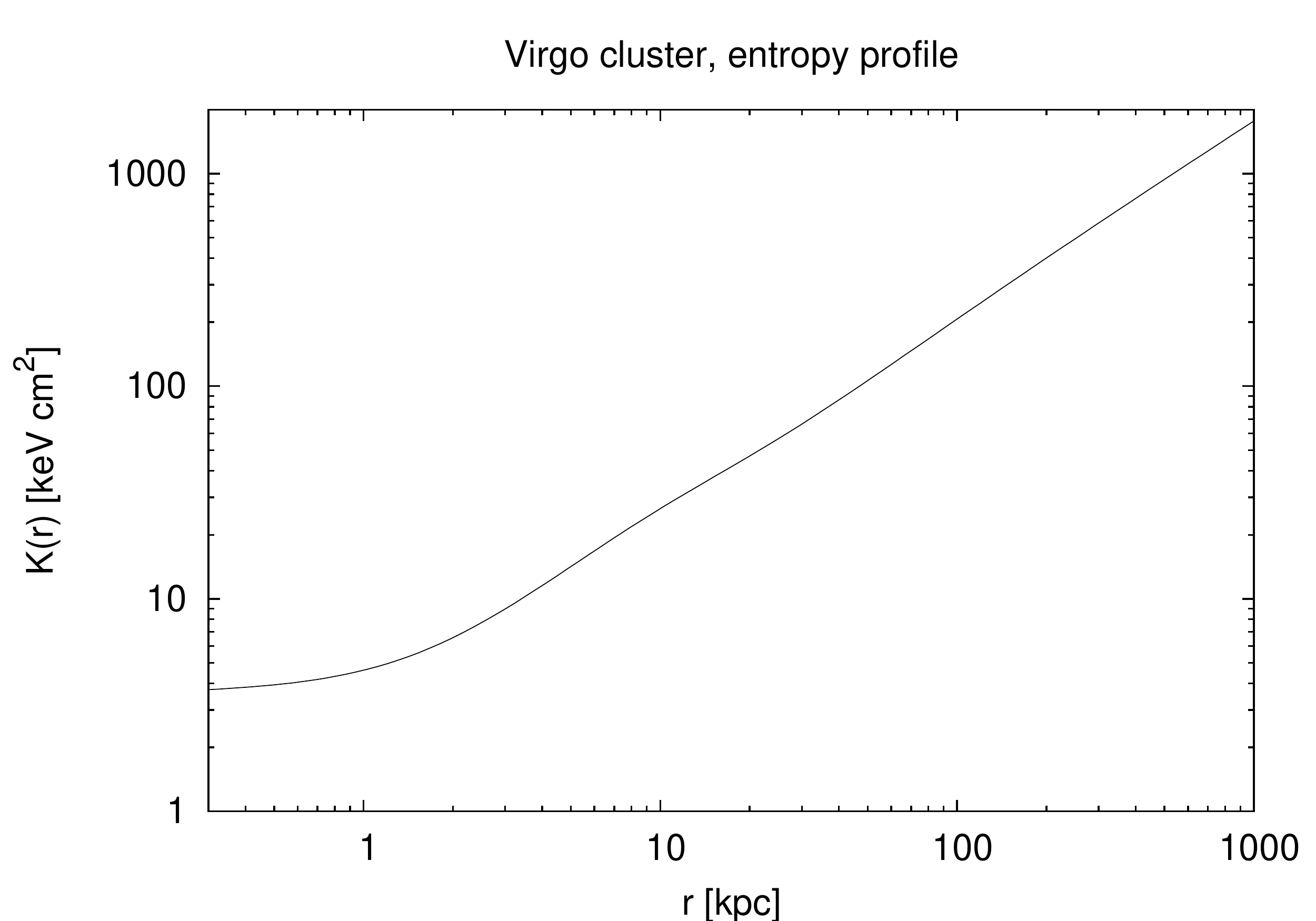}\\
\hspace{0.1in}\includegraphics[width=1.9in, angle=-90]{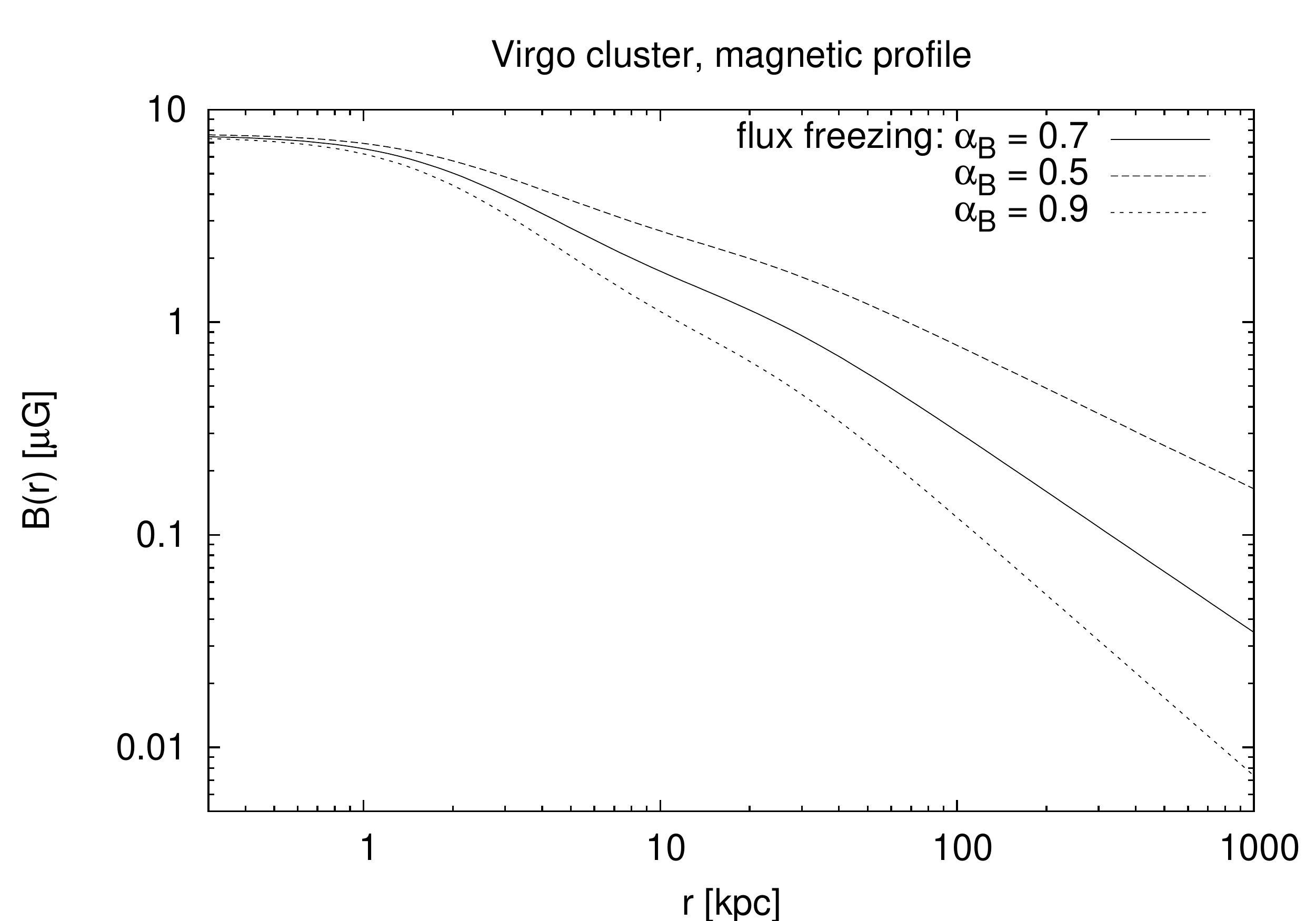}\\
\end{center}
\caption{{\bf Profiles of gas properties and the magnetic field of Virgo.}  {\em
    Top, middle:} profiles of the electron density (described by a double-beta
  model where we show both components with dotted lines) and entropy variable
  $K=kT_\rmn{x} n_\e^{-2/3}$ of the Virgo cluster as inferred from X-ray
  data\cite{2002A&A...386...77M}. Note the asymptotic behaviour of the entropy
  variable $K\propto r^{0.9}$ for large radii. {\em Bottom:} illustrative
  profile of a magnetic field model with a central field strength of
  $B_0=8\,\mu$G and an assumed scaling with the gas density of $B\propto
  n^{\alpha_B}$. We show plausible ranges for the scaling parameter as
  determined from cosmological cluster simulations\cite{1999A&A...348..351D,
    2001A&A...378..777D} and Faraday rotation
  measurements\cite{2005A&A...434...67V} and note the choice $\alpha_B=0.7$
  corresponds to the flux freezing condition.}
\label{fig:profiles}
\end{figure}

In which category of clusters does the Virgo cluster fall into -- is it a
cooling core cluster (CC) or a non-cooling core (NCC)? It turns out that this
question is not as simple to answer as one might think.  Due to its proximity,
already ROSAT and later on XMM-Newton resolved a small cooling core at radii $<
14$~kpc\cite{1994Natur.368..828B, 2002A&A...386...77M}. But how does its overall
appearance compare to the clusters in the Chandra
sample\cite{2008ApJ...683L.107C}, in particular if we placed it to higher
redshifts?

First we note that there is a discrepancy in the literature about the virial
mass and radius of the Virgo cluster. The X-ray community converged on a value
for its virial mass of $M_{200} = 2.1\times 10^{14}\, \rmn{M}_\odot$ resulting
in a virial radius of $R_{200} = 0.95$~Mpc\cite{1999A&A...343..420S}.  We define
the virial mass $M_\Delta$ and virial radius $R_\Delta$ as the mass and radius
of a sphere enclosing a mean density that is $\Delta=200$ times the critical
density of the Universe, $\rho_\rmn{crit} = 3 H_0^2/ (8\pi G)$. The
corresponding virial temperature of the halo is
\begin{equation}
  \label{eq:T200}
  kT_{200} = \frac{G\, M_{200}\,\mu\,m_\rmn{p}}{2\,R_{200}} \simeq 2.8\mbox{ keV},
\end{equation}
where $\mu\simeq 0.6$ denotes the mean molecular weight. This corresponds well
with the deprojected temperature profiles from X-ray
observations\cite{1994Natur.368..828B, 2002A&A...386...77M} that observed a
maximum temperature of $\simeq 3$~keV at a radius of 200~kpc (corresponding to
40 arcmin). Conversely, applying a relativistic Tolman-Bondi model of the Virgo
cluster to a sample of 183 galaxies with measured distances within a radius of 8
degrees from M87, Virgo's virial mass is found to be $M_{200} = 7\times
10^{14}\, \rmn{M}_\odot$\cite{2001A&A...375..770F}. A similar large mass
estimate of $M_{200} = 1.3\times 10^{15}\, \rmn{M}_\odot$ is supported by
modelling of the velocity field of the local super-cluster, assuming a
mass-to-light ratio of $150~\rmn{M}_\odot/\rmn{L}_\odot$ in the field, but 1000
for the Virgo cluster\cite{1999elss.conf..296T}. For the purpose of this section
we use however the virial estimates from the X-ray measurements.

We use the deprojected density profile\cite{2002A&A...386...77M} (described by a
double-beta model) as well as a fit to the temperature
profile\cite{2004A&A...413...17P}\footnote{The functional form of the
  temperature profile was chosen to follow the universal temperature
  profile\cite{2001MNRAS.328L..37A} which assumes an isothermal cluster outside
  a central cooling region. We note that this is not only in conflict with ROSAT
  data\cite{1994Natur.368..828B} but also with the cluster temperature profile
  of recent cosmological simulations\cite{2007MNRAS.378..385P,
    2007ApJ...668....1N}. However, since we focus here on the central part to
  assess angular resolution effects, this should have little influence on our
  results.} in order to construct the entropy profile. To this end, we use the
X-ray entropy variable\cite{1999Natur.397..135P} $K=kT_\rmn{x} n_\e^{-2/3}$ that
is derived from the equation for the adiabatic index, $K\propto P\rho^{-5/3}$.
For radii $r<2$~kpc we observe an entropy core, outside a power-law behaviour
that asymptotes $K\propto r^{0.9}$. This is somewhat shallower compared to that
of most clusters of the Chandra sample\cite{2008ApJ...683L.107C}, $K\propto
r^{1.1-1.2}$. This discrepancy will be reinforced when adopting a declining
temperature profile in the outskirts; a behaviour possibly caused by the low
X-ray counts in these regions due to the large angular extent of seven degrees
on the sky.

We now determine the entropy core value of Virgo if we placed it successively at
higher redshifts, i.e. we address angular resolution effects. From $K\propto
r^{0.9}$ outside $r=2$~kpc, we conclude any weighting (over volume or emission)
will produce the same radial behaviour, i.e. the inferred value for the central
entropy $\tilde{K}_0$ is always dominated by the value of the entropy profile at
the radius corresponding to the angular resolution $\theta$ of an X-ray
instrument or the binning algorithm used for the
analysis\cite{2008ApJ...683L.107C}.  The physical radius $r(z) = D_\rmn{ang}(z)
\tan\theta$, corresponding to $\theta=5''$ increases as a function of redshift
with the angular diameter distance $D_\rmn{ang}$; and so does
$\tilde{K}_0$. Adopting the entropy profile of Fig.~\ref{fig:profiles}, we solve
the equation $\tilde{K}[r(z)]> 30 \,\mbox{keV cm}^2$ for $z$. We find that Virgo
-- placed at a redshift $z>0.13$ -- would show a high-entropy core in agreement
with NCCs.  The very low H$\alpha$ luminosity\cite{1999MNRAS.306..857C}
$L_{\rmn{H}\alpha}\simeq 10^{39} \,\rmn{erg/s}$, being below most of the upper
limits in the Chandra sample\cite{2008ApJ...683L.107C}, confirms this picture
that Virgo might be on its transition to a cool core at the centre but still
shows all signs of a NCC on large scales.

\bibliography{bibtex/si}

\begin{thebibliography}{10}
\expandafter\ifx\csname url\endcsname\relax
  \def\url#1{\texttt{#1}}\fi
\expandafter\ifx\csname urlprefix\endcsname\relax\def\urlprefix{URL }\fi
\providecommand{\bibinfo}[2]{#2}
\providecommand{\eprint}[2][]{\url{#2}}

\bibitem{2004AJ....127.3375V}
\bibinfo{author}{{Vollmer}, B.}, \bibinfo{author}{{Beck}, R.},
  \bibinfo{author}{{Kenney}, J.~D.~P.} \& \bibinfo{author}{{van Gorkom}, J.~H.}
\newblock \bibinfo{title}{{Radio Continuum Observations of the Virgo Cluster
  Spiral NGC 4522: The Signature of Ram Pressure}}.
\newblock \emph{\bibinfo{journal}{\aj}} \textbf{\bibinfo{volume}{127}},
  \bibinfo{pages}{3375--3381} (\bibinfo{year}{2004}).

\bibitem{2007A&A...464L..37V}
\bibinfo{author}{{Vollmer}, B.} \emph{et~al.}
\newblock \bibinfo{title}{{The characteristic polarized radio continuum
  distribution of cluster spiral galaxies}}.
\newblock \emph{\bibinfo{journal}{\aap}} \textbf{\bibinfo{volume}{464}},
  \bibinfo{pages}{L37--L40} (\bibinfo{year}{2007}).

\bibitem{2007A&A...471...93W}
\bibinfo{author}{{We{\.z}gowiec}, M.} \emph{et~al.}
\newblock \bibinfo{title}{{The magnetic fields of large Virgo Cluster
  spirals}}.
\newblock \emph{\bibinfo{journal}{\aap}} \textbf{\bibinfo{volume}{471}},
  \bibinfo{pages}{93--102} (\bibinfo{year}{2007}).

\bibitem{2010A&A...512A..36V}
\bibinfo{author}{{Vollmer}, B.} \emph{et~al.}
\newblock \bibinfo{title}{{The influence of the cluster environment on the
  large-scale radio continuum emission of 8 Virgo cluster spirals}}.
\newblock \emph{\bibinfo{journal}{\aap}} \textbf{\bibinfo{volume}{512}},
  \bibinfo{pages}{A36} (\bibinfo{year}{2010}).

\bibitem{2001SSRv...99..243B}
\bibinfo{author}{{Beck}, R.}
\newblock \bibinfo{title}{{Galactic and Extragalactic Magnetic Fields}}.
\newblock \emph{\bibinfo{journal}{Space Science Reviews}}
  \textbf{\bibinfo{volume}{99}}, \bibinfo{pages}{243--260}
  (\bibinfo{year}{2001}).

\bibitem{2005RMP...77...207V}
\bibinfo{author}{Voit, G.~M.}
\newblock \bibinfo{title}{Tracing cosmic evolution with clusters of galaxies}.
\newblock \emph{\bibinfo{journal}{Reviews of Modern Physics}}
  \textbf{\bibinfo{volume}{77}}, \bibinfo{pages}{207--258}
  (\bibinfo{year}{2005}).

\bibitem{2004mmis.book.....W}
\bibinfo{author}{{Winterhalter}, D.}, \bibinfo{author}{{Acu{\'n}a}, M.} \&
  \bibinfo{author}{{Zakharov}, A.}
\newblock \emph{\bibinfo{title}{{Mars' Magnetism and its Interaction with the
  Solar Wind.}}} (\bibinfo{year}{2004}).

\bibitem{2004inco.book.....B}
\bibinfo{author}{{Brandt}, J.~C.} \& \bibinfo{author}{{Chapman}, R.~D.}
\newblock \emph{\bibinfo{title}{{Introduction to Comets}}}
  (\bibinfo{year}{2004}).

\bibitem{2005JGRA..11001209B}
\bibinfo{author}{{Bertucci}, C.}, \bibinfo{author}{{Mazelle}, C.},
  \bibinfo{author}{{Acu{\~n}a}, M.~H.}, \bibinfo{author}{{Russell}, C.~T.} \&
  \bibinfo{author}{{Slavin}, J.~A.}
\newblock \bibinfo{title}{{Structure of the magnetic pileup boundary at Mars
  and Venus}}.
\newblock \emph{\bibinfo{journal}{Journal of Geophysical Research (Space
  Physics)}} \textbf{\bibinfo{volume}{110}}, \bibinfo{pages}{A01209}
  (\bibinfo{year}{2005}).

\bibitem{2005AnGeo..23..885C}
\bibinfo{author}{{Coleman}, I.~J.}
\newblock \bibinfo{title}{{A multi-spacecraft survey of magnetic field line
  draping in the dayside magnetosheath}}.
\newblock \emph{\bibinfo{journal}{Annales Geophysicae}}
  \textbf{\bibinfo{volume}{23}}, \bibinfo{pages}{885--900}
  (\bibinfo{year}{2005}).

\bibitem{2006JGRA..11110220N}
\bibinfo{author}{{Neubauer}, F.~M.} \emph{et~al.}
\newblock \bibinfo{title}{{Titan's near magnetotail from magnetic field and
  electron plasma observations and modeling: Cassini flybys TA, TB, and T3}}.
\newblock \emph{\bibinfo{journal}{Journal of Geophysical Research (Space
  Physics)}} \textbf{\bibinfo{volume}{111}}, \bibinfo{pages}{A10220}
  (\bibinfo{year}{2006}).

\bibitem{2006JGRA..11109108L}
\bibinfo{author}{{Liu}, Y.}, \bibinfo{author}{{Richardson}, J.~D.},
  \bibinfo{author}{{Belcher}, J.~W.}, \bibinfo{author}{{Kasper}, J.~C.} \&
  \bibinfo{author}{{Skoug}, R.~M.}
\newblock \bibinfo{title}{{Plasma depletion and mirror waves ahead of
  interplanetary coronal mass ejections}}.
\newblock \emph{\bibinfo{journal}{Journal of Geophysical Research (Space
  Physics)}} \textbf{\bibinfo{volume}{111}}, \bibinfo{pages}{A09108}
  (\bibinfo{year}{2006}).

\bibitem{lyutikovdraping}
\bibinfo{author}{{Lyutikov}, M.}
\newblock \bibinfo{title}{{Magnetic draping of merging cores and radio bubbles
  in clusters of galaxies}}.
\newblock \emph{\bibinfo{journal}{Mon. Not. R. Astron. Soc}}
  \textbf{\bibinfo{volume}{373}}, \bibinfo{pages}{73--78}
  (\bibinfo{year}{2006}).

\bibitem{2008ApJ...677..993D}
\bibinfo{author}{{Dursi}, L.~J.} \& \bibinfo{author}{{Pfrommer}, C.}
\newblock \bibinfo{title}{{Draping of Cluster Magnetic Fields over Bullets and
  Bubbles-Morphology and Dynamic Effects}}.
\newblock \emph{\bibinfo{journal}{\apj}} \textbf{\bibinfo{volume}{677}},
  \bibinfo{pages}{993--1018} (\bibinfo{year}{2008}).

\bibitem{2007ApJ...670..221D}
\bibinfo{author}{{Dursi}, L.~J.}
\newblock \bibinfo{title}{{Bubble Wrap for Bullets: The Stability Imparted by a
  Thin Magnetic Layer}}.
\newblock \emph{\bibinfo{journal}{\apj}} \textbf{\bibinfo{volume}{670}},
  \bibinfo{pages}{221--230} (\bibinfo{year}{2007}).

\bibitem{1999ApJ...525..357S}
\bibinfo{author}{{Slane}, P.} \emph{et~al.}
\newblock \bibinfo{title}{{Nonthermal X-Ray Emission from the Shell-Type
  Supernova Remnant G347.3-0.5}}.
\newblock \emph{\bibinfo{journal}{\apj}} \textbf{\bibinfo{volume}{525}},
  \bibinfo{pages}{357--367} (\bibinfo{year}{1999}).

\bibitem{2006ApJ...648L..33V}
\bibinfo{author}{{Vink}, J.} \emph{et~al.}
\newblock \bibinfo{title}{{The X-Ray Synchrotron Emission of RCW 86 and the
  Implications for Its Age}}.
\newblock \emph{\bibinfo{journal}{\apjl}} \textbf{\bibinfo{volume}{648}},
  \bibinfo{pages}{L33--L37} (\bibinfo{year}{2006}).

\bibitem{2009ApJS..180..265G}
\bibinfo{author}{{Gold}, B.} \emph{et~al.}
\newblock \bibinfo{title}{{Five-Year Wilkinson Microwave Anisotropy Probe
  Observations: Galactic Foreground Emission}}.
\newblock \emph{\bibinfo{journal}{\apjs}} \textbf{\bibinfo{volume}{180}},
  \bibinfo{pages}{265--282} (\bibinfo{year}{2009}).

\bibitem{2008A&A...483...89V}
\bibinfo{author}{{Vollmer}, B.} \emph{et~al.}
\newblock \bibinfo{title}{{Pre-peak ram pressure stripping in the Virgo cluster
  spiral galaxy NGC 4501}}.
\newblock \emph{\bibinfo{journal}{\aap}} \textbf{\bibinfo{volume}{483}},
  \bibinfo{pages}{89--106} (\bibinfo{year}{2008}).

\bibitem{1985A&A...147L...6D}
\bibinfo{author}{{de Jong}, T.}, \bibinfo{author}{{Klein}, U.},
  \bibinfo{author}{{Wielebinski}, R.} \& \bibinfo{author}{{Wunderlich}, E.}
\newblock \bibinfo{title}{{Radio continuum and far-infrared emission from
  spiral galaxies - A close correlation}}.
\newblock \emph{\bibinfo{journal}{\aap}} \textbf{\bibinfo{volume}{147}},
  \bibinfo{pages}{L6--L9} (\bibinfo{year}{1985}).

\bibitem{1985ApJ...298L...7H}
\bibinfo{author}{{Helou}, G.}, \bibinfo{author}{{Soifer}, B.~T.} \&
  \bibinfo{author}{{Rowan-Robinson}, M.}
\newblock \bibinfo{title}{{Thermal infrared and nonthermal radio - Remarkable
  correlation in disks of galaxies}}.
\newblock \emph{\bibinfo{journal}{\apjl}} \textbf{\bibinfo{volume}{298}},
  \bibinfo{pages}{L7--L11} (\bibinfo{year}{1985}).

\bibitem{2009ApJ...694.1435M}
\bibinfo{author}{{Murphy}, E.~J.}, \bibinfo{author}{{Kenney}, J.~D.~P.},
  \bibinfo{author}{{Helou}, G.}, \bibinfo{author}{{Chung}, A.} \&
  \bibinfo{author}{{Howell}, J.~H.}
\newblock \bibinfo{title}{{Environmental Effects in Clusters: Modified
  Far-Infrared-Radio Relations within Virgo Cluster Galaxies}}.
\newblock \emph{\bibinfo{journal}{\apj}} \textbf{\bibinfo{volume}{694}},
  \bibinfo{pages}{1435--1451} (\bibinfo{year}{2009}).

\bibitem{2000ApJ...534..420B}
\bibinfo{author}{{Balbus}, S.~A.}
\newblock \bibinfo{title}{{Stability, Instability, and ``Backward'' Transport
  in Stratified Fluids}}.
\newblock \emph{\bibinfo{journal}{\apj}} \textbf{\bibinfo{volume}{534}},
  \bibinfo{pages}{420--427} (\bibinfo{year}{2000}).

\bibitem{1965RvPP....1..205B}
\bibinfo{author}{{Braginskii}, S.~I.}
\newblock \bibinfo{title}{{Transport Processes in a Plasma}}.
\newblock \emph{\bibinfo{journal}{Reviews of Plasma Physics}}
  \textbf{\bibinfo{volume}{1}}, \bibinfo{pages}{205--311}
  (\bibinfo{year}{1965}).

\bibitem{2007ApJ...664..135P}
\bibinfo{author}{{Parrish}, I.~J.} \& \bibinfo{author}{{Stone}, J.~M.}
\newblock \bibinfo{title}{{Saturation of the Magnetothermal Instability in
  Three Dimensions}}.
\newblock \emph{\bibinfo{journal}{\apj}} \textbf{\bibinfo{volume}{664}},
  \bibinfo{pages}{135--148} (\bibinfo{year}{2007}).

\bibitem{2008ApJ...688..905P}
\bibinfo{author}{{Parrish}, I.~J.}, \bibinfo{author}{{Stone}, J.~M.} \&
  \bibinfo{author}{{Lemaster}, N.}
\newblock \bibinfo{title}{{The Magnetothermal Instability in the Intracluster
  Medium}}.
\newblock \emph{\bibinfo{journal}{\apj}} \textbf{\bibinfo{volume}{688}},
  \bibinfo{pages}{905--917} (\bibinfo{year}{2008}).

\bibitem{2006MNRAS.367..113P}
\bibinfo{author}{{Pfrommer}, C.}, \bibinfo{author}{{Springel}, V.},
  \bibinfo{author}{{En{\ss}lin}, T.~A.} \& \bibinfo{author}{{Jubelgas}, M.}
\newblock \bibinfo{title}{{Detecting shock waves in cosmological smoothed
  particle hydrodynamics simulations}}.
\newblock \emph{\bibinfo{journal}{\mnras}} \textbf{\bibinfo{volume}{367}},
  \bibinfo{pages}{113--131} (\bibinfo{year}{2006}).

\bibitem{2004MNRAS.351..423J}
\bibinfo{author}{{Jubelgas}, M.}, \bibinfo{author}{{Springel}, V.} \&
  \bibinfo{author}{{Dolag}, K.}
\newblock \bibinfo{title}{{Thermal conduction in cosmological SPH
  simulations}}.
\newblock \emph{\bibinfo{journal}{\mnras}} \textbf{\bibinfo{volume}{351}},
  \bibinfo{pages}{423--435} (\bibinfo{year}{2004}).

\bibitem{2004ApJ...606L..97D}
\bibinfo{author}{{Dolag}, K.}, \bibinfo{author}{{Jubelgas}, M.},
  \bibinfo{author}{{Springel}, V.}, \bibinfo{author}{{Borgani}, S.} \&
  \bibinfo{author}{{Rasia}, E.}
\newblock \bibinfo{title}{{Thermal Conduction in Simulated Galaxy Clusters}}.
\newblock \emph{\bibinfo{journal}{\apjl}} \textbf{\bibinfo{volume}{606}},
  \bibinfo{pages}{L97--L100} (\bibinfo{year}{2004}).

\bibitem{2001MNRAS.328L..37A}
\bibinfo{author}{{Allen}, S.~W.}, \bibinfo{author}{{Schmidt}, R.~W.} \&
  \bibinfo{author}{{Fabian}, A.~C.}
\newblock \bibinfo{title}{{The X-ray virial relations for relaxed lensing
  clusters observed with Chandra}}.
\newblock \emph{\bibinfo{journal}{\mnras}} \textbf{\bibinfo{volume}{328}},
  \bibinfo{pages}{L37--L41} (\bibinfo{year}{2001}).

\bibitem{2008ApJ...673..758Q}
\bibinfo{author}{{Quataert}, E.}
\newblock \bibinfo{title}{{Buoyancy Instabilities in Weakly Magnetized
  Low-Collisionality Plasmas}}.
\newblock \emph{\bibinfo{journal}{\apj}} \textbf{\bibinfo{volume}{673}},
  \bibinfo{pages}{758--762} (\bibinfo{year}{2008}).

\bibitem{2008ApJ...677L...9P}
\bibinfo{author}{{Parrish}, I.~J.} \& \bibinfo{author}{{Quataert}, E.}
\newblock \bibinfo{title}{{Nonlinear Simulations of the Heat-Flux-driven
  Buoyancy Instability and Its Implications for Galaxy Clusters}}.
\newblock \emph{\bibinfo{journal}{\apjl}} \textbf{\bibinfo{volume}{677}},
  \bibinfo{pages}{L9--L12} (\bibinfo{year}{2008}).

\bibitem{2008ApJ...688..208R}
\bibinfo{author}{{Randall}, S.} \emph{et~al.}
\newblock \bibinfo{title}{{Chandra's View of the Ram Pressure Stripped Galaxy
  M86}}.
\newblock \emph{\bibinfo{journal}{\apj}} \textbf{\bibinfo{volume}{688}},
  \bibinfo{pages}{208--223} (\bibinfo{year}{2008}).

\bibitem{2009ApJS..182...12C}
\bibinfo{author}{{Cavagnolo}, K.~W.}, \bibinfo{author}{{Donahue}, M.},
  \bibinfo{author}{{Voit}, G.~M.} \& \bibinfo{author}{{Sun}, M.}
\newblock \bibinfo{title}{{Intracluster Medium Entropy Profiles for a Chandra
  Archival Sample of Galaxy Clusters}}.
\newblock \emph{\bibinfo{journal}{\apjs}} \textbf{\bibinfo{volume}{182}},
  \bibinfo{pages}{12--32} (\bibinfo{year}{2009}).

\bibitem{2009MNRAS.395..764S}
\bibinfo{author}{{Sanderson}, A.~J.~R.}, \bibinfo{author}{{O'Sullivan}, E.} \&
  \bibinfo{author}{{Ponman}, T.~J.}
\newblock \bibinfo{title}{{A statistically selected Chandra sample of 20 galaxy
  clusters - II. Gas properties and cool core/non-cool core bimodality}}.
\newblock \emph{\bibinfo{journal}{\mnras}} \textbf{\bibinfo{volume}{395}},
  \bibinfo{pages}{764--776} (\bibinfo{year}{2009}).

\bibitem{2008ApJ...681L...5V}
\bibinfo{author}{{Voit}, G.~M.} \emph{et~al.}
\newblock \bibinfo{title}{{Conduction and the Star Formation Threshold in
  Brightest Cluster Galaxies}}.
\newblock \emph{\bibinfo{journal}{\apjl}} \textbf{\bibinfo{volume}{681}},
  \bibinfo{pages}{L5--L8} (\bibinfo{year}{2008}).

\bibitem{2008ApJ...688..859G}
\bibinfo{author}{{Guo}, F.}, \bibinfo{author}{{Oh}, S.~P.} \&
  \bibinfo{author}{{Ruszkowski}, M.}
\newblock \bibinfo{title}{{A Global Stability Analysis of Clusters of Galaxies
  with Conduction and AGN Feedback Heating}}.
\newblock \emph{\bibinfo{journal}{\apj}} \textbf{\bibinfo{volume}{688}},
  \bibinfo{pages}{859--874} (\bibinfo{year}{2008}).

\bibitem{2001ApJ...554..261C}
\bibinfo{author}{{Churazov}, E.}, \bibinfo{author}{{Br{\"u}ggen}, M.},
  \bibinfo{author}{{Kaiser}, C.~R.}, \bibinfo{author}{{B{\"o}hringer}, H.} \&
  \bibinfo{author}{{Forman}, W.}
\newblock \bibinfo{title}{{Evolution of Buoyant Bubbles in M87}}.
\newblock \emph{\bibinfo{journal}{\apj}} \textbf{\bibinfo{volume}{554}},
  \bibinfo{pages}{261--273} (\bibinfo{year}{2001}).

\bibitem{2007MNRAS.378..385P}
\bibinfo{author}{{Pfrommer}, C.}, \bibinfo{author}{{En{\ss}lin}, T.~A.},
  \bibinfo{author}{{Springel}, V.}, \bibinfo{author}{{Jubelgas}, M.} \&
  \bibinfo{author}{{Dolag}, K.}
\newblock \bibinfo{title}{{Simulating cosmic rays in clusters of galaxies - I.
  Effects on the Sunyaev-Zel'dovich effect and the X-ray emission}}.
\newblock \emph{\bibinfo{journal}{\mnras}} \textbf{\bibinfo{volume}{378}},
  \bibinfo{pages}{385--408} (\bibinfo{year}{2007}).

\bibitem{2007ApJ...668....1N}
\bibinfo{author}{{Nagai}, D.}, \bibinfo{author}{{Kravtsov}, A.~V.} \&
  \bibinfo{author}{{Vikhlinin}, A.}
\newblock \bibinfo{title}{{Effects of Galaxy Formation on Thermodynamics of the
  Intracluster Medium}}.
\newblock \emph{\bibinfo{journal}{\apj}} \textbf{\bibinfo{volume}{668}},
  \bibinfo{pages}{1--14} (\bibinfo{year}{2007}).

\bibitem{2009AJ....138.1741C}
\bibinfo{author}{{Chung}, A.}, \bibinfo{author}{{van Gorkom}, J.~H.},
  \bibinfo{author}{{Kenney}, J.~D.~P.}, \bibinfo{author}{{Crowl}, H.} \&
  \bibinfo{author}{{Vollmer}, B.}
\newblock \bibinfo{title}{{VLA Imaging of Virgo Spirals in Atomic Gas (VIVA).
  I. The Atlas and the H I Properties}}.
\newblock \emph{\bibinfo{journal}{\aj}} \textbf{\bibinfo{volume}{138}},
  \bibinfo{pages}{1741--1816} (\bibinfo{year}{2009}).

\bibitem{StoneEtAl2008}
\bibinfo{author}{{Stone}, J.~M.}, \bibinfo{author}{{Gardiner}, T.~A.},
  \bibinfo{author}{{Teuben}, P.}, \bibinfo{author}{{Hawley}, J.~F.} \&
  \bibinfo{author}{{Simon}, J.~B.}
\newblock \bibinfo{title}{{Athena: A New Code for Astrophysical MHD}}.
\newblock \emph{\bibinfo{journal}{\apjs}} \textbf{\bibinfo{volume}{178}},
  \bibinfo{pages}{137--177} (\bibinfo{year}{2008}).

\bibitem{2005JCoPh.205..509G}
\bibinfo{author}{{Gardiner}, T.~A.} \& \bibinfo{author}{{Stone}, J.~M.}
\newblock \bibinfo{title}{{An unsplit Godunov method for ideal MHD via
  constrained transport}}.
\newblock \emph{\bibinfo{journal}{Journal of Computational Physics}}
  \textbf{\bibinfo{volume}{205}}, \bibinfo{pages}{509--539}
  (\bibinfo{year}{2005}).

\bibitem{2008JCoPh.227.4123G}
\bibinfo{author}{{Gardiner}, T.~A.} \& \bibinfo{author}{{Stone}, J.~M.}
\newblock \bibinfo{title}{{An unsplit Godunov method for ideal MHD via
  constrained transport in three dimensions}}.
\newblock \emph{\bibinfo{journal}{Journal of Computational Physics}}
  \textbf{\bibinfo{volume}{227}}, \bibinfo{pages}{4123--4141}
  (\bibinfo{year}{2008}).

\bibitem{1994Natur.368..828B}
\bibinfo{author}{{B{\"o}hringer}, H.} \emph{et~al.}
\newblock \bibinfo{title}{{The structure of the Virgo cluster of galaxies from
  Rosat X-ray images}}.
\newblock \emph{\bibinfo{journal}{\nat}} \textbf{\bibinfo{volume}{368}},
  \bibinfo{pages}{828--831} (\bibinfo{year}{1994}).

\end{thebibliography}


\begin{thebibliography}{10}
\expandafter\ifx\csname url\endcsname\relax
  \def\url#1{\texttt{#1}}\fi
\expandafter\ifx\csname urlprefix\endcsname\relax\def\urlprefix{URL }\fi
\providecommand{\bibinfo}[2]{#2}
\providecommand{\eprint}[2][]{\url{#2}}

\bibitem{GardinerStone2005}
\bibinfo{author}{{Gardiner}, T.~A.} \& \bibinfo{author}{{Stone}, J.~M.}
\newblock \bibinfo{title}{{An unsplit Godunov method for ideal MHD via
  constrained transport}}.
\newblock \emph{\bibinfo{journal}{Journal of Computational Physics}}
  \textbf{\bibinfo{volume}{205}}, \bibinfo{pages}{509--539}
  (\bibinfo{year}{2005}).

\bibitem{GardinerStone2008}
\bibinfo{author}{{Gardiner}, T.~A.} \& \bibinfo{author}{{Stone}, J.~M.}
\newblock \bibinfo{title}{{An unsplit Godunov method for ideal MHD via
  constrained transport in three dimensions}}.
\newblock \emph{\bibinfo{journal}{Journal of Computational Physics}}
  \textbf{\bibinfo{volume}{227}}, \bibinfo{pages}{4123--4141}
  (\bibinfo{year}{2008}).

\bibitem{StoneEtAl2008}
\bibinfo{author}{{Stone}, J.~M.}, \bibinfo{author}{{Gardiner}, T.~A.},
  \bibinfo{author}{{Teuben}, P.}, \bibinfo{author}{{Hawley}, J.~F.} \&
  \bibinfo{author}{{Simon}, J.~B.}
\newblock \bibinfo{title}{{Athena: A New Code for Astrophysical MHD}}.
\newblock \emph{\bibinfo{journal}{\apjs}} \textbf{\bibinfo{volume}{178}},
  \bibinfo{pages}{137--177} (\bibinfo{year}{2008}).

\bibitem{2008ApJ...677..993D}
\bibinfo{author}{{Dursi}, L.~J.} \& \bibinfo{author}{{Pfrommer}, C.}
\newblock \bibinfo{title}{{Draping of Cluster Magnetic Fields over Bullets and
  Bubbles-Morphology and Dynamic Effects}}.
\newblock \emph{\bibinfo{journal}{\apj}} \textbf{\bibinfo{volume}{677}},
  \bibinfo{pages}{993--1018} (\bibinfo{year}{2008}).

\bibitem{flashcode}
\bibinfo{author}{{Fryxell}, B.} \emph{et~al.}
\newblock \bibinfo{title}{{FLASH: An Adaptive Mesh Hydrodynamics Code for
  Modeling Astrophysical Thermonuclear Flashes}}.
\newblock \emph{\bibinfo{journal}{\apjs}} \textbf{\bibinfo{volume}{131}},
  \bibinfo{pages}{273--334} (\bibinfo{year}{2000}).

\bibitem{flashvalidation}
\bibinfo{author}{{Calder}, A.~C.} \emph{et~al.}
\newblock \bibinfo{title}{{On Validating an Astrophysical Simulation Code}}.
\newblock \emph{\bibinfo{journal}{\apjs}} \textbf{\bibinfo{volume}{143}},
  \bibinfo{pages}{201--229} (\bibinfo{year}{2002}).

\bibitem{2002ARA&A..40..319C}
\bibinfo{author}{{Carilli}, C.~L.} \& \bibinfo{author}{{Taylor}, G.~B.}
\newblock \bibinfo{title}{{Cluster Magnetic Fields}}.
\newblock \emph{\bibinfo{journal}{\araa}} \textbf{\bibinfo{volume}{40}},
  \bibinfo{pages}{319--348} (\bibinfo{year}{2002}).

\bibitem{2002RvMP...74..775W}
\bibinfo{author}{{Widrow}, L.~M.}
\newblock \bibinfo{title}{{Origin of galactic and extragalactic magnetic
  fields}}.
\newblock \emph{\bibinfo{journal}{Reviews of Modern Physics}}
  \textbf{\bibinfo{volume}{74}}, \bibinfo{pages}{775--823}
  (\bibinfo{year}{2002}).

\bibitem{2004IJMPD..13.1549G}
\bibinfo{author}{{Govoni}, F.} \& \bibinfo{author}{{Feretti}, L.}
\newblock \bibinfo{title}{{Magnetic Fields in Clusters of Galaxies}}.
\newblock \emph{\bibinfo{journal}{International Journal of Modern Physics D}}
  \textbf{\bibinfo{volume}{13}}, \bibinfo{pages}{1549--1594}
  (\bibinfo{year}{2004}).

\bibitem{2005A&A...434...67V}
\bibinfo{author}{{Vogt}, C.} \& \bibinfo{author}{{En{\ss}lin}, T.~A.}
\newblock \bibinfo{title}{{A Bayesian view on Faraday rotation maps Seeing the
  magnetic power spectra in galaxy clusters}}.
\newblock \emph{\bibinfo{journal}{\aap}} \textbf{\bibinfo{volume}{434}},
  \bibinfo{pages}{67--76} (\bibinfo{year}{2005}).

\bibitem{2009arXiv0912.3930K}
\bibinfo{author}{{Kuchar}, P.} \& \bibinfo{author}{{Ensslin}, T.~A.}
\newblock \bibinfo{title}{{Magnetic power spectra from Faraday rotation maps -
  REALMAF and its use on Hydra A}}.
\newblock \emph{\bibinfo{journal}{ArXiv e-prints}}  (\bibinfo{year}{2009}).
\newblock \eprint{0912.3930}.

\bibitem{2002A&A...386...77M}
\bibinfo{author}{{Matsushita}, K.}, \bibinfo{author}{{Belsole}, E.},
  \bibinfo{author}{{Finoguenov}, A.} \& \bibinfo{author}{{B{\"o}hringer}, H.}
\newblock \bibinfo{title}{{XMM-Newton observation of M 87. I. Single-phase
  temperature structure of intracluster medium}}.
\newblock \emph{\bibinfo{journal}{\aap}} \textbf{\bibinfo{volume}{386}},
  \bibinfo{pages}{77--96} (\bibinfo{year}{2002}).

\bibitem{1999A&A...348..351D}
\bibinfo{author}{{Dolag}, K.}, \bibinfo{author}{{Bartelmann}, M.} \&
  \bibinfo{author}{{Lesch}, H.}
\newblock \bibinfo{title}{Sph simulations of magnetic fields in galaxy
  clusters}.
\newblock \emph{\bibinfo{journal}{\aap}} \textbf{\bibinfo{volume}{348}},
  \bibinfo{pages}{351--363} (\bibinfo{year}{1999}).

\bibitem{2001A&A...378..777D}
\bibinfo{author}{{Dolag}, K.}, \bibinfo{author}{{Schindler}, S.},
  \bibinfo{author}{{Govoni}, F.} \& \bibinfo{author}{{Feretti}, L.}
\newblock \bibinfo{title}{{Correlation of the magnetic field and the
  intra-cluster gas density in galaxy clusters}}.
\newblock \emph{\bibinfo{journal}{\aap}} \textbf{\bibinfo{volume}{378}},
  \bibinfo{pages}{777--786} (\bibinfo{year}{2001}).

\bibitem{2006PhPl...13e6501S}
\bibinfo{author}{{Schekochihin}, A.~A.} \& \bibinfo{author}{{Cowley}, S.~C.}
\newblock \bibinfo{title}{{Turbulence, magnetic fields, and plasma physics in
  clusters of galaxies}}.
\newblock \emph{\bibinfo{journal}{Physics of Plasmas}}
  \textbf{\bibinfo{volume}{13}}, \bibinfo{pages}{056501}
  (\bibinfo{year}{2006}).

\bibitem{1980Ge&Ae..19..671B}
\bibinfo{author}{{Bernikov}, L.~V.} \& \bibinfo{author}{{Semenov}, V.~S.}
\newblock \bibinfo{title}{{Problem of MHD flow around the magnetosphere}}.
\newblock \emph{\bibinfo{journal}{Geomagnetizm i Aeronomiia}}
  \textbf{\bibinfo{volume}{19}}, \bibinfo{pages}{671--675}
  (\bibinfo{year}{1980}).

\bibitem{lyutikovdraping}
\bibinfo{author}{{Lyutikov}, M.}
\newblock \bibinfo{title}{{Magnetic draping of merging cores and radio bubbles
  in clusters of galaxies}}.
\newblock \emph{\bibinfo{journal}{\mnras}} \textbf{\bibinfo{volume}{373}},
  \bibinfo{pages}{73--78} (\bibinfo{year}{2006}).

\bibitem{D&P}
\bibinfo{author}{{Dursi}, L.~J.} \& \bibinfo{author}{{Pfrommer}, C.}
\newblock \emph{\bibinfo{journal}{in prep.}}  (\bibinfo{year}{2010}).

\bibitem{2007MNRAS.378..662R}
\bibinfo{author}{{Ruszkowski}, M.}, \bibinfo{author}{{En{\ss}lin}, T.~A.},
  \bibinfo{author}{{Br{\"u}ggen}, M.}, \bibinfo{author}{{Heinz}, S.} \&
  \bibinfo{author}{{Pfrommer}, C.}
\newblock \bibinfo{title}{{Impact of tangled magnetic fields on fossil radio
  bubbles}}.
\newblock \emph{\bibinfo{journal}{\mnras}} \textbf{\bibinfo{volume}{378}},
  \bibinfo{pages}{662--672} (\bibinfo{year}{2007}).

\bibitem{Ruszkowski2010}
\bibinfo{author}{{Ruszkowski}, M.}
\newblock \emph{\bibinfo{journal}{private communication}}
  (\bibinfo{year}{2010}).

\bibitem{2005AnGeo..23..885C}
\bibinfo{author}{{Coleman}, I.~J.}
\newblock \bibinfo{title}{{A multi-spacecraft survey of magnetic field line
  draping in the dayside magnetosheath}}.
\newblock \emph{\bibinfo{journal}{Annales Geophysicae}}
  \textbf{\bibinfo{volume}{23}}, \bibinfo{pages}{885--900}
  (\bibinfo{year}{2005}).

\bibitem{chandra}
\bibinfo{author}{Chandrasekhar, S.}
\newblock \emph{\bibinfo{title}{Hydrodynamic and Hydromagnetic Stability}}
  (\bibinfo{publisher}{Dover}, \bibinfo{address}{New York},
  \bibinfo{year}{1981}).

\bibitem{2007ApJ...670..221D}
\bibinfo{author}{{Dursi}, L.~J.}
\newblock \bibinfo{title}{{Bubble Wrap for Bullets: The Stability Imparted by a
  Thin Magnetic Layer}}.
\newblock \emph{\bibinfo{journal}{\apj}} \textbf{\bibinfo{volume}{670}},
  \bibinfo{pages}{221--230} (\bibinfo{year}{2007}).

\bibitem{RybickiLightman}
\bibinfo{author}{{Rybicki}, G.~B.} \& \bibinfo{author}{{Lightman}, A.~P.}
\newblock \emph{\bibinfo{title}{{Radiative Processes in Astrophysics}}}
  (\bibinfo{year}{1986}).

\bibitem{2004AJ....127.3375V}
\bibinfo{author}{{Vollmer}, B.}, \bibinfo{author}{{Beck}, R.},
  \bibinfo{author}{{Kenney}, J.~D.~P.} \& \bibinfo{author}{{van Gorkom}, J.~H.}
\newblock \bibinfo{title}{{Radio Continuum Observations of the Virgo Cluster
  Spiral NGC 4522: The Signature of Ram Pressure}}.
\newblock \emph{\bibinfo{journal}{\aj}} \textbf{\bibinfo{volume}{127}},
  \bibinfo{pages}{3375--3381} (\bibinfo{year}{2004}).

\bibitem{1987PhR...154....1B}
\bibinfo{author}{{Blandford}, R.} \& \bibinfo{author}{{Eichler}, D.}
\newblock \bibinfo{title}{{Particle Acceleration at Astrophysical Shocks - a
  Theory of Cosmic-Ray Origin}}.
\newblock \emph{\bibinfo{journal}{\physrep}} \textbf{\bibinfo{volume}{154}},
  \bibinfo{pages}{1--75} (\bibinfo{year}{1987}).

\bibitem{2007A&A...473...41E}
\bibinfo{author}{{En{\ss}lin}, T.~A.}, \bibinfo{author}{{Pfrommer}, C.},
  \bibinfo{author}{{Springel}, V.} \& \bibinfo{author}{{Jubelgas}, M.}
\newblock \bibinfo{title}{{Cosmic ray physics in calculations of cosmological
  structure formation}}.
\newblock \emph{\bibinfo{journal}{\aap}} \textbf{\bibinfo{volume}{473}},
  \bibinfo{pages}{41--57} (\bibinfo{year}{2007}).

\bibitem{1999ApJ...520..529S}
\bibinfo{author}{{Sarazin}, C.~L.}
\newblock \bibinfo{title}{The energy spectrum of primary cosmic-ray electrons
  in clusters of galaxies and inverse compton emission}.
\newblock \emph{\bibinfo{journal}{\apj}} \textbf{\bibinfo{volume}{520}},
  \bibinfo{pages}{529--547} (\bibinfo{year}{1999}).

\bibitem{1977ApJ...212....1J}
\bibinfo{author}{{Jaffe}, W.~J.}
\newblock \bibinfo{title}{Origin and transport of electrons in the halo radio
  source in the coma cluster}.
\newblock \emph{\bibinfo{journal}{\apj}} \textbf{\bibinfo{volume}{212}},
  \bibinfo{pages}{1--7} (\bibinfo{year}{1977}).

\bibitem{1987A&A...182...21S}
\bibinfo{author}{{Schlickeiser}, R.}, \bibinfo{author}{{Sievers}, A.} \&
  \bibinfo{author}{{Thiemann}, H.}
\newblock \bibinfo{title}{The diffuse radio emission from the coma cluster}.
\newblock \emph{\bibinfo{journal}{\aap}} \textbf{\bibinfo{volume}{182}},
  \bibinfo{pages}{21--35} (\bibinfo{year}{1987}).

\bibitem{2001MNRAS.320..365B}
\bibinfo{author}{{Brunetti}, G.}, \bibinfo{author}{{Setti}, G.},
  \bibinfo{author}{{Feretti}, L.} \& \bibinfo{author}{{Giovannini}, G.}
\newblock \bibinfo{title}{Particle reacceleration in the coma cluster: radio
  properties and hard x-ray emission}.
\newblock \emph{\bibinfo{journal}{\mnras}} \textbf{\bibinfo{volume}{320}},
  \bibinfo{pages}{365--378} (\bibinfo{year}{2001}).

\bibitem{2002ApJ...577..658O}
\bibinfo{author}{{Ohno}, H.}, \bibinfo{author}{{Takizawa}, M.} \&
  \bibinfo{author}{{Shibata}, S.}
\newblock \bibinfo{title}{{Radio Halo Formation through Magnetoturbulent
  Particle Acceleration in Clusters of Galaxies}}.
\newblock \emph{\bibinfo{journal}{\apj}} \textbf{\bibinfo{volume}{577}},
  \bibinfo{pages}{658--667} (\bibinfo{year}{2002}).

\bibitem{2004MNRAS.350.1174B}
\bibinfo{author}{{Brunetti}, G.}, \bibinfo{author}{{Blasi}, P.},
  \bibinfo{author}{{Cassano}, R.} \& \bibinfo{author}{{Gabici}, S.}
\newblock \bibinfo{title}{{Alfv{\' e}nic reacceleration of relativistic
  particles in galaxy clusters: MHD waves, leptons and hadrons}}.
\newblock \emph{\bibinfo{journal}{\mnras}} \textbf{\bibinfo{volume}{350}},
  \bibinfo{pages}{1174--1194} (\bibinfo{year}{2004}).

\bibitem{2007MNRAS.378..245B}
\bibinfo{author}{{Brunetti}, G.} \& \bibinfo{author}{{Lazarian}, A.}
\newblock \bibinfo{title}{{Compressible turbulence in galaxy clusters: physics
  and stochastic particle re-acceleration}}.
\newblock \emph{\bibinfo{journal}{\mnras}} \textbf{\bibinfo{volume}{378}},
  \bibinfo{pages}{245--275} (\bibinfo{year}{2007}).

\bibitem{Vollmer2009}
\bibinfo{author}{{Vollmer}, B.}
\newblock \emph{\bibinfo{journal}{private communication}}
  (\bibinfo{year}{2009}).

\bibitem{1999ApJ...525..357S}
\bibinfo{author}{{Slane}, P.} \emph{et~al.}
\newblock \bibinfo{title}{{Nonthermal X-Ray Emission from the Shell-Type
  Supernova Remnant G347.3-0.5}}.
\newblock \emph{\bibinfo{journal}{\apj}} \textbf{\bibinfo{volume}{525}},
  \bibinfo{pages}{357--367} (\bibinfo{year}{1999}).

\bibitem{2006ApJ...648L..33V}
\bibinfo{author}{{Vink}, J.} \emph{et~al.}
\newblock \bibinfo{title}{{The X-Ray Synchrotron Emission of RCW 86 and the
  Implications for Its Age}}.
\newblock \emph{\bibinfo{journal}{\apjl}} \textbf{\bibinfo{volume}{648}},
  \bibinfo{pages}{L33--L37} (\bibinfo{year}{2006}).

\bibitem{2006Natur.439..695A}
\bibinfo{author}{{Aharonian}, F.} \emph{et~al.}
\newblock \bibinfo{title}{{Discovery of very-high-energy {$\gamma$}-rays from
  the Galactic Centre ridge}}.
\newblock \emph{\bibinfo{journal}{\nat}} \textbf{\bibinfo{volume}{439}},
  \bibinfo{pages}{695--698} (\bibinfo{year}{2006}).

\bibitem{2003A&A...399..409E}
\bibinfo{author}{{En{\ss}lin}, T.~A.}
\newblock \bibinfo{title}{{On the escape of cosmic rays from radio galaxy
  cocoons}}.
\newblock \emph{\bibinfo{journal}{\aap}} \textbf{\bibinfo{volume}{399}},
  \bibinfo{pages}{409--420} (\bibinfo{year}{2003}).

\bibitem{2009ApJS..180..265G}
\bibinfo{author}{{Gold}, B.} \emph{et~al.}
\newblock \bibinfo{title}{{Five-Year Wilkinson Microwave Anisotropy Probe
  Observations: Galactic Foreground Emission}}.
\newblock \emph{\bibinfo{journal}{\apjs}} \textbf{\bibinfo{volume}{180}},
  \bibinfo{pages}{265--282} (\bibinfo{year}{2009}).

\bibitem{2008A&A...477..573S}
\bibinfo{author}{{Sun}, X.~H.}, \bibinfo{author}{{Reich}, W.},
  \bibinfo{author}{{Waelkens}, A.} \& \bibinfo{author}{{En{\ss}lin}, T.~A.}
\newblock \bibinfo{title}{{Radio observational constraints on Galactic
  3D-emission models}}.
\newblock \emph{\bibinfo{journal}{\aap}} \textbf{\bibinfo{volume}{477}},
  \bibinfo{pages}{573--592} (\bibinfo{year}{2008}).

\bibitem{2009A&A...495..697W}
\bibinfo{author}{{Waelkens}, A.}, \bibinfo{author}{{Jaffe}, T.},
  \bibinfo{author}{{Reinecke}, M.}, \bibinfo{author}{{Kitaura}, F.~S.} \&
  \bibinfo{author}{{En{\ss}lin}, T.~A.}
\newblock \bibinfo{title}{{Simulating polarized Galactic synchrotron emission
  at all frequencies. The Hammurabi code}}.
\newblock \emph{\bibinfo{journal}{\aap}} \textbf{\bibinfo{volume}{495}},
  \bibinfo{pages}{697--706} (\bibinfo{year}{2009}).

\bibitem{2002cra..book.....S}
\bibinfo{author}{{Schlickeiser}, R.}
\newblock \emph{\bibinfo{title}{{Cosmic Ray Astrophysics}}}
  (\bibinfo{year}{2002}).

\bibitem{1998ApJ...498..541K}
\bibinfo{author}{{Kennicutt}, R.~C., Jr.}
\newblock \bibinfo{title}{{The Global Schmidt Law in Star-forming Galaxies}}.
\newblock \emph{\bibinfo{journal}{\apj}} \textbf{\bibinfo{volume}{498}},
  \bibinfo{pages}{541--552} (\bibinfo{year}{1998}).

\bibitem{2006ApJ...645..186T}
\bibinfo{author}{{Thompson}, T.~A.}, \bibinfo{author}{{Quataert}, E.},
  \bibinfo{author}{{Waxman}, E.}, \bibinfo{author}{{Murray}, N.} \&
  \bibinfo{author}{{Martin}, C.~L.}
\newblock \bibinfo{title}{{Magnetic Fields in Starburst Galaxies and the Origin
  of the FIR-Radio Correlation}}.
\newblock \emph{\bibinfo{journal}{\apj}} \textbf{\bibinfo{volume}{645}},
  \bibinfo{pages}{186--198} (\bibinfo{year}{2006}).

\bibitem{1994hea..book.....L}
\bibinfo{author}{{Longair}, M.~S.}
\newblock \emph{\bibinfo{title}{{High energy astrophysics. Vol.2: Stars, the
  galaxy and the interstellar medium}}} (\bibinfo{year}{1994}).

\bibitem{2001SSRv...99..243B}
\bibinfo{author}{{Beck}, R.}
\newblock \bibinfo{title}{{Galactic and Extragalactic Magnetic Fields}}.
\newblock \emph{\bibinfo{journal}{Space Science Reviews}}
  \textbf{\bibinfo{volume}{99}}, \bibinfo{pages}{243--260}
  (\bibinfo{year}{2001}).

\bibitem{1966MNRAS.133...67B}
\bibinfo{author}{{Burn}, B.~J.}
\newblock \bibinfo{title}{{On the depolarization of discrete radio sources by
  Faraday dispersion}}.
\newblock \emph{\bibinfo{journal}{\mnras}} \textbf{\bibinfo{volume}{133}},
  \bibinfo{pages}{67--83} (\bibinfo{year}{1966}).

\bibitem{2009MNRAS.393.1073B}
\bibinfo{author}{{Battaglia}, N.}, \bibinfo{author}{{Pfrommer}, C.},
  \bibinfo{author}{{Sievers}, J.~L.}, \bibinfo{author}{{Bond}, J.~R.} \&
  \bibinfo{author}{{En{\ss}lin}, T.~A.}
\newblock \bibinfo{title}{{Exploring the magnetized cosmic web through
  low-frequency radio emission}}.
\newblock \emph{\bibinfo{journal}{\mnras}} \textbf{\bibinfo{volume}{393}},
  \bibinfo{pages}{1073--1089} (\bibinfo{year}{2009}).

\bibitem{Vollmer2008}
\bibinfo{author}{{Vollmer}, B.} \emph{et~al.}
\newblock \bibinfo{title}{{Pre-peak ram pressure stripping in the Virgo cluster
  spiral galaxy NGC 4501}}.
\newblock \emph{\bibinfo{journal}{A\&Ap}} \textbf{\bibinfo{volume}{483}},
  \bibinfo{pages}{89--106} (\bibinfo{year}{2008}).

\bibitem{2009ApJ...694.1435M}
\bibinfo{author}{{Murphy}, E.~J.}, \bibinfo{author}{{Kenney}, J.~D.~P.},
  \bibinfo{author}{{Helou}, G.}, \bibinfo{author}{{Chung}, A.} \&
  \bibinfo{author}{{Howell}, J.~H.}
\newblock \bibinfo{title}{{Environmental Effects in Clusters: Modified
  Far-Infrared-Radio Relations within Virgo Cluster Galaxies}}.
\newblock \emph{\bibinfo{journal}{\apj}} \textbf{\bibinfo{volume}{694}},
  \bibinfo{pages}{1435--1451} (\bibinfo{year}{2009}).

\bibitem{2009ApJ...694..789T}
\bibinfo{author}{{Tonnesen}, S.} \& \bibinfo{author}{{Bryan}, G.~L.}
\newblock \bibinfo{title}{{Gas Stripping in Simulated Galaxies with a
  Multiphase Interstellar Medium}}.
\newblock \emph{\bibinfo{journal}{\apj}} \textbf{\bibinfo{volume}{694}},
  \bibinfo{pages}{789--804} (\bibinfo{year}{2009}).

\bibitem{2007A&A...464L..37V}
\bibinfo{author}{{Vollmer}, B.} \emph{et~al.}
\newblock \bibinfo{title}{{The characteristic polarized radio continuum
  distribution of cluster spiral galaxies}}.
\newblock \emph{\bibinfo{journal}{\aap}} \textbf{\bibinfo{volume}{464}},
  \bibinfo{pages}{L37--L40} (\bibinfo{year}{2007}).

\bibitem{2007A&A...471...93W}
\bibinfo{author}{{We{\.z}gowiec}, M.} \emph{et~al.}
\newblock \bibinfo{title}{{The magnetic fields of large Virgo Cluster
  spirals}}.
\newblock \emph{\bibinfo{journal}{\aap}} \textbf{\bibinfo{volume}{471}},
  \bibinfo{pages}{93--102} (\bibinfo{year}{2007}).

\bibitem{2010arXiv1001.3597V}
\bibinfo{author}{{Vollmer}, B.} \emph{et~al.}
\newblock \bibinfo{title}{{The influence of the cluster environment on the
  large-scale radio continuum emission of 8 Virgo cluster spirals}}.
\newblock \emph{\bibinfo{journal}{\aap~accepted}}  (\bibinfo{year}{2010}).
\newblock \eprint{arXiv:1001.3597}.

\bibitem{2003A&A...402..879O}
\bibinfo{author}{{Otmianowska-Mazur}, K.} \& \bibinfo{author}{{Vollmer}, B.}
\newblock \bibinfo{title}{{Magnetic field evolution in galaxies interacting
  with the intracluster medium. 3D numerical simulations}}.
\newblock \emph{\bibinfo{journal}{\aap}} \textbf{\bibinfo{volume}{402}},
  \bibinfo{pages}{879--889} (\bibinfo{year}{2003}).

\bibitem{2006A&A...458..727S}
\bibinfo{author}{{Soida}, M.}, \bibinfo{author}{{Otmianowska-Mazur}, K.},
  \bibinfo{author}{{Chy{\.z}y}, K.} \& \bibinfo{author}{{Vollmer}, B.}
\newblock \bibinfo{title}{{NGC 4654: polarized radio continuum emission as a
  diagnostic tool for a galaxy-cluster interaction. Models versus
  observations}}.
\newblock \emph{\bibinfo{journal}{\aap}} \textbf{\bibinfo{volume}{458}},
  \bibinfo{pages}{727--739} (\bibinfo{year}{2006}).

\bibitem{2009AJ....138.1741C}
\bibinfo{author}{{Chung}, A.}, \bibinfo{author}{{van Gorkom}, J.~H.},
  \bibinfo{author}{{Kenney}, J.~D.~P.}, \bibinfo{author}{{Crowl}, H.} \&
  \bibinfo{author}{{Vollmer}, B.}
\newblock \bibinfo{title}{{VLA Imaging of Virgo Spirals in Atomic Gas (VIVA).
  I. The Atlas and the H I Properties}}.
\newblock \emph{\bibinfo{journal}{\aj}} \textbf{\bibinfo{volume}{138}},
  \bibinfo{pages}{1741--1816} (\bibinfo{year}{2009}).

\bibitem{1999ApJ...520..111V}
\bibinfo{author}{{Veilleux}, S.}, \bibinfo{author}{{Bland-Hawthorn}, J.},
  \bibinfo{author}{{Cecil}, G.}, \bibinfo{author}{{Tully}, R.~B.} \&
  \bibinfo{author}{{Miller}, S.~T.}
\newblock \bibinfo{title}{{Galactic-Scale Outflow and Supersonic Ram-Pressure
  Stripping in the Virgo Cluster Galaxy NGC 4388}}.
\newblock \emph{\bibinfo{journal}{\apj}} \textbf{\bibinfo{volume}{520}},
  \bibinfo{pages}{111--123} (\bibinfo{year}{1999}).

\bibitem{1991A&A...249...43H}
\bibinfo{author}{{Hummel}, E.} \& \bibinfo{author}{{Saikia}, D.~J.}
\newblock \bibinfo{title}{{The anomalous radio features in NGC 4388 and NGC
  4438}}.
\newblock \emph{\bibinfo{journal}{\aap}} \textbf{\bibinfo{volume}{249}},
  \bibinfo{pages}{43--56} (\bibinfo{year}{1991}).

\bibitem{2007ApJ...659L.115C}
\bibinfo{author}{{Chung}, A.}, \bibinfo{author}{{van Gorkom}, J.~H.},
  \bibinfo{author}{{Kenney}, J.~D.~P.} \& \bibinfo{author}{{Vollmer}, B.}
\newblock \bibinfo{title}{{Virgo Galaxies with Long One-sided H I Tails}}.
\newblock \emph{\bibinfo{journal}{\apjl}} \textbf{\bibinfo{volume}{659}},
  \bibinfo{pages}{L115--L119} (\bibinfo{year}{2007}).

\bibitem{1995ApJ...453..154P}
\bibinfo{author}{{Phookun}, B.} \& \bibinfo{author}{{Mundy}, L.~G.}
\newblock \bibinfo{title}{{NGC 4654: A Virgo Cluster Spiral Interacting with
  the Intracluster Medium}}.
\newblock \emph{\bibinfo{journal}{\apj}} \textbf{\bibinfo{volume}{453}},
  \bibinfo{pages}{154--161} (\bibinfo{year}{1995}).

\bibitem{2007ApJ...664..135P}
\bibinfo{author}{{Parrish}, I.~J.} \& \bibinfo{author}{{Stone}, J.~M.}
\newblock \bibinfo{title}{{Saturation of the Magnetothermal Instability in
  Three Dimensions}}.
\newblock \emph{\bibinfo{journal}{\apj}} \textbf{\bibinfo{volume}{664}},
  \bibinfo{pages}{135--148} (\bibinfo{year}{2007}).

\bibitem{2008ApJ...688..905P}
\bibinfo{author}{{Parrish}, I.~J.}, \bibinfo{author}{{Stone}, J.~M.} \&
  \bibinfo{author}{{Lemaster}, N.}
\newblock \bibinfo{title}{{The Magnetothermal Instability in the Intracluster
  Medium}}.
\newblock \emph{\bibinfo{journal}{\apj}} \textbf{\bibinfo{volume}{688}},
  \bibinfo{pages}{905--917} (\bibinfo{year}{2008}).

\bibitem{2005ApJ...628..655V}
\bibinfo{author}{{Vikhlinin}, A.} \emph{et~al.}
\newblock \bibinfo{title}{{Chandra Temperature Profiles for a Sample of Nearby
  Relaxed Galaxy Clusters}}.
\newblock \emph{\bibinfo{journal}{\apj}} \textbf{\bibinfo{volume}{628}},
  \bibinfo{pages}{655--672} (\bibinfo{year}{2005}).

\bibitem{2007MNRAS.378..385P}
\bibinfo{author}{{Pfrommer}, C.}, \bibinfo{author}{{En{\ss}lin}, T.~A.},
  \bibinfo{author}{{Springel}, V.}, \bibinfo{author}{{Jubelgas}, M.} \&
  \bibinfo{author}{{Dolag}, K.}
\newblock \bibinfo{title}{{Simulating cosmic rays in clusters of galaxies - I.
  Effects on the Sunyaev-Zel'dovich effect and the X-ray emission}}.
\newblock \emph{\bibinfo{journal}{\mnras}} \textbf{\bibinfo{volume}{378}},
  \bibinfo{pages}{385--408} (\bibinfo{year}{2007}).

\bibitem{2004MNRAS.351..423J}
\bibinfo{author}{{Jubelgas}, M.}, \bibinfo{author}{{Springel}, V.} \&
  \bibinfo{author}{{Dolag}, K.}
\newblock \bibinfo{title}{{Thermal conduction in cosmological SPH
  simulations}}.
\newblock \emph{\bibinfo{journal}{\mnras}} \textbf{\bibinfo{volume}{351}},
  \bibinfo{pages}{423--435} (\bibinfo{year}{2004}).

\bibitem{2004ApJ...606L..97D}
\bibinfo{author}{{Dolag}, K.}, \bibinfo{author}{{Jubelgas}, M.},
  \bibinfo{author}{{Springel}, V.}, \bibinfo{author}{{Borgani}, S.} \&
  \bibinfo{author}{{Rasia}, E.}
\newblock \bibinfo{title}{{Thermal Conduction in Simulated Galaxy Clusters}}.
\newblock \emph{\bibinfo{journal}{\apjl}} \textbf{\bibinfo{volume}{606}},
  \bibinfo{pages}{L97--L100} (\bibinfo{year}{2004}).

\bibitem{2006MNRAS.367..113P}
\bibinfo{author}{{Pfrommer}, C.}, \bibinfo{author}{{Springel}, V.},
  \bibinfo{author}{{En{\ss}lin}, T.~A.} \& \bibinfo{author}{{Jubelgas}, M.}
\newblock \bibinfo{title}{{Detecting shock waves in cosmological smoothed
  particle hydrodynamics simulations}}.
\newblock \emph{\bibinfo{journal}{\mnras}} \textbf{\bibinfo{volume}{367}},
  \bibinfo{pages}{113--131} (\bibinfo{year}{2006}).

\bibitem{2008MNRAS.385.1211P}
\bibinfo{author}{{Pfrommer}, C.}, \bibinfo{author}{{En{\ss}lin}, T.~A.} \&
  \bibinfo{author}{{Springel}, V.}
\newblock \bibinfo{title}{{Simulating cosmic rays in clusters of galaxies - II.
  A unified scheme for radio haloes and relics with predictions of the
  {$\gamma$}-ray emission}}.
\newblock \emph{\bibinfo{journal}{\mnras}} \textbf{\bibinfo{volume}{385}},
  \bibinfo{pages}{1211--1241} (\bibinfo{year}{2008}).

\bibitem{2005RMP...77...207V}
\bibinfo{author}{Voit, G.~M.}
\newblock \bibinfo{title}{Tracing cosmic evolution with clusters of galaxies}.
\newblock \emph{\bibinfo{journal}{Reviews of Modern Physics}}
  \textbf{\bibinfo{volume}{77}}, \bibinfo{pages}{207--258}
  (\bibinfo{year}{2005}).

\bibitem{1994Natur.368..828B}
\bibinfo{author}{{B{\"o}hringer}, H.} \emph{et~al.}
\newblock \bibinfo{title}{{The structure of the Virgo cluster of galaxies from
  Rosat X-ray images}}.
\newblock \emph{\bibinfo{journal}{\nat}} \textbf{\bibinfo{volume}{368}},
  \bibinfo{pages}{828--831} (\bibinfo{year}{1994}).

\bibitem{2008ApJ...683L.107C}
\bibinfo{author}{{Cavagnolo}, K.~W.}, \bibinfo{author}{{Donahue}, M.},
  \bibinfo{author}{{Voit}, G.~M.} \& \bibinfo{author}{{Sun}, M.}
\newblock \bibinfo{title}{{An Entropy Threshold for Strong H{$\alpha$} and
  Radio Emission in the Cores of Galaxy Clusters}}.
\newblock \emph{\bibinfo{journal}{\apjl}} \textbf{\bibinfo{volume}{683}},
  \bibinfo{pages}{L107--L110} (\bibinfo{year}{2008}).

\bibitem{1999A&A...343..420S}
\bibinfo{author}{{Schindler}, S.}, \bibinfo{author}{{Binggeli}, B.} \&
  \bibinfo{author}{{B{\"o}hringer}, H.}
\newblock \bibinfo{title}{{Morphology of the Virgo cluster: Gas versus
  galaxies}}.
\newblock \emph{\bibinfo{journal}{\aap}} \textbf{\bibinfo{volume}{343}},
  \bibinfo{pages}{420--438} (\bibinfo{year}{1999}).

\bibitem{2001A&A...375..770F}
\bibinfo{author}{{Fouqu{\'e}}, P.}, \bibinfo{author}{{Solanes}, J.~M.},
  \bibinfo{author}{{Sanchis}, T.} \& \bibinfo{author}{{Balkowski}, C.}
\newblock \bibinfo{title}{{Structure, mass and distance of the Virgo cluster
  from a Tolman-Bondi model}}.
\newblock \emph{\bibinfo{journal}{\aap}} \textbf{\bibinfo{volume}{375}},
  \bibinfo{pages}{770--780} (\bibinfo{year}{2001}).

\bibitem{1999elss.conf..296T}
\bibinfo{author}{{Tully}, B.} \& \bibinfo{author}{{Shaya}, E.}
\newblock \bibinfo{title}{{Antibiasing: high mass-to-light ratios in dense
  clusters}}.
\newblock In \bibinfo{editor}{{Banday}, A.~J.}, \bibinfo{editor}{{Sheth},
  R.~K.} \& \bibinfo{editor}{{da Costa}, L.~N.} (eds.)
  \emph{\bibinfo{booktitle}{Evolution of Large Scale Structure : From
  Recombination to Garching}}, \bibinfo{pages}{296} (\bibinfo{year}{1999}).

\bibitem{2004A&A...413...17P}
\bibinfo{author}{{Pfrommer}, C.} \& \bibinfo{author}{{En{\ss}lin}, T.~A.}
\newblock \bibinfo{title}{{Constraining the population of cosmic ray protons in
  cooling flow clusters with {$\gamma$}-ray and radio observations: Are radio
  mini-halos of hadronic origin?}}
\newblock \emph{\bibinfo{journal}{\aap}} \textbf{\bibinfo{volume}{413}},
  \bibinfo{pages}{17--36} (\bibinfo{year}{2004}).

\bibitem{2001MNRAS.328L..37A}
\bibinfo{author}{{Allen}, S.~W.}, \bibinfo{author}{{Schmidt}, R.~W.} \&
  \bibinfo{author}{{Fabian}, A.~C.}
\newblock \bibinfo{title}{{The X-ray virial relations for relaxed lensing
  clusters observed with Chandra}}.
\newblock \emph{\bibinfo{journal}{\mnras}} \textbf{\bibinfo{volume}{328}},
  \bibinfo{pages}{L37--L41} (\bibinfo{year}{2001}).

\bibitem{2007ApJ...668....1N}
\bibinfo{author}{{Nagai}, D.}, \bibinfo{author}{{Kravtsov}, A.~V.} \&
  \bibinfo{author}{{Vikhlinin}, A.}
\newblock \bibinfo{title}{{Effects of Galaxy Formation on Thermodynamics of the
  Intracluster Medium}}.
\newblock \emph{\bibinfo{journal}{\apj}} \textbf{\bibinfo{volume}{668}},
  \bibinfo{pages}{1--14} (\bibinfo{year}{2007}).

\bibitem{1999Natur.397..135P}
\bibinfo{author}{{Ponman}, T.~J.}, \bibinfo{author}{{Cannon}, D.~B.} \&
  \bibinfo{author}{{Navarro}, J.~F.}
\newblock \bibinfo{title}{{The thermal imprint of galaxy formation on X-ray
  clusters}}.
\newblock \emph{\bibinfo{journal}{\nat}} \textbf{\bibinfo{volume}{397}},
  \bibinfo{pages}{135--137} (\bibinfo{year}{1999}).

\bibitem{1999MNRAS.306..857C}
\bibinfo{author}{{Crawford}, C.~S.}, \bibinfo{author}{{Allen}, S.~W.},
  \bibinfo{author}{{Ebeling}, H.}, \bibinfo{author}{{Edge}, A.~C.} \&
  \bibinfo{author}{{Fabian}, A.~C.}
\newblock \bibinfo{title}{{The ROSAT Brightest Cluster Sample - III. Optical
  spectra of the central cluster galaxies}}.
\newblock \emph{\bibinfo{journal}{\mnras}} \textbf{\bibinfo{volume}{306}},
  \bibinfo{pages}{857--896} (\bibinfo{year}{1999}).

\end{thebibliography}

\end{document}